\documentclass{aa}

\usepackage{graphicx}
\usepackage{txfonts}
\usepackage{float}
\usepackage{mathtools}
\usepackage{subcaption}
\usepackage[version=4]{mhchem}
\usepackage{gensymb}
\usepackage{tikzsymbols}
\usepackage{rotating}
\usepackage{multirow}
\usepackage{makecell}
\usepackage{mleftright}
\usepackage{ulem}

     \def\aap{A\&A}
     \def\apjl{ApJL}
     \def\apjs{ApJS}

\newcommand{\lc}[1]{\textcolor{magenta}{#1}}

\newcommand{\ds}[1]{\textcolor{black}{#1}}

\begin{document} 

   \title{WASP-39b: exo-Saturn with patchy cloud composition, moderate metallicity, and underdepleted S/O}

   \author{
   Ludmila Carone\inst{1}
   \and
   David A. Lewis\inst{1,2}
   \and 
   Dominic Samra\inst{1}
   \and 
   Aaron D. Schneider\inst{3,4}
  \and
   Christiane Helling\inst{1,2}
   } 
   
   \institute{
      Space Research Institute, Austrian Academy of Sciences, Schmiedlstrasse 6, A-8042 Graz, Austria\\
      \email{Ludmila.Carone@oeaw.ac.at}
            \and
            Institute for Theoretical Physics and Computational Physics, Graz University of Technology, Petersgasse 16
8010 Graz
\and
Institute of Astronomy, KU Leuven, Celestijnenlaan 200D, 3001, Leuven, Belgium
\and
Center for ExoLife Sciences, Niels Bohr Institute, Øster Voldgade 5, 1350 Copenhagen, Denmark
             }

      \date{\today, Received September 15, 2996; accepted March 16, 2997}

\abstract
    {WASP-39b is one of the first extrasolar giant gas planet that have been observed within the JWST ERS program. Data interpretation by different retrieval approaches diverge.
    Fundamental properties that may enable the link to exoplanet formation differ amongst methods, for example metallicity and mineral ratios. The retrieval of these values impact the results for individual element abundances as well as the presence or absence of chemical tracer species. This challenge is eminent for all JWST targets.}
    {The formation of clouds in the atmosphere of WASP-39b is explored to investigate how inhomogeneous cloud properties (particle sizes, material composition, opacity) may be for this intermediately warm gaseous exoplanet.}
     {1D profiles extracted from the 3D GCM expeRT/MITgcm results are used as input for a kinetic. non-equilibrium cloud model. Resulting cloud particle sizes, number densities and material volume fractions are the input for opacity calculations. }
    {WASP-39b's atmosphere has a comparable day-night temperature median with sufficiently low temperatures that clouds may form globally.
    The presence of clouds on WASP-39b can explain observations without resorting to a high ($>100\times$ solar) metallicity atmosphere for a reduced vertical mixing efficiency. The assessment of mineral ratios shows an under-depletion of S/O due to condensation compared to C/O, Mg/O, Si/O, Fe/O ratios. Vertical patchiness due to heterogeneous cloud composition challenges simple cloud models. An equal mixture of silicates and metal oxides is expected to characterise the cloud top. Further, optical properties of Fe and Mg silicates in the mid-infrared differ significantly which will impact the interpretation of JWST observations.}
    {
    WASP-39b's atmosphere contains clouds and the underdepletion of S/O by atmospheric condensation processes suggest the use of sulphur gas species as a possible link to primordial element abundances. Over-simplified cloud models do not capture the complex nature of mixed-condensate clouds in exoplanet atmospheres. The clouds in the observable upper atmosphere of WASP-39b are 
    a mixture of different silicates and metal oxides. The use of constant particles sizes and/or one-material cloud particles alone to interpret spectra  may not be sufficient to capture the full complexity available through JWST observations.} 

  \maketitle
  
\section{Introduction}\label{s:intro}

WASP-39b was one of the first extrasolar planets for which the James Webb Space Telescope (JWST) observations were released to the community. Similar to WASP-96b, it was observed in transmission using the NIRSpec instrument as part of the Early Release Science Programme (ERS) \citep{JWST_39b_CO2}. WASP-39b has a mass of $M_p=0.28~{\rm M_{Jup}}$, a radius of $R_p=1.27~{\rm R_{Jup}}$, an equilibrium temperature of $T_{\rm eq}\sim1100$ K, and orbits a G-type star with a period of 4.055 days \citep{39b_discovery}. WASP-96b has a mass of 0.48~$\pm$~0.03~M$_{\rm Jup}$, a radius of 1.2~$\pm$~0.06~R$_{\rm Jup}$, an equilibrium temperature $T_{\rm eq}\sim1300$ K, and orbits a G-type star with a period 3.4 days \citep{14HeAnCa}.


\begin{figure}[!ht]
    \includegraphics[width=0.93\linewidth]{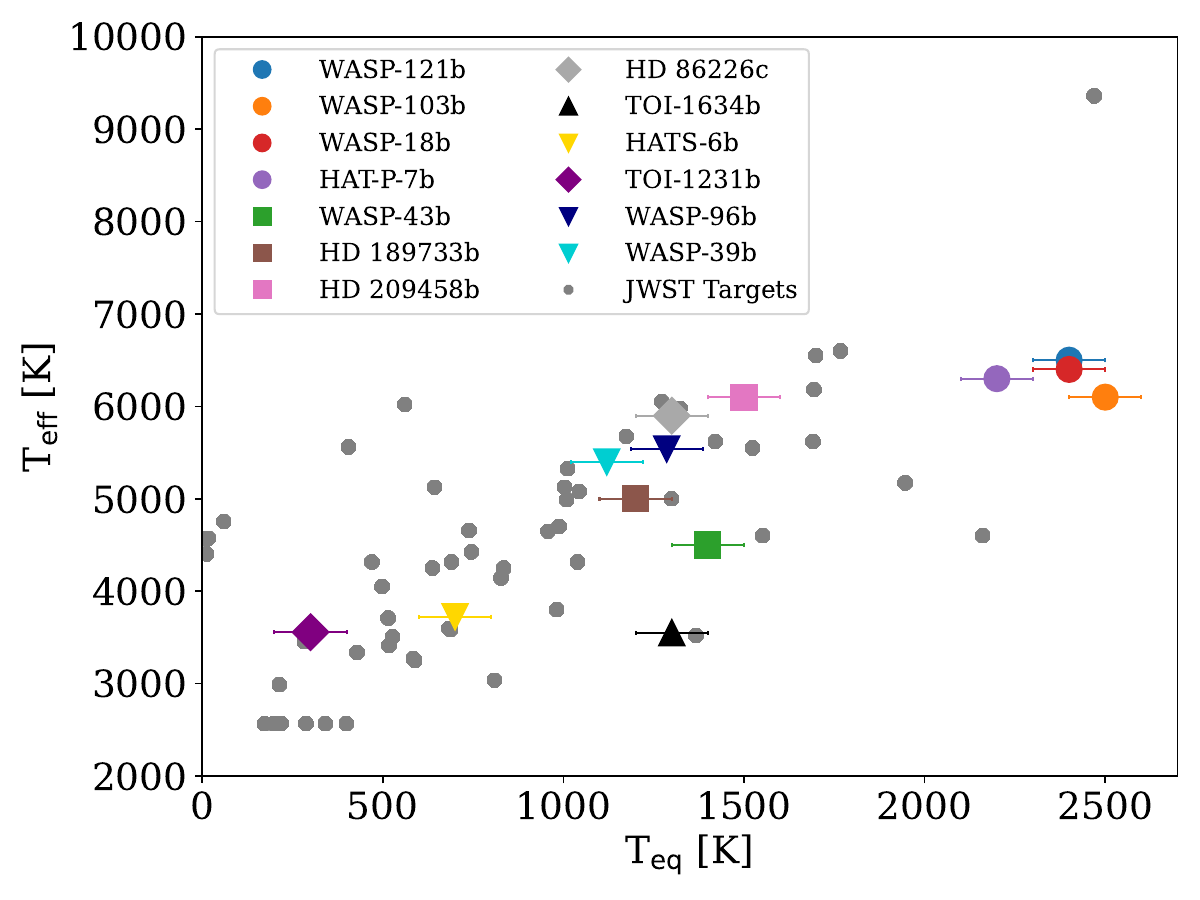}\\
    \includegraphics[width=0.93\linewidth]{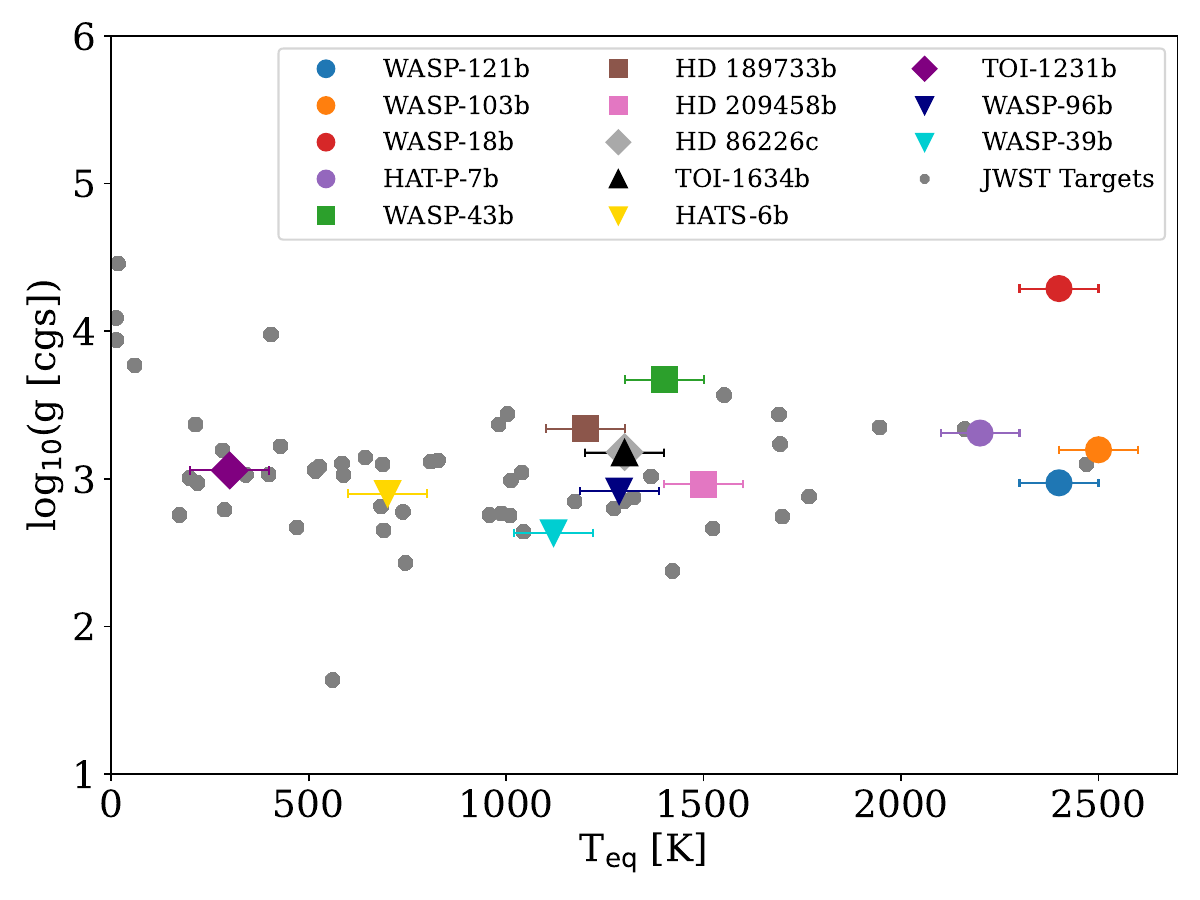}\\
    \includegraphics[width=0.93\linewidth]{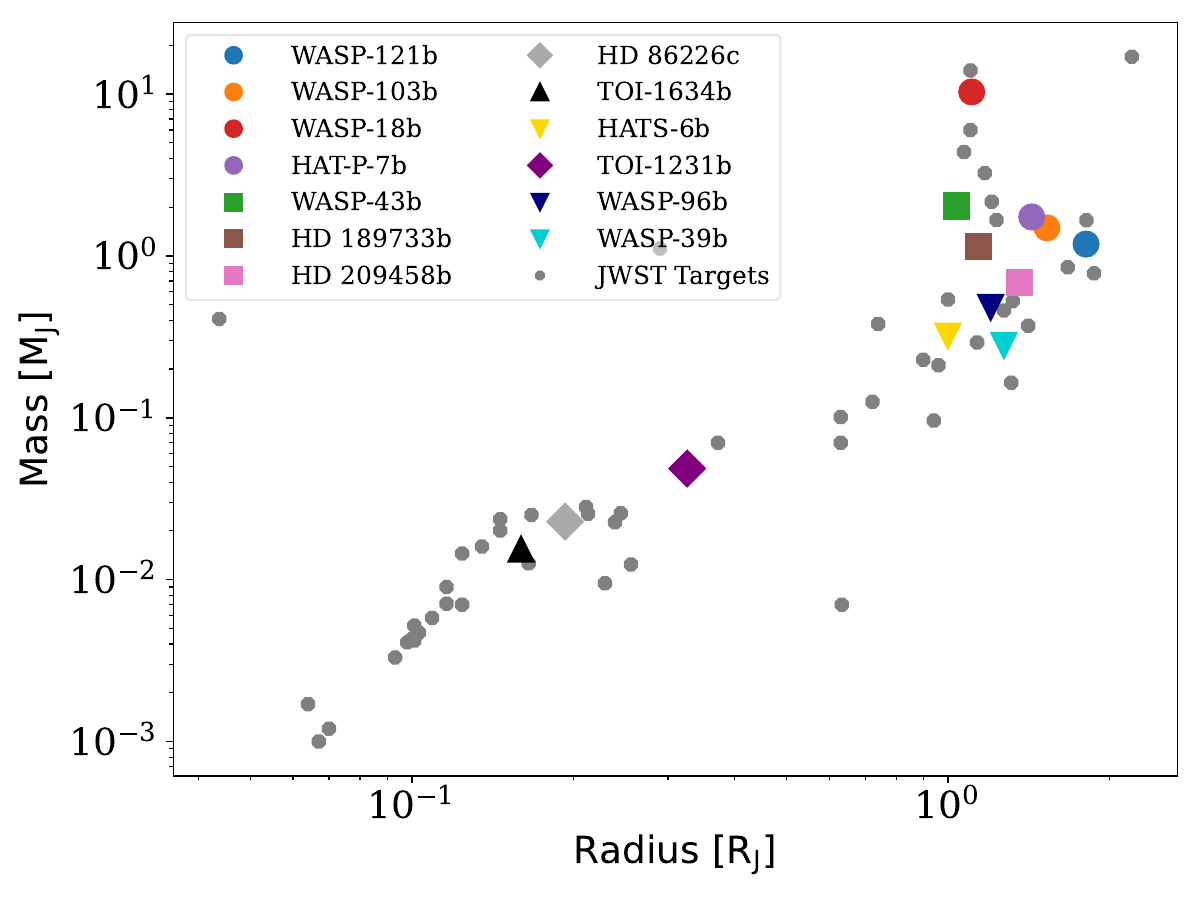}
    \caption{The JWST ESR targets WASP-39b (light-blue triangle) and WASP-96b (purple triangle) comfortably share the T$_{\rm eq}$, log(g) and T$_{\rm eff}$ parameter ranges with the exoplanet subclass of hot Jupiters like HD\,189773b. Comparing known exoplanets in the (R$_{\rm P}$, M$_{\rm P}$)-plane shows both sharing the parameter space with the warm Saturn HATS-6b. The grey symbols indicate the presently known JWST exoplanet targets.}
    \label{fig:global_paramters}
\end{figure}


The WASP-39b JWST ERS observation, covering the $3-5~{\rm\mu m}$ wavelength range, suggest a strong \ce{CO2} absorption feature at $4.3~{\rm \mu m}$ \citep{JWST_39b_CO2}. Further observations with JWST instruments have also reported the detection of \ce{CO2} \citep{https://doi.org/10.48550/arxiv.2211.10487,https://doi.org/10.48550/arxiv.2211.10489}. Previous observations with HST/WFC3, HST/STIS, and the VLT, as well as recent observations with several instruments on JWST have ascertained the presence of water, sodium, and potassium in the atmosphere of WASP-39b \citep{2016ApJ...827...19F,Nikolov_2016,2018AJ....155...29W,https://doi.org/10.48550/arxiv.2211.10487,https://doi.org/10.48550/arxiv.2211.10488,https://doi.org/10.48550/arxiv.2211.10489,Feinstein2022_JWST_NIRISS}. Photochemical production has been proposed as the source of the \ce{SO2} detection in the WASP-39b atmosphere \citep{https://doi.org/10.48550/arxiv.2211.10490}. The observability of spectral features, in particular the pressure broadened Na and K lines lead 
\citet{2016Natur.529...59S}  to suggest that  the atmosphere of WASP–39b may be relatively cloud free. \citet{2016ApJ...827...19F} use HST/STIS in combination with the Spitzer/IRAC photometry to suggest a clear-sky WAPS-39b. Their best fist to the data suggest a \ce{H2} dominated atmosphere with either a clear atmosphere of $0.1\times$ to solar metallicity, or a weak haze layer with solar abundances. \cite{2018AJ....155...29W} suggest asymmetric limbs as a possibilities to fit their observations of WASP-39b. Analysis of observations with JWST/NIRCam, JWST/NIRSpec G395H, JWST/NIRSpec PRISM, and JWST/NIRISS, favour a cloudy atmosphere \citep{https://doi.org/10.48550/arxiv.2211.10487,https://doi.org/10.48550/arxiv.2211.10488,https://doi.org/10.48550/arxiv.2211.10489,Feinstein2022_JWST_NIRISS}. Retrieval studies often use assumptions about atmospheric clouds, including clouds being homogeneous in wavelength (`grey' clouds), in composition as well as in particle size \citep{Barstow2017,2018AJ....155...29W}. These assumptions could be problematic as evidenced by disparate results for WASP-39b and planets of similar mass and temperature ($T_{\rm eq} \lesssim 1300$~K). 




\cite{2021A&A...646A.168C} studied the atmosphere of WASP-117b which is similar to WASP-39b in mass and radius. A muted water feature was detected in WASP-117b using HST/WFC3. There was no conclusive ($>3\sigma$) detection of Na and K in the high-resolution VLT/ESPRESSO data. It was shown that the retrieval process can lead to bifurcating results since two models would be consistent with the observations. The first, a 1D, isothermal atmosphere model with a uniform cloud deck and in equilibrium chemistry suggests 
preference for a high atmospheric metallicity [Fe/H] = $2.58\pm0.26$ but clear skies in Bayesian analysis. The data are, however, also consistent with a lower metallicity $\sim0.37\times \varepsilon_{\rm solar}$ ($\varepsilon_{\rm solar}$ - solar metallicity),  [Fe/H] < 1.75 and a cloud deck between 10$^{-2.2}\,\ldots\,10^{-5.1}$ bar.

\citet{2018AJ....155...29W} report a cloud-free and very high metallicity of more than $100 \times \varepsilon_{\rm solar}$ for WASP-39b based on a combination of HST/WFC3, VLT/FORS2, HST/STIS and \textit{Spitzer} observations. This result is degenerate with C/O ratio and 
cloud coverage. As was found for WASP-117b, a lower metallicity - more in line with Solar System Saturn of $\sim10\times\varepsilon_{\rm solar}$ \citep{FLETCHER2011510,Atreya2016} - can be fitted to the data if a cloud coverage is allowed within the retrieval approach. Further, asymmetrically cloudy limbs would mimick a high metallicity atmosphere if 1D retrieval models are assumed \citep{LineParmentier2016}. 


Statistical trends in exoplanet atmosphere metallicity suggest that also exoplanets follow a similar metallicity-mass trend as known for the gas planets in the Solar System \citep{Chachan2019,Welbanks2019}: lower mass gas planets are more metal-rich than more massive gas planet. This trend is also in-line with planet formation models for planets that have migrated in the proto-planetary disk \citep{Schneider2021b,Knierim2022}. Thus, for Saturn-mass objects like WASP-39b, a moderately increased atmosphere metallicity ($10 \times \mathrm{solar}$) similar to Saturn in the Solar System can be expected 
\citep{Thorngren2019,2022arXiv220504100G}.

This paper addresses the question of cloud formation in the atmosphere of WASP-39b by applying a microphysical model that self-consistently treats the formation of cloud condensation nuclei, their growth to macroscopic particles from multiple condensing species, element depletion, and the feedback of gravitational settling on these processes.

Combining the cloud model with the output of a 3D General Circulation Model (GCM) for the 3D thermodynamic atmosphere structure, shows that the cloud properties vary are vertically heterogeneous however they maintain a relatively homogeneous horizontal distribution, in contrast to ultra-hot Jupiters, such as HAT-P-7b or WASP-121b \citep{Helling2021}. The dominant gas phase species after condensation are explored showing that \ce{H2S} is less strongly affected by cloud formation and the local thermodynamics than \ce{CO2}. The impact of atmospheric metallicity on cloud formation is assessed, illustrating that high metallicity leads to an increased cloud mass. The potential for using the S/O ratio as a link to planet formation processes is discussed. The potential of using the complex cloud model to link to the currently available observational data for WASP-39b is explored. For WASP-39b, similar to WASP-96b, a reduced mixing efficiency in the cloud model is required to produce a cloud deck between $p_{\rm gas} = 10^{-2}~{\rm and}~5\times 10^{-3}$ bar as implied by the observations. The potentially misleading effects of over-simplified cloud parameterisations (for example, constant particle sizes or homogeneous cloud particle composition) is demonstrated.



\begin{figure*}
\includegraphics[page=1,width=0.55\linewidth]{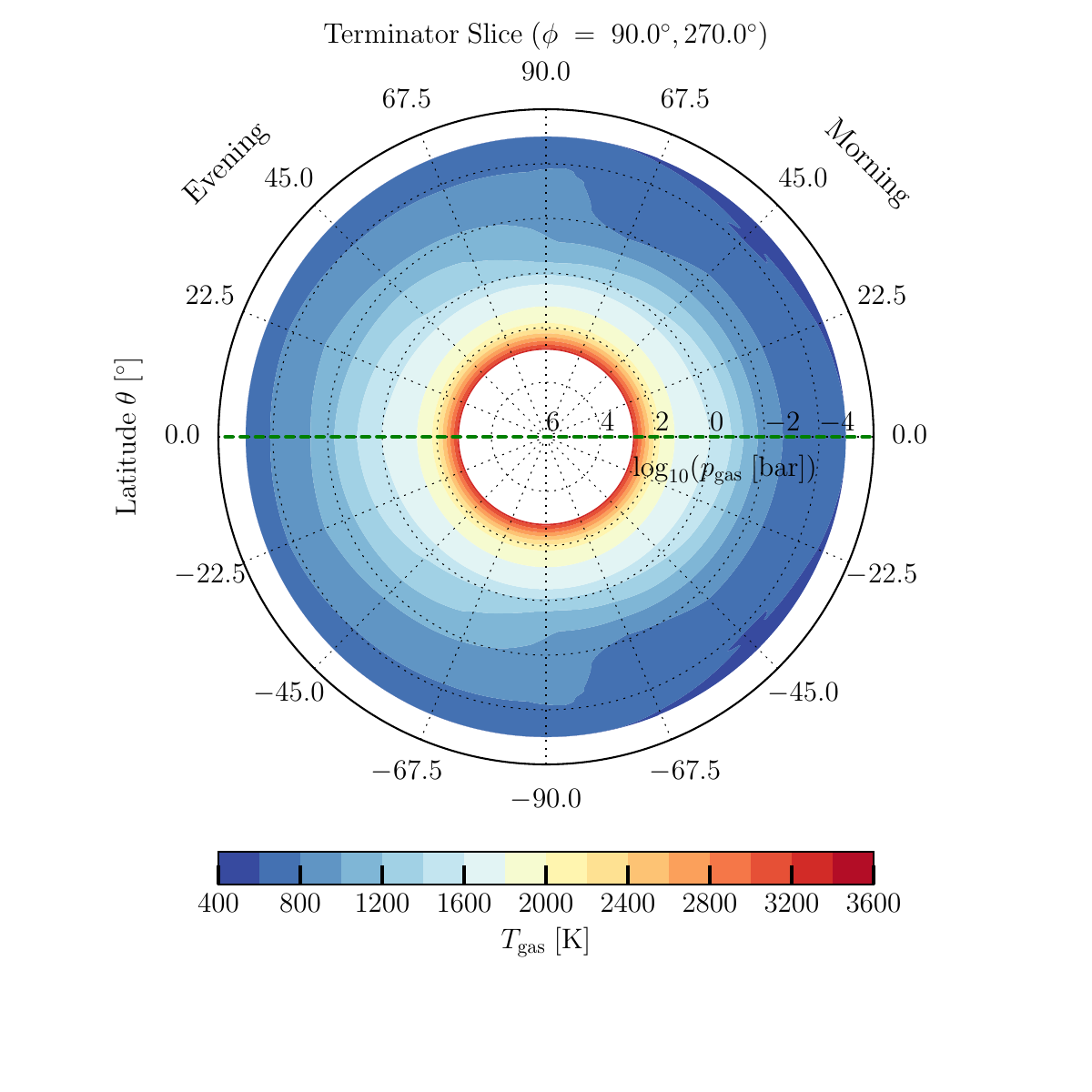}
\hspace*{-1.5cm}\includegraphics[page=4,width=0.55\linewidth]{Figures/Slice_Plots_WASP039b_termonly_SAgs.pdf}\\*[-1cm]
\includegraphics[page=2,width=0.55\linewidth]{Figures/Slice_Plots_WASP039b_termonly_SAgs.pdf}
\hspace*{-1.5cm}\includegraphics[page=3,width=0.55\linewidth]{Figures/Slice_Plots_WASP039b_termonly_SAgs.pdf}
\caption{WASP-39b 2D slices showing  atmosphere and cloud structure terminator maps. {\bf Top Left:} Local atmospheric gas temperature and gas pressure (T$_{\rm gas}$, p$_{\rm gas}$). {\bf Top Right:} Total nucleation rate, $J_*=\sum_i J_{\rm i}$ [cm$^{-3}$ s$^{-1}$] (i=TiO$_2$, SiO, NaCl, KCl). {\bf Bottom left:} Dust-to-gas mass ratio $\rho_{\rm d}/ \rho$. {\bf Bottom right:} Surface averaged mean cloud particle radius $\langle a \rangle_{A}$ [$\mu$m]. }
  \label{TpNuc}
\end{figure*}

\begin{figure*}
\includegraphics[page=1,width=0.55\linewidth]{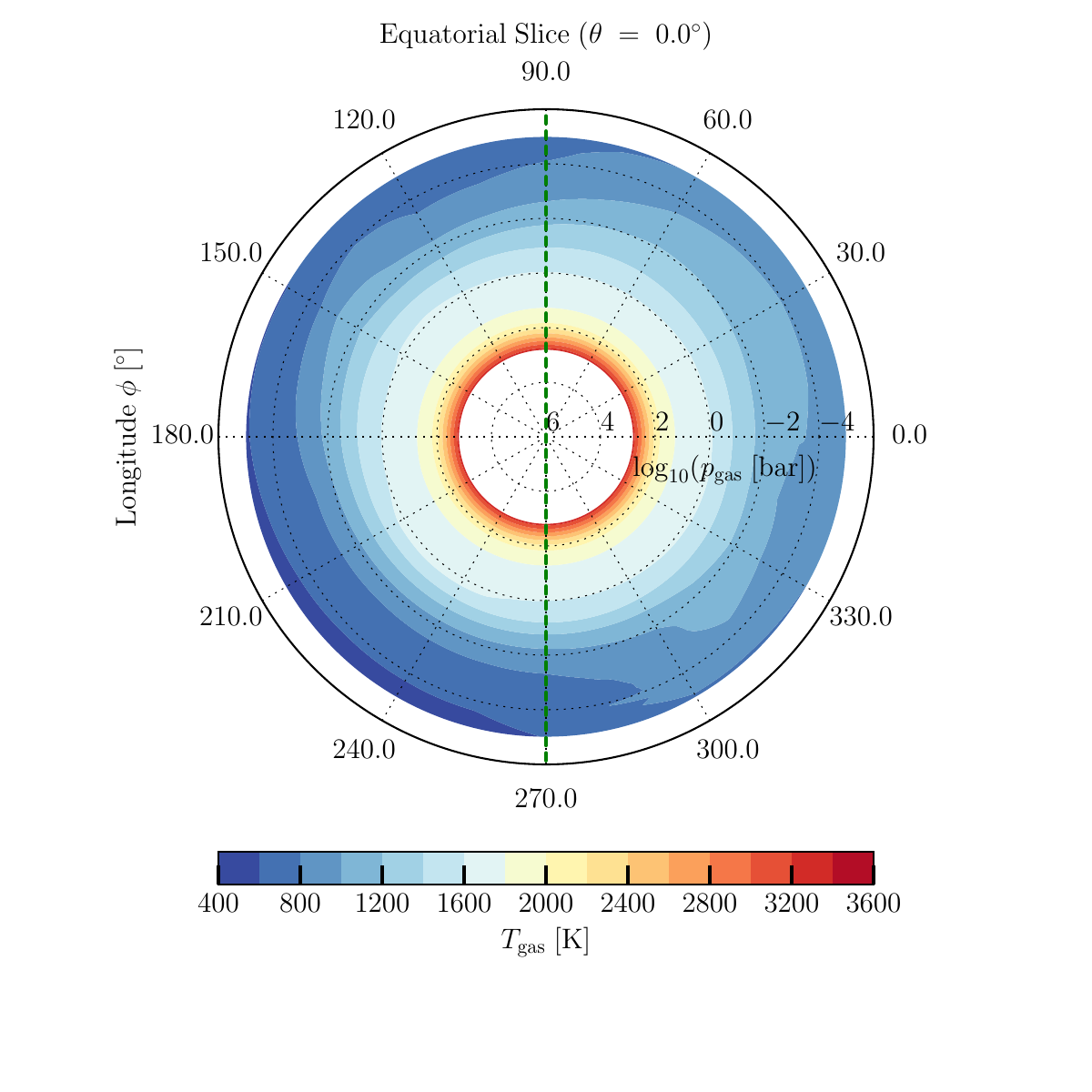}
\hspace*{-1.5cm}\includegraphics[page=4,width=0.55\linewidth]{Figures/Slice_Plots_WASP039b_eqonly.pdf}\\*[-1cm]
\includegraphics[page=2,width=0.55\linewidth]{Figures/Slice_Plots_WASP039b_eqonly.pdf}
\hspace*{-1.5cm}\includegraphics[page=3,width=0.55\linewidth]{Figures/Slice_Plots_WASP039b_eqonly.pdf}
\caption{WASP-39b 2D slices showing  atmosphere and cloud structure equatorial maps. {\bf Top Left:} Local atmospheric gas temperature and gas pressure (T$_{\rm gas}$, p$_{\rm gas}$). {\bf Top Right:} Total nucleation rate, $J_*=\sum_i J_{\rm i}$ [cm$^{-3}$ s$^{-1}$] (i=TiO$_2$, SiO, NaCl, KCl). {\bf Bottom left:} Dust-to-gas mass ratio $\rho_{\rm d}/ \rho$. {\bf Bottom right:} Surface averaged mean cloud particle radius $\langle a \rangle_{A}$ [$\mu$m]. }
  \label{TpNuc_eq}
\end{figure*}

\begin{figure*}
\hspace*{-0.7cm}  
    \includegraphics[width=0.5\linewidth]{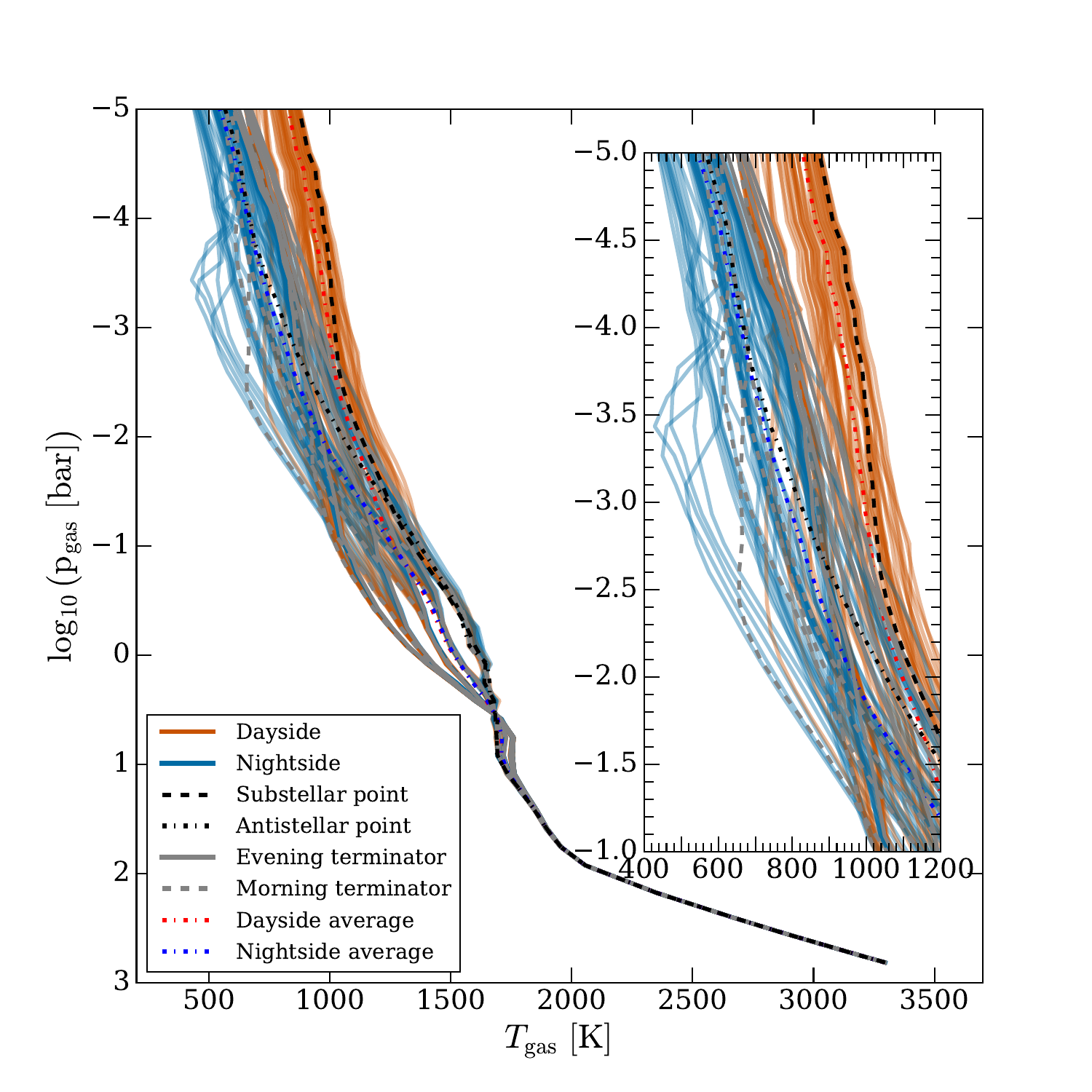}  
    \includegraphics[width=0.5\linewidth]{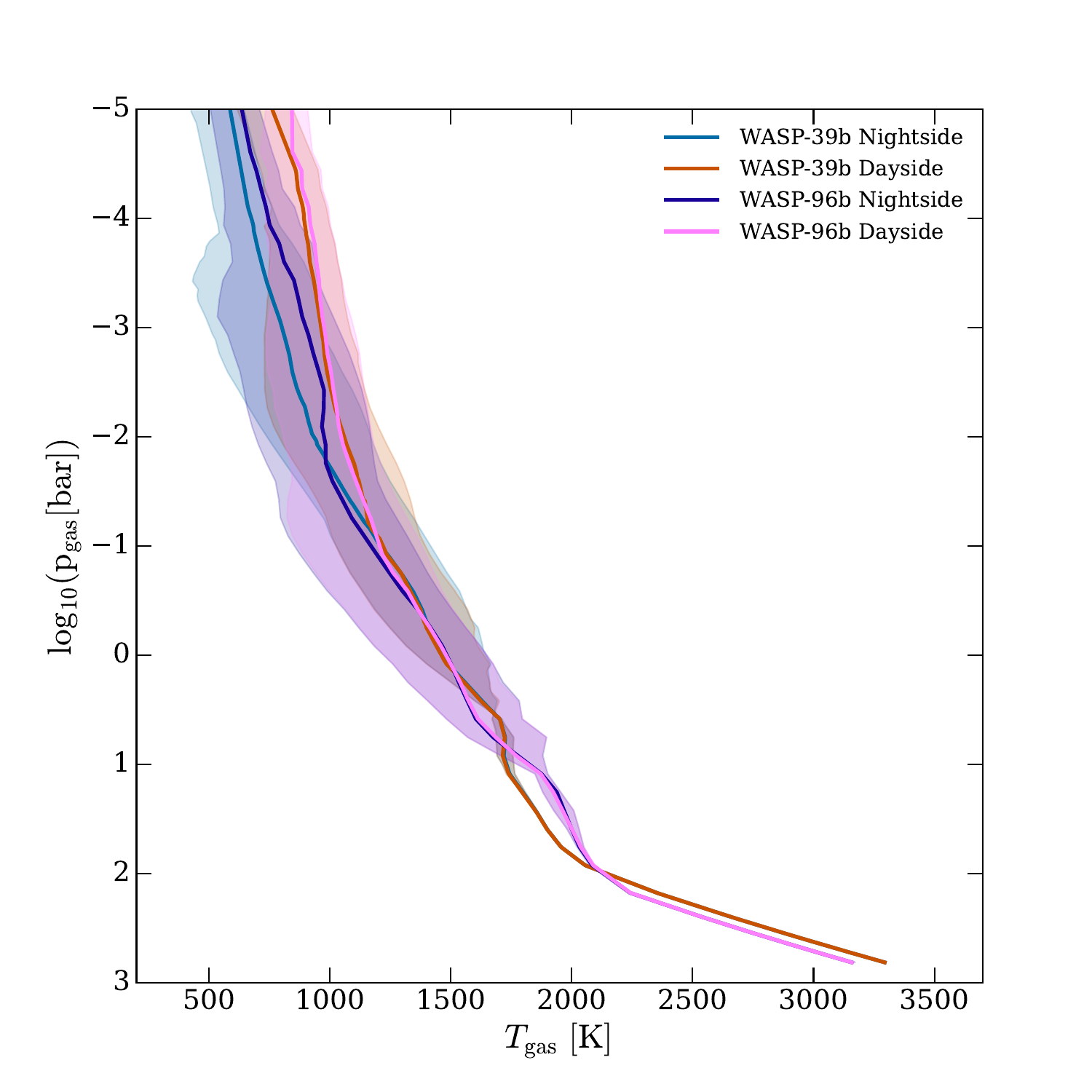}  
    \caption{ (T$_{\rm gas}$, p$_{\rm gas}$) - profiles extracted from the 3D GCM in the non-grey version presented in \cite{2022arXiv220209183S}. \textbf{Left:} The 120 1D profiles extracted from a WASP-39b 3D GCM. The inset highlights the region of the nightside where Rossby vortices form at $\theta\sim\pm68^{\circ}$ (see also Fig.~\ref{fig:gcm_maps}) \textbf{Right:} WASP-39b and WASP-96b  day- and nightside median (T$_{\rm gas}$, p$_{\rm gas}$)  profiles with maximum and minimum temperature envelopes. The dayside and the nightside of WASP-96b are on average slightly hotter than WASP-39b. }
    \label{fig:1dprofiles}
\end{figure*}

\begin{figure*}
    \centering
    \includegraphics[width=0.45\linewidth]{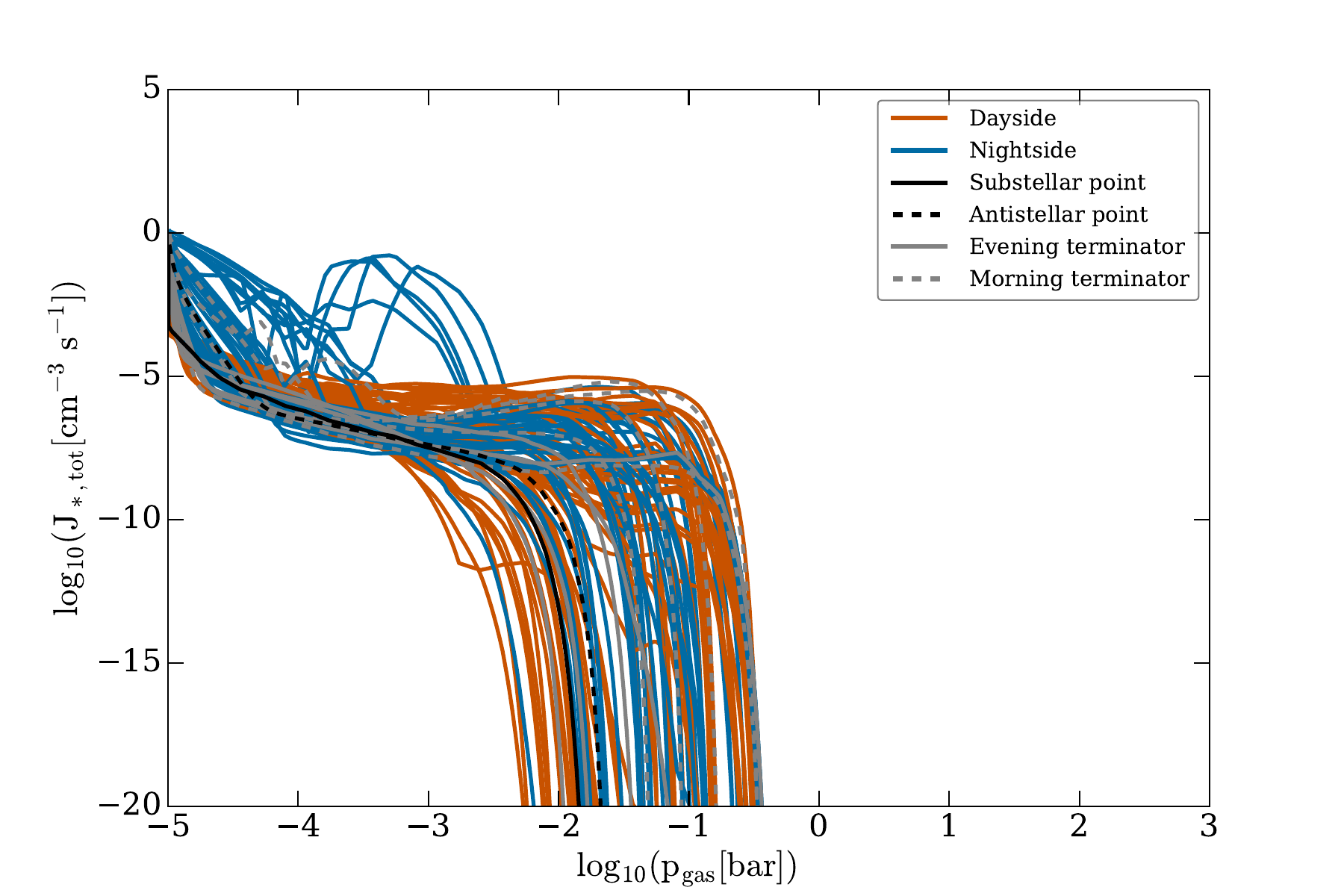}
    \includegraphics[width=0.45\linewidth]{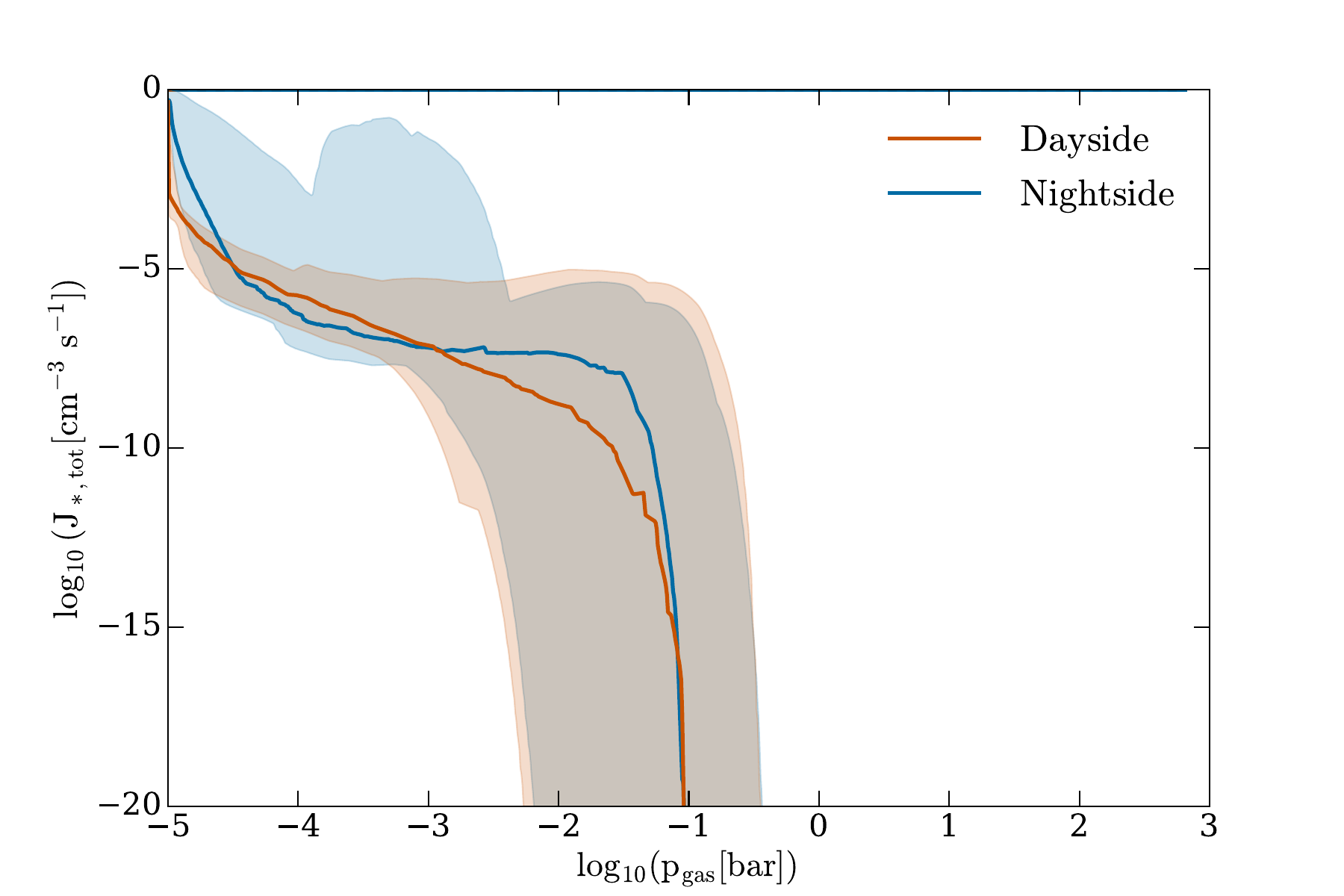}\\
    \includegraphics[width=0.45\linewidth]{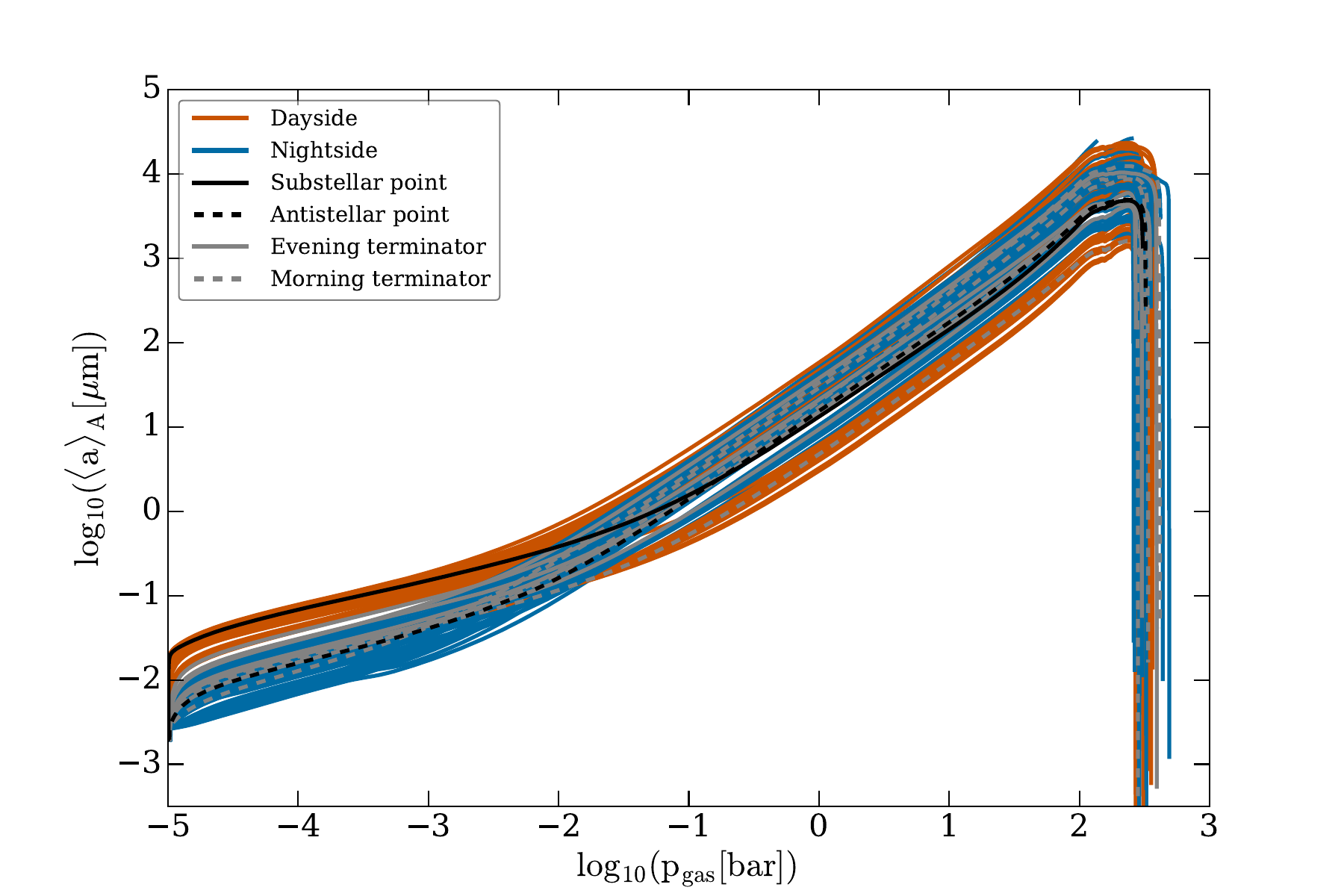}
    \includegraphics[width=0.45\linewidth]{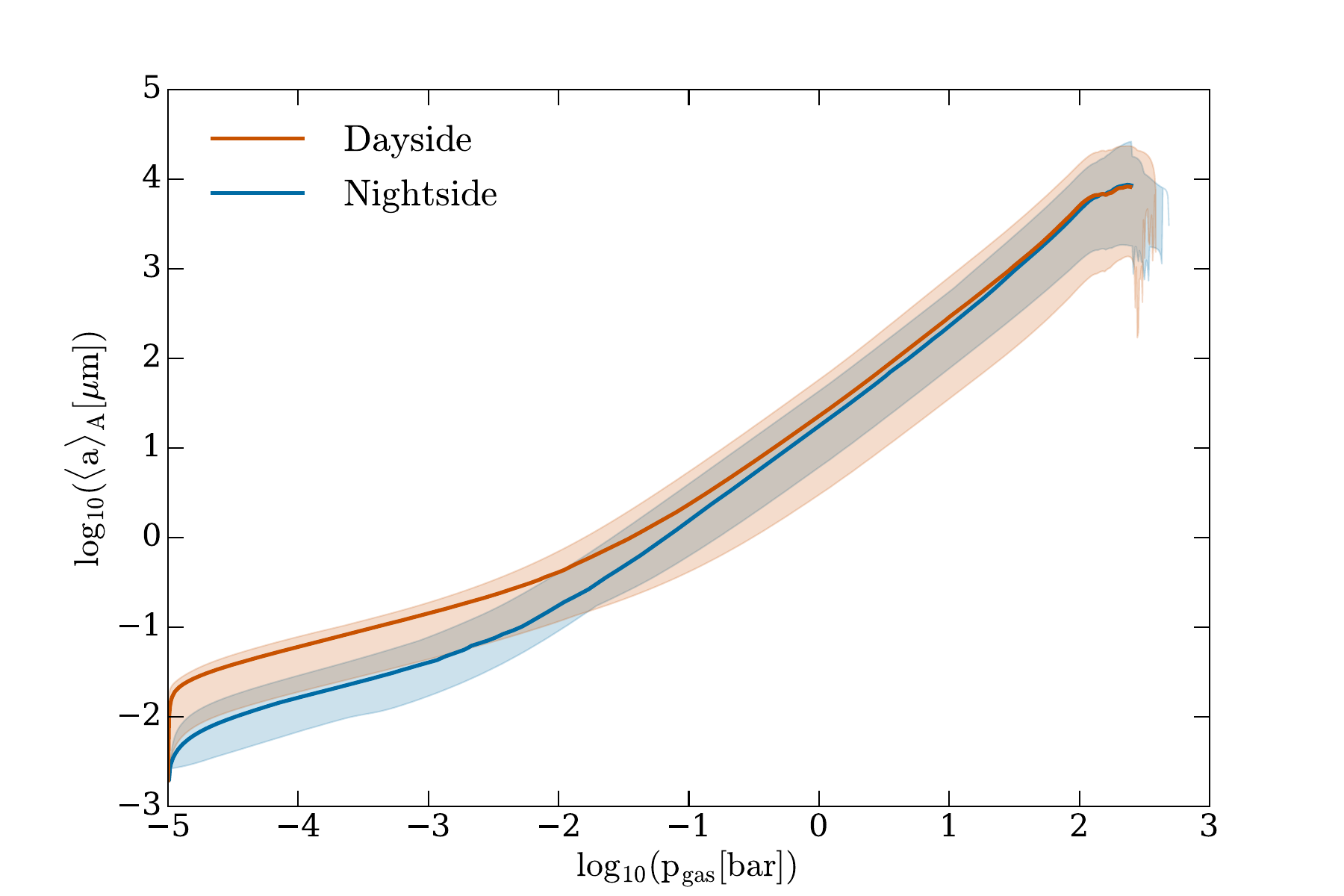}\\
    \includegraphics[width=0.45\linewidth]{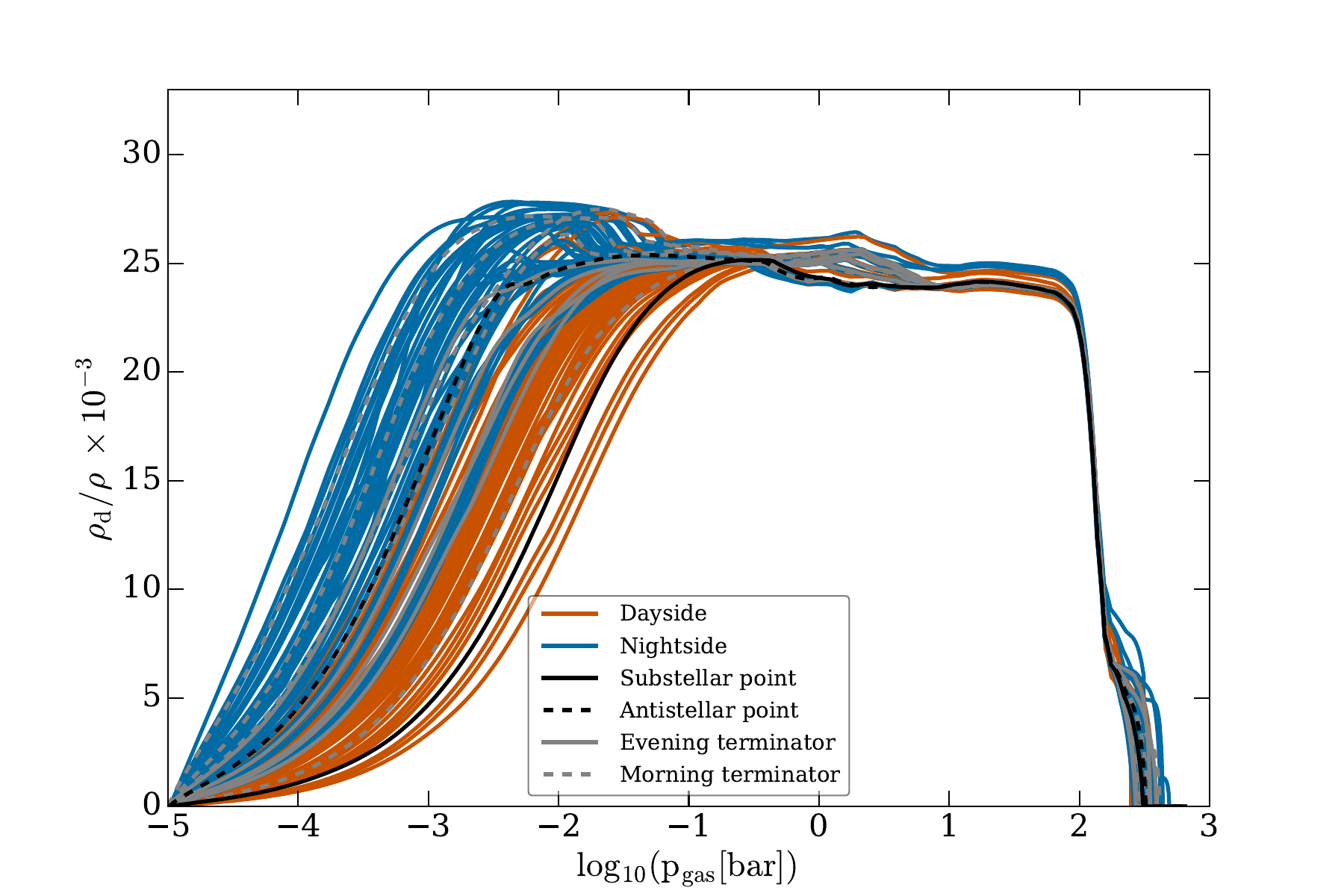}
    \includegraphics[width=0.45\linewidth]{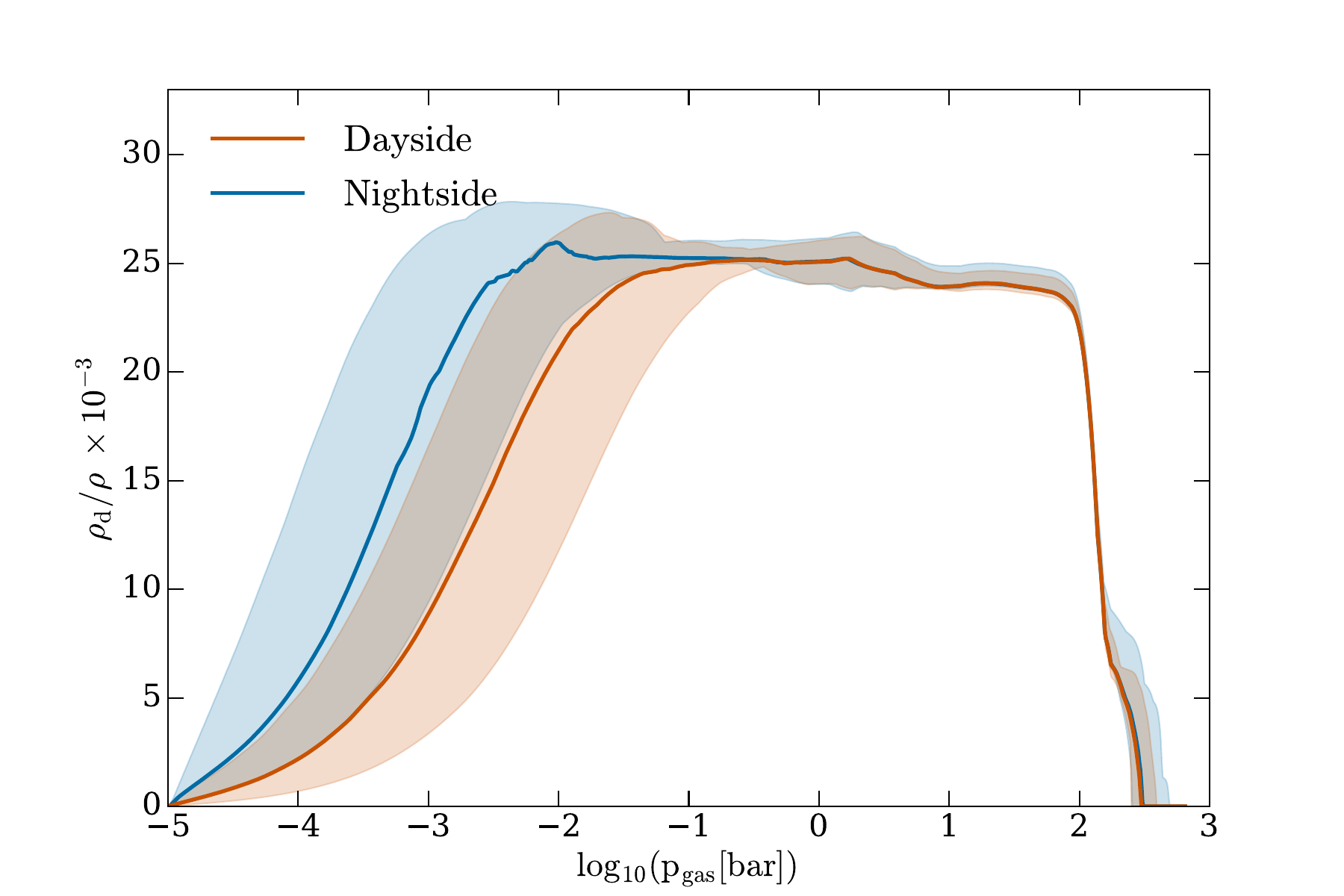}

    \caption{Microphysical cloud properties of WASP-39 b. \textbf{Left column:} Individual 1D profiles which describe the local properties of the cloud. \textbf{Right column:} Median dayside and nightside profiles with maximum and minimum planet wide value envelopes. {\bf Top row:} Total nucleation rate $J_{\rm *,~tot}$. {\bf Middle row:} Surface averaged mean cloud particle radius $\langle a \rangle_{A}$. {\bf Bottom row}: Dust-to-gas mass ratio $\rho_{\rm d}/ \rho$.}

    \label{fig:spaghetti_clouds}
\end{figure*}

\section{Approach}\label{s:ap}

The cloud structure on the JWST ERS target WASP-39b is examined by adopting a hierarchical approach similar to works on another JWST ERS target WASP-96b \citep{Samra2022_96b}, the canonical hot Jupiters HD\,189733b and HD\,209458b (\citealt{Lee2015,2016MNRAS.460..855H}), and the ultra-hot Jupiters WASP-18b (\citealt{Helling2019_18b}) and HAT-P-7b (\citealt{2019A&A...631A..79H,Mol2019}). The first modelling step produces a cloud-free 3D GCM representing WASP-39b. 
These results are used as input for the second modelling step, which is a kinetic cloud formation model consistently combined with equilibrium gas-chemistry calculations. 120 1D ($T_{\rm gas}$(z), $p_{\rm gas}$(z), $v_{\rm z}(z)$)-profiles are utilised for WASP-39b similar to our previous works. $T_{\rm gas}$(z) is the local gas temperature [K], $p_{\rm gas}$(z) is the local gas pressure [bar], and $v_{\rm z}$(z) is the local vertical velocity component [cm s$^{-1}$]. 

 This hierarchical approach is limited by not explicitly taking into account the potential effect of horizontal winds on cloud formation. However, processes governing the formation of mineral clouds are mainly determined by local thermodynamic properties which result from the 3D GCM. 
The temperature structure may change if the cloud particle opacity is fully taken into account in the solution of the radiative transfer. This may change the precise location of the cloud in pressure space but not the principle result of clouds forming in WASP-39b. In the following, the individual modelling steps are described in more detail.

\paragraph{3D atmosphere modelling:}
The 3D GCM  expeRT/MITgcm \citep{2019arXiv190413334C, 2021MNRAS.505.5603B} is utilised to model WASP-39b. The code was used by \cite{2022A&A...666L..11S} to demonstrate that inflation of extrasolar giant gas planets is probably not caused by vertically advected heat. The expeRT/MITgcm builds on the dynamical core of MITgcm \citep{Adcroft2004} and has been adapted to model tidally locked gas giants. Recent extensions in \cite{2022arXiv220209183S} include non-grey radiative transfer coupling. The model parameters used for the GCM, representative of the hot Saturn WASP-39b, are: $R_{\rm p} = 9.07\times10^{9}\,{\rm cm}$, $P_{\rm rot} = 4.06\,{\rm days}$, $\log_{10}(g\,{\rm [cm\,s^{-2}]}) = 2.63$, and the substellar point irradiation temperature $T_{\rm irr} = 1580\,{\rm K}$ or an equilibrium temperature assuming full heat distribution and zero albedo of  $T_{\rm eq} = 1117.4\,{\rm K}$ (Eq. 20 \citealt{2022arXiv220209183S}). The atmosphere of WASP-39b is assumed to have a metallicity of $10\times\varepsilon_{\rm solar}$. The model is run for $700\,{\rm days}$. 
Additional details for the 3D GCM setup can be found in Table~\ref{tab:GCM_params}.

\paragraph{Kinetic cloud formation:}  The kinetic cloud formation model (nucleation, growth, evaporation, gravitational settling, element consumption and replenishment) and equilibrium gas-phase calculations are applied following a similar approach as taken in \cite{Helling2022}. The undepleted gas phase element abundances are set to $10\times\varepsilon_{\rm solar}$ by increasing all element abundances of metals. A constant mean molecular weight\footnote{This value is derived from the specific gas constant $R$ used in the GCM (Table~\ref{tab:GCM_params}).} of $\mu = 2.4$ is used to reflect the higher value expected compared to a solar abundance H$_2$/He dominated atmosphere. This constant value is a reasonable assumption given that the thermodynamic structure of the atmosphere does not cause the gas composition to deviate from a H$_2$-dominated gas. 

In total, 31 ODEs are solved to describe the formation of cloud condensation nuclei ($J_*(z) = \sum_i J_{\rm i}$, i=TiO$_2$, SiO, KCl, NaCl) which grow to macroscopic sized cloud particles comprised of a difference condensate species which change depending on the local atmospheric gas temperature and gas pressure. The 16 condensate species considered are \ce{TiO2}[s], \ce{Mg2SiO4}[s], \ce{MgSiO3}[s], MgO[s], SiO[s], \ce{SiO2}[s], Fe[s], FeO[s], FeS[s], \ce{Fe2O3}[s], \ce{Fe2SiO4}[s], \ce{Al2O3}[s], \ce{CaTiO3}[s], \ce{CaSiO3}[s], \ce{NaCl}[s], \ce{KCl}[s] which form from 11 elements (Mg, Si, Ti, O, Fe, Al, Ca, S, K, Cl, Na) by 132 surface reactions. The vertical mixing is based on $v_{\rm z}$(z) and calculated according to Appendix B.1. in \cite{Helling2022} mimicking a diffusive flux across computational cells.


\paragraph{Deriving cloud properties:} In Section~\ref{ss:cloudglobal}, the clouds are quantified in terms of the surface averaged mean particle size $\langle a \rangle_{A}$ [$\mu$m] of the particles that make up the clouds, their material volume fractions $V_{\rm s}/V_{\rm tot}$, and the dust-to-gas mass ratio, $\rho_{\rm d}/\rho$ which represents the cloud mass load. The surface averaged mean particle size $\langle a \rangle_{A}$ is defined as
\begin{equation}
    \centering
    \langle a\rangle_{\rm A} = \sqrt[3]{\frac{3}{4\pi}}\, \frac{L_3}{L_2},
    \label{eq:surf_size}
\end{equation}
where $L_{2}$ and $L_{3}$ are the second and third dust moments \citep[Eq.A.1 in][]{helling2020mineral}. \\

In Section~\ref{sec:column_int_props}, column integrated properties are discussed. As outlined in previous works \citep{helling2020mineral,Helling2021,Helling2022} 
the column integrated total nucleation rate is
\begin{equation}
    \int_{z_{min}}^{z_{max}} J_{*,{\rm ~tot}}(z) dz ~~ {\rm [cm^{-2}~s^{-1}]}.
\end{equation}
It quantifies the total amount of 
cloud condensation nuclei 
that form 
along the atmosphere column. The mass that makes up this column of cloud condensation nuclei is 
\begin{equation}
\label{eq:J*tot}
    \dot\Sigma = \int_{z_{min}}^{z_{max}} \sum_{i} m_{i} J_{*,i}(z) dz ~~ {\rm [g~cm^{-2}~s^{-1}]},
\end{equation}
taking into account the four individual nucleation species ($i=$\ce{TiO2}, \ce{SiO}, \ce{NaCl}, \ce{KCl}), with the mass of individual cloud condensation nuclei $m_{i}$ and their respective nucleation rates $J_{*,i}$.\\

The column integrated, number density weighted, surface averaged mean particle size is
\begin{equation}
\label{eq:aa}
\langle \langle a \rangle_{\rm A} \rangle = \frac{\int_{z_{min}}^{z_{max}} n_{d}(z)\langle a \rangle_{\rm A}(z) dz }
{ \int_{z_{min}}^{z_{max}} n_{d}(z) dz} \quad \mbox{with}\quad n_{\rm d}(z) = \frac{\rho(z) L_3(z)}{4\pi  \langle a(z)\rangle_{\rm A}^3/3}.
\end{equation}
The number density of cloud particles (that result from the nucleation rate, Eq.~\ref{eq:J*tot}) is a weighting factor in Eq.~\ref{eq:aa} such that the average accounts for differing numbers of particles of different sizes through the atmosphere.

\section{Cloud properties on the hot Saturn WASP-39b}\label{sec:cps}

The similarity of WASP-39b with WASP-96b in mass and temperature allows to explore the consistency of the cloud model employed here for WASP-39b and in \citealt{Samra2022_96b} for WASP-96b. For both planets, ample observational data are available, including JWST data from the ERS programme for WASP-39b, which can be used to further constrain cloud model parameters like vertical mixing. Further, the diverging retrieval results (Sect.~\ref{s:intro})  of fundamental properties like metallicity for WASP-39b poses a challenge for planet formation and evolution studies, which may be resolved by taking into account cloud formation.  Section~\ref{ss:3Dres} presents the WASP-39b atmosphere thermodynamic structure as a base for the global cloud results in Sect.~\ref{ss:cloudglobal} and all following sections. In Sect.~\ref{ss:cloudglobal} the global distribution of the cloud properties is presented which indicates the clouds on WASP-39b are rather homogeneous in nature. It, however, becomes clear in the following sections that, for example, morning and evening terminator differences (Sect.~\ref{sec:column_int_props}) and changing material composition (Sect.~\ref{Vs}) that determine the remaining gas-phase abundances (Sect.~\ref{s:gas}) and the cloud opacity (Sect.~\ref{ss:os}) get lost in simplifications.


\subsection{The 3D GCM atmosphere structure}\label{ss:3Dres}

The 3D atmosphere structures of gas giants like WASP-39b and WASP-96b that share a global temperature of T$_{\rm eq}\sim 1100\,\ldots\,1300$K undergo moderate and relatively smooth day-night temperature changes compared to ultra-hot Jupiters with temperatures T$_{\rm eq}\gtrsim 2000$ K, for example, HAT-P-7b \citep{2019A&A...631A..79H}. This is shown in Fig.~\ref{fig:1dprofiles} (left) which displays the 120 1D $(T_{\rm gas}, p_{\rm gas})$-profiles which were extracted from the 3D GCM for WASP-39b. 
The maximum temperature difference between the dayside and the nightside is $\Delta T_{\rm day-night}\sim500$ K. The flow of hot gas across the dayside results in the evening terminator being 100-200 K hotter than the morning terminator for a given pressure level where $p_{\rm gas}\leq10^{-2}$ bar (see Fig.~\ref{TpNuc}, top left). 

Figure~\ref{fig:1dprofiles} (right) encapsulates the complexity of the 3D $(T_{\rm gas}, p_{\rm gas})$-profiles in terms of dayside and nightside median profiles to facilitate comparison or application in retrieval approaches. The maximum error that would occur when only using the median profiles is shown as (red/blue) envelop which represent the maximum deviation from the median amongst all 120 1D profiles. 
This maximum deviation from the median is determined by a few profiles on the night side that form a vortex cold-trap (Fig.~\ref{fig:1dprofiles}, inset). These Rossby vorticies appear at latitude $\theta\approx\pm68^{\circ}$ as shown in Fig.~\ref{fig:gcm_maps}. This cold-trap is sampled by the profiles $(\phi=-165.0^{\degree},\theta=68.0^{\degree})$, $(\phi=-150.0^{\degree},\theta=68.0^{\degree})$, $(\phi=-135.0^{\degree},\theta=68.0^{\degree})$, $(\phi=-120.0^{\degree},\theta=45.0^{\degree})$, $(\phi=-120.0^{\degree},\theta=68.0^{\degree})$, and $(\phi=-105.0^{\degree},\theta=68.0^{\degree})$ on the nightside at $p_{\rm gas}\sim10^{-3.5}~{\rm bar}$. 

The median $(T_{\rm gas}, p_{\rm gas})$-profiles also highlight that WASP-39b and WASP-96b differ only moderately with respect to their atmosphere temperatures. Figure~\ref{TpNuc} (top left) further demonstrates in 2D terminator slices that the WASP-39b terminator temperature distribution is only very mildly asymmetric within the present 3D GCM modelling domain, again similar to WASP-96b. 
Since cloud formation is determined by the local thermodynamical conditions in the collision dominated part of any atmosphere, the cloud distribution will be similarly symmetric globally. However, a local cold trap, like the cold Rossby vortices on the night side, may amplify the cloud formation efficiency locally. Such details may be lost when representing the WASP-39b atmosphere profiles in terms of median day and night profiles.
It may, however, be reasonable to cast the complexity of physically self-consistent temperature profiles, as used here, in terms of median profiles for better comparison with temperature profiles from retrieval frameworks.







\subsection{The global properties of mineral clouds on WASP-39b}\label{ss:cloudglobal}

Figure~\ref{TpNuc} (top right) demonstrates where cloud formation is triggered  by the formation of cloud condensation nuclei. The local thermodynamic conditions (Fig.~\ref{TpNuc}, top left) trigger the formation of cloud particles generally at pressures $p_{\rm gas}<10^{-1}$ bar on the dayside and $p_{\rm gas}<10^{-2}$ bar on the nightside. The extension of the global cloud layers reach considerably higher pressures as cloud particles grow and gravitationally settle into the deeper layers where they evaporate. Hence, the largest particles of $\langle a \rangle_{A}\sim10^4\mu$m  (Fig.~\ref{TpNuc}, lower right) appear at the cloud base of $p_{\rm gas} \sim 10^{2.5}$ bar. The dust-to-gas ratio (Fig.~\ref{TpNuc}, lower left)  demonstrates  the rather symmetric cloud mass load of the WASP-39b atmosphere.

Figure~\ref{fig:spaghetti_clouds} (left) visualises the detailed cloud property results for the total nucleation rate, $J_*$ (top), the surface averaged mean particle size, $\langle a \rangle_{A}$ (middle), and the dust-to-gas mass ratio, $\rho_{\rm d}/\rho$ (bottom), for the 120 1D profiles that represent the WASP-39b atmosphere. Figure~\ref{fig:spaghetti_clouds} (right) presents the median and maximum deviation values for the day (orange) and the night (blue) side. The opacity relevant surface averaged mean particle size, $\langle a \rangle_{A}$ (middle) appears well represented by the median values and the maxima deviations appear very moderate. The nucleation rate, $J_*$  (top), however, has a few order of magnitudes differences between the median profiles for the dayside and the nightside in the upper atmosphere. Furthermore, the spread in deviation away from the nighside median is much larger on the nightside in the upper atmosphere. This affects the cloud particle number density and therefore translates in considerable differences between the median profiles for  $\rho_{\rm d}/\rho$ (bottom). The peak in the maximum deviation from the nightside median profile for $J_*$, between $p_{\rm gas}\sim10^{-4}-10^{-3.5}$ bar, is due to the few atmosphere profiles that constitute a cold trap due to the Rossby vorticies. 


Similarly to the cloud model performed for WASP-96b \citep{Samra2022_96b}, also for WASP-39b the cloud properties appear to be relatively horizontally homogeneous for a given hemisphere, except for the Rossby cold traps. The latter will be on the night side and will thus not be observable with transmission spectroscopy. However, to be able to interpret transmission spectra derived by JWST that probe the planetary limbs, particle sizes and material compositions at the morning and evening terminator have to be investigated in more detail. The local thermodynamic conditions of the morning terminator are similar to that of the night side and thus exhibits night side nucleation and the local thermodynamic conditions of the evening terminator are similar to the day side and exhibit correspondingly different nucleation.

\begin{figure}
    \centering
    \includegraphics[width=\linewidth]{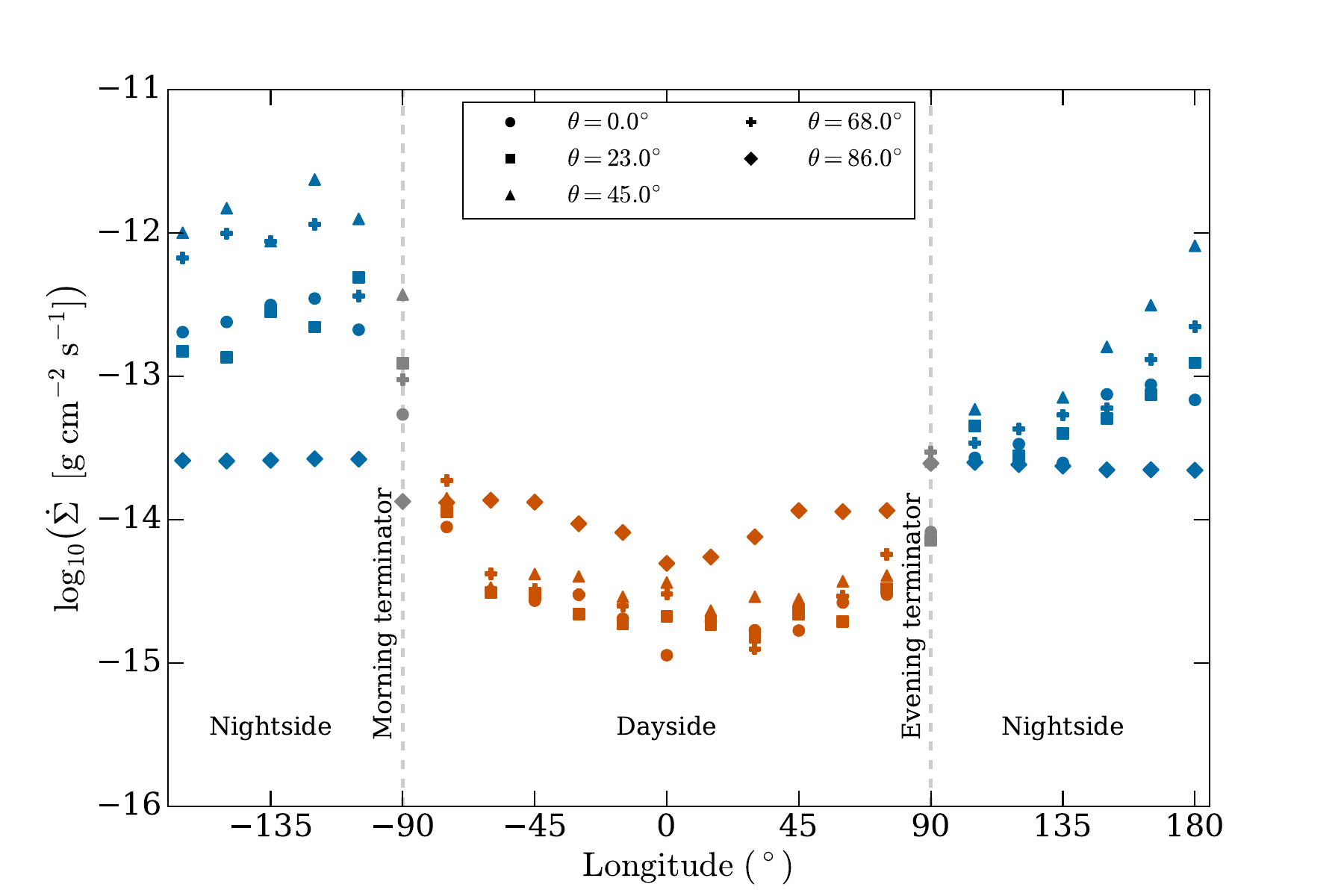}\\    
    \includegraphics[width=\linewidth]{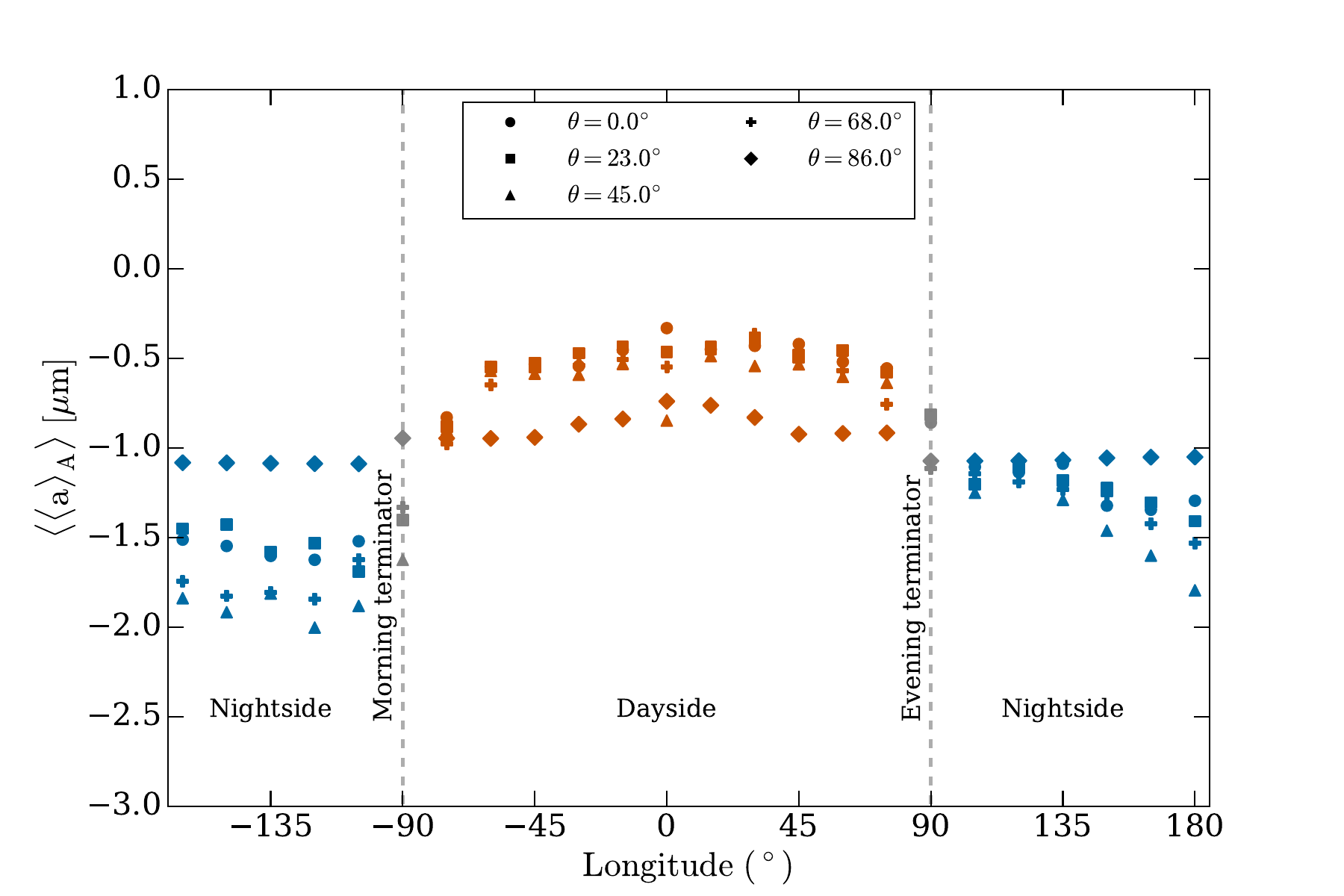}
    \caption{Column integrated cloud properties for WASP-39b. \textbf{Top:} Column integrated mass nucleation rate, $\dot\Sigma$. \textbf{Bottom:} Column integrated, number density weighted surface averaged mean, particle size, $\langle \langle a \rangle_{\rm A} \rangle$}
    \label{fig:CI_nuc_rate}
\end{figure}

\begin{figure*}
    \centering
    \includegraphics[width=0.45\linewidth]{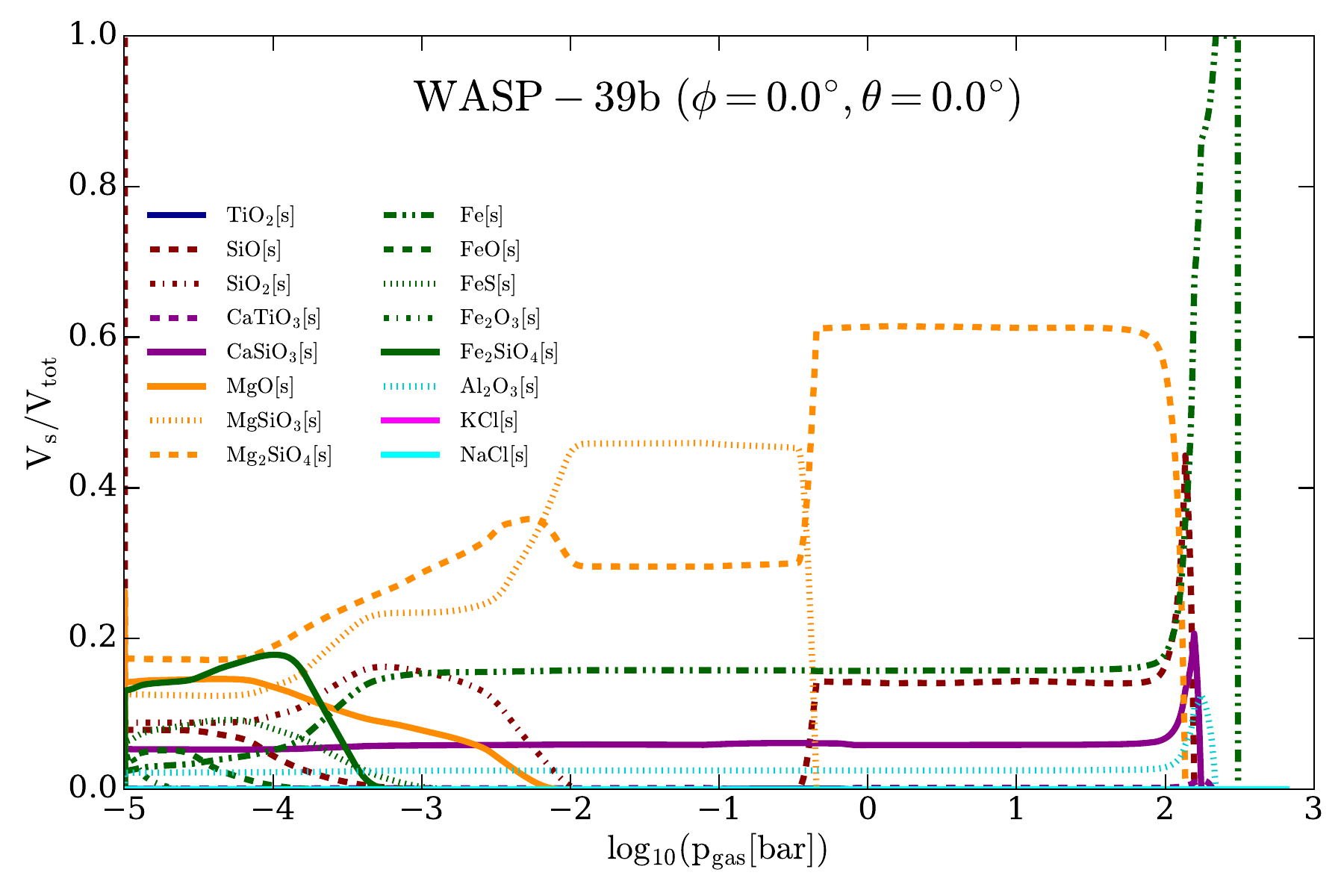}
    \includegraphics[width=0.45\linewidth]{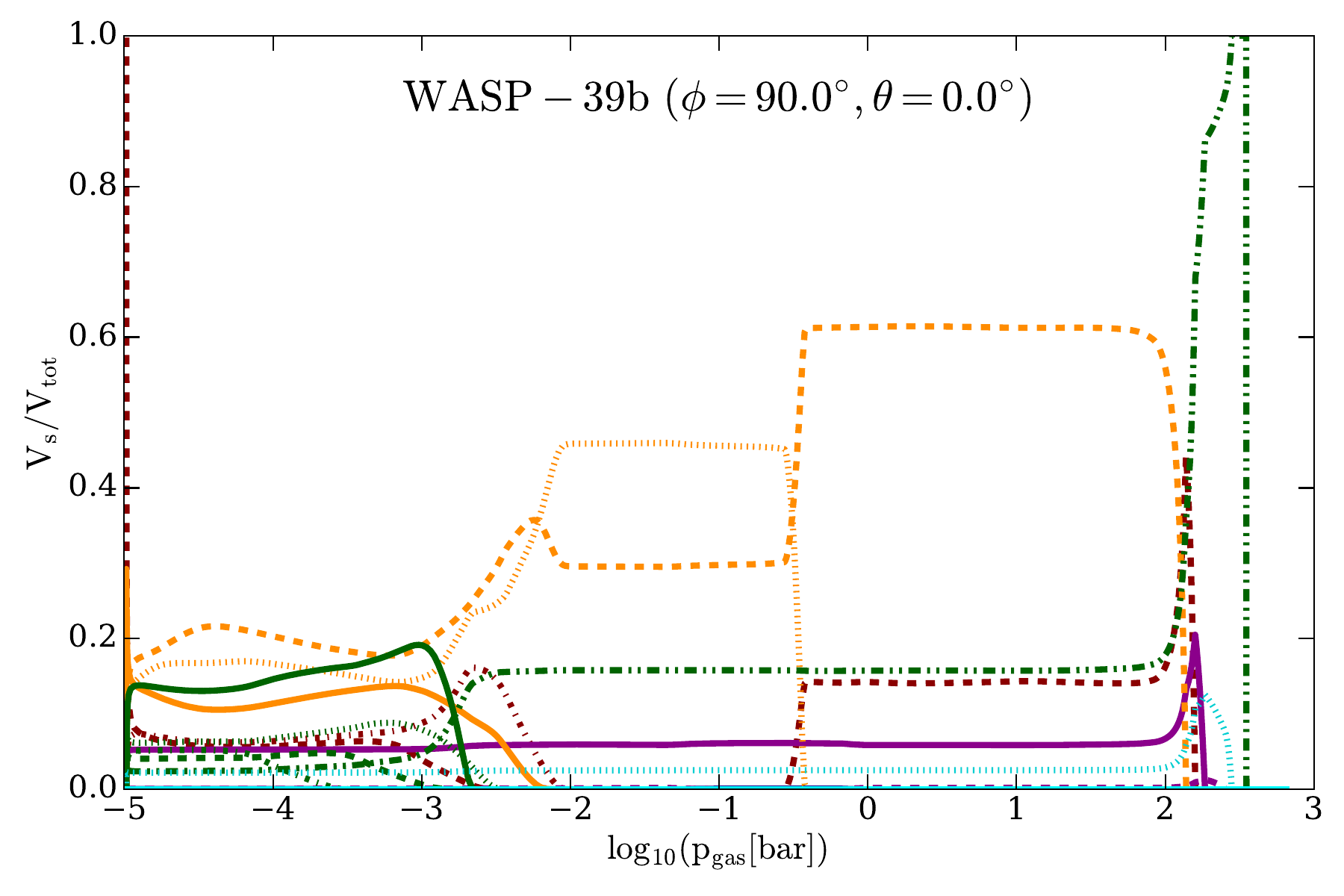}\\
    \includegraphics[width=0.45\linewidth]{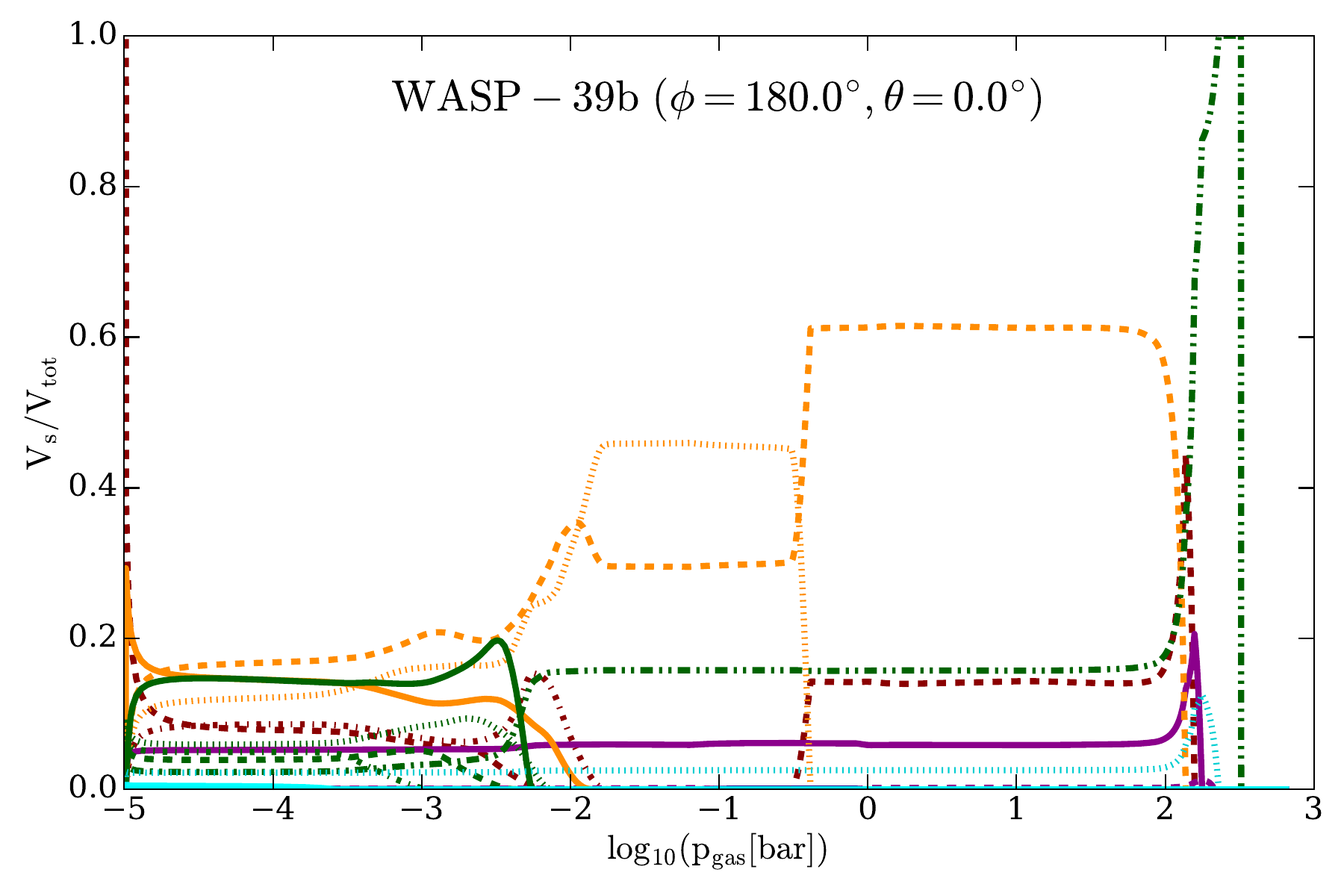}
    \includegraphics[width=0.45\linewidth]{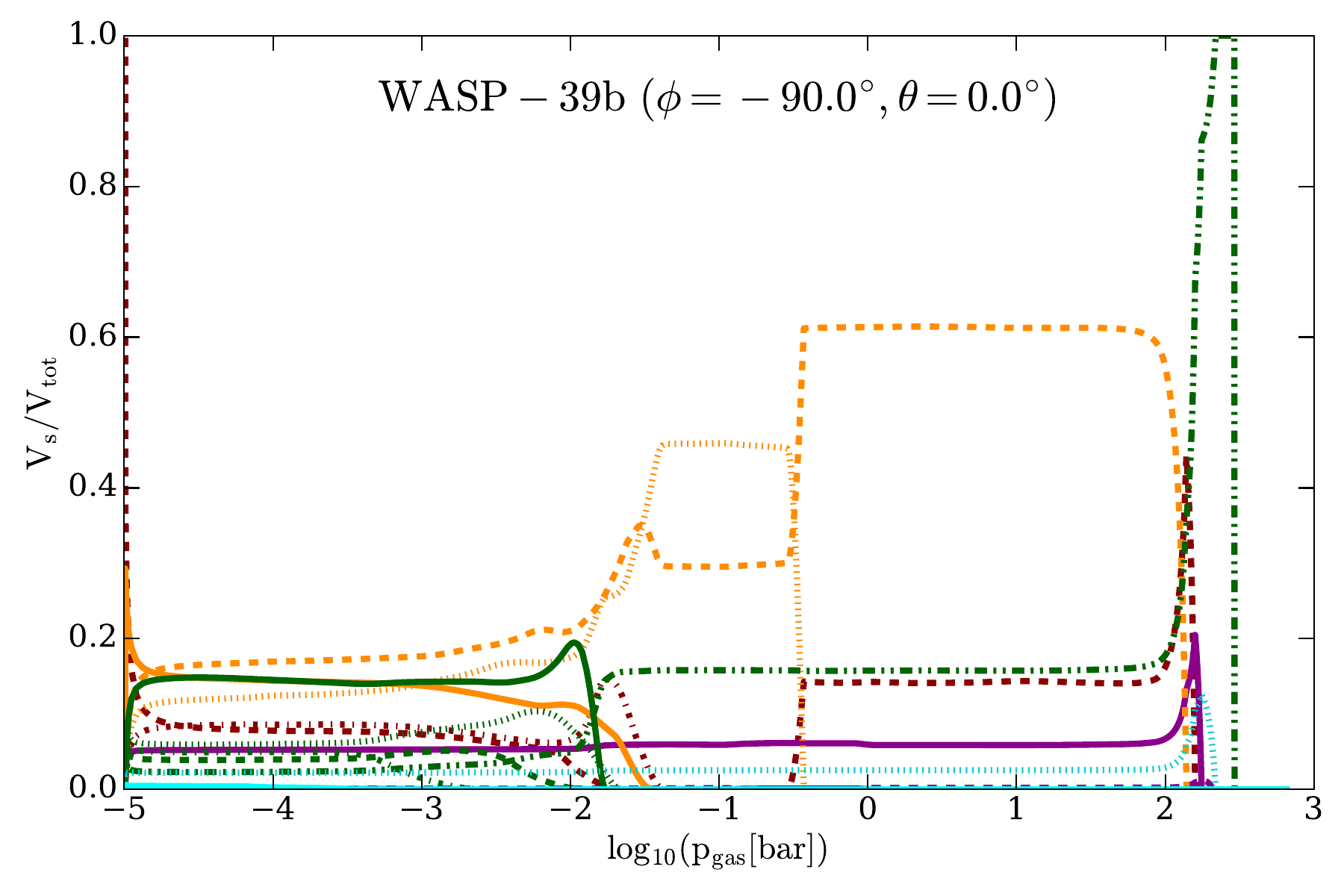}
    \caption{Volume fractions, $V_{\rm s}/V_{\rm tot}$, of individual cloud material condensates at the substellar, antistellar, and equatorial morning and evening terminators for WASP-39b.}
    \label{fig:mat_comp_39b_indivdiual}
\end{figure*}

\subsection{Column integrated properties to reveal differences at the terminators}\label{sec:column_int_props}

\begin{figure*}
\includegraphics[page=1,width=0.55\linewidth]{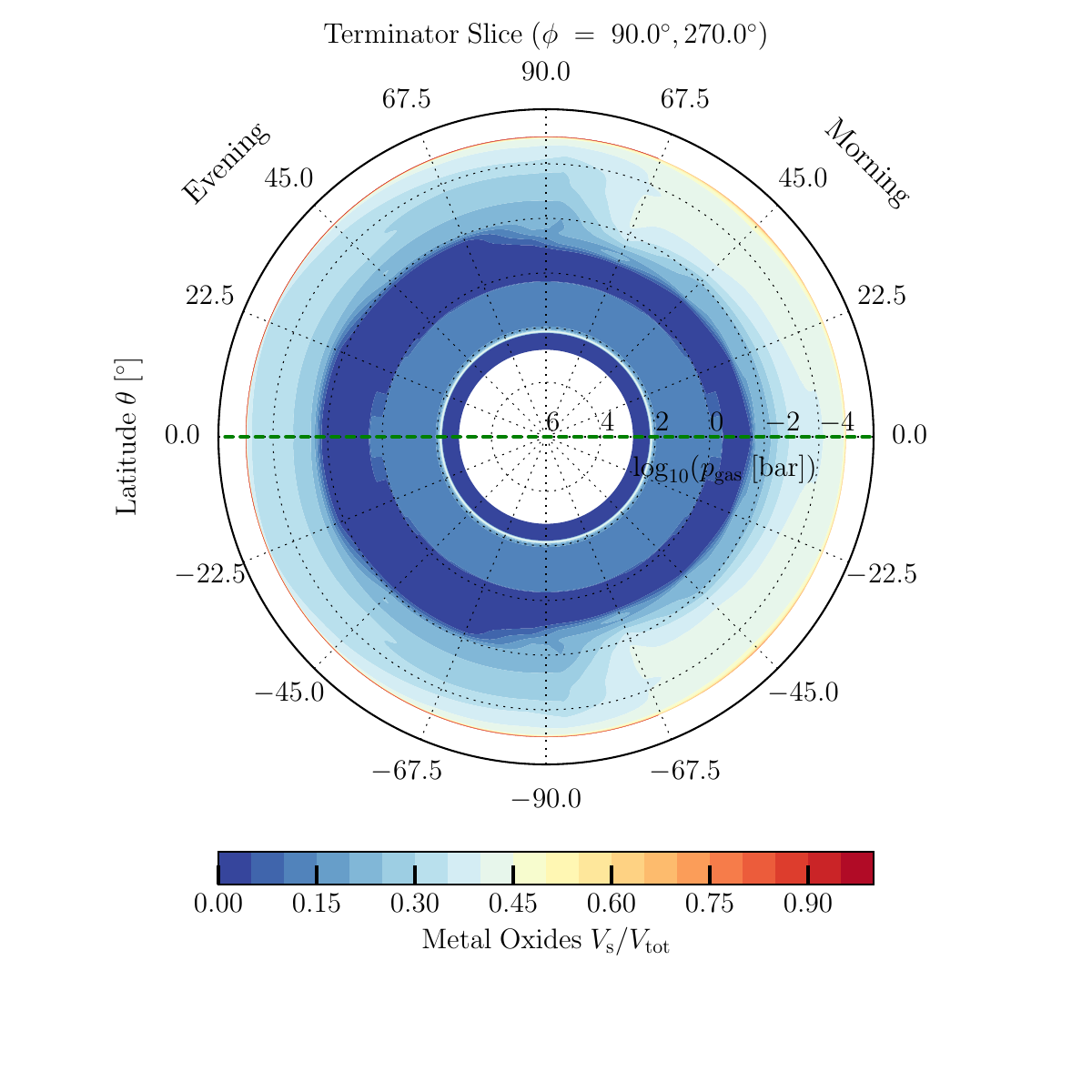}
\hspace*{-1.5cm}\includegraphics[page=2,width=0.55\linewidth]{Figures/Slice_Plots_MaterialComps_WASP039b_termonly.pdf}\\*[-1cm]
\includegraphics[page=3,width=0.55\linewidth]{Figures/Slice_Plots_MaterialComps_WASP039b_termonly.pdf}
\hspace*{-1.5cm}\includegraphics[page=4,width=0.55\linewidth]{Figures/Slice_Plots_MaterialComps_WASP039b_termonly.pdf}
\caption{WASP-39b 2D terminator slices showing the bulk material composition of cloud particles. The materials are grouped as in \citep{Helling2021}: {\bf Top Left:} Metal oxides ($s=$\ce{SiO}[s], \ce{SiO2}[s], \ce{MgO}[s], \ce{FeO}[s], \ce{Fe2O3}[s]), {\bf Top Right:} Silicates ($s=$\ce{MgSiO3}[s], \ce{Mg2SiO4}[s], \ce{Fe2SiO4}[s], \ce{CaSiO3}[s]). 
{\bf Bottom left} High temperature condensates ($s=$\ce{TiO2}[s], \ce{Fe}[s], \ce{FeS}[s], \ce{Al2O3}[s], \ce{CaTiO3}[s]), {\bf Bottom right} Salts ($s=$\ce{KCl}[s], \ce{NaCl}[s]).  
}
  \label{VsVtot}
\end{figure*}


Column integrated properties (definitions in Sect.~\ref{s:ap}) provide additional insights at the terminators. The column integrated values are less affected by extreme events like the Rossby vortices which determine the maximum deviations from the median values in Fig.~\ref{fig:spaghetti_clouds}. Further, they enable the comparison with results from the ARCiS \citep{18OrMi.arcis,20MiOrCh.arcis} retrieval framework, which incorporates a self-consistent cloud model, too.

The column integrated nucleation rate mass, $\dot\Sigma$, and the column integrated, number density weighted, surface averaged mean particle size, $\langle \langle a \rangle_{\rm A} \rangle$, are shown in Fig.~\ref{fig:CI_nuc_rate}. Figure~\ref{fig:CI_nuc_rate} highlights that differences between the morning and evening terminators of WASP-39b become more apparent in the integrated properties.

The integrated nucleation rates are higher on the nightside with a range of $\dot\Sigma\sim10^{-11.5}\,\ldots\,10^{-13.5}~{\rm g~cm^{-2}~s^{-1}}$ compared to the dayside with a range of $\dot\Sigma\sim10^{-13.5}\,\ldots\,10^{-15.5}~{\rm g~cm^{-2}~s^{-1}}$. Correspondingly, the values of $\langle \langle a \rangle_{\rm A} \rangle$ are larger on the dayside, ranging from $\langle \langle a \rangle_{\rm A} \rangle\sim10^{-1}\,\ldots\,10^{-0.5}~{\rm \mu m}$, than on the nighside where $\langle \langle a \rangle_{\rm A} \rangle\sim10^{-2}\,\ldots\,10^{-1}~{\rm \mu m}$. The evening terminator inherits similar local thermodynamic conditions to the dayside, whereas the morning terminator is similar to the nightside. Hence, nucleation on the evening terminator is less efficient than the morning terminator ($\dot\Sigma_{\rm evening}<\dot\Sigma_{\rm morning}$) resulting in a larger average particle size at the evening terminator compared to the morning terminator. The evening terminator appears more homogeneous in nucleation efficiency which results into a smaller variation in particle sizes across the limb. Both, $\dot\Sigma$ and $\langle \langle a \rangle_{\rm A} \rangle$ show a moderate variation with latitude. These notable differences in column integrated cloud properties are caused by the moderate differences in the local thermodynamic conditions e.g. the (T$_{\rm gas}$, p$_{\rm gas})$-profiles.  

The values presented in Fig.~\ref{fig:CI_nuc_rate} compare well with the values derived by \citet{20MiOrCh.arcis} using ARCiS to perform retrieval on observations of WASP-39b (without JWST data). They derived $\log_{10}\dot\Sigma = -12.77^{+3.68}_{-2.93}$ [g cm$^{-2}$ s$^{-1}$] from pre-JWST observations, which is consistent with the values that are derived here with self-consistent forward modelling. Similarly, \citet{Samra2022_96b} noted that their column integrated nucleation rate also matched within $1\sigma$ ARCiS results for the exo-Saturn WASP-96b. Thus, forward modelling and retrieval can complement each other if retrieved cloud model properties are complex enough.

\subsection{Non-homogeneous vertical cloud material composition}\label{Vs}

WASP-39b's limbs have notable differences in particle size at a given pressure level as result of variations in the local atmospheric density structures. The particle sizes do also strongly vary in the vertical direction. Therefore, the change of the vertical thermodynamic structure within the atmosphere of WASP-39b causes the cloud properties to be non-homogeneous in size and number, but also in material composition. The changing atmosphere density affects the collisional rates, but the temperature affects the thermal stability (and to a lesser extent the collisional rates) which then results in a changing composition of the cloud particle within the atmosphere. The detailed material compositions of the cloud particles are presented for the substellar (dayside) and the antistellar (nightside) points, as well as for the morning and the evening terminator in the equatorial plane in Sect.~\ref{ss:patchy}. Triggered by the pre-JWST large values ($150\times\epsilon_{\rm solar}$) of inferred metallicities for WASP-39b, Sect.~\ref{ss:metal} explores the effect of increasing amounts of heavy elements (heavier than He) on the cloud results.

\subsubsection{Patchy clouds due to changing thermal stability}\label{ss:patchy}

The composition of the particles that compose the clouds in the atmosphere of WASP-39b varies throughout the atmosphere due to the changing thermal stability of condensing materials  in response to changing local thermodynamic conditions.
Figure~\ref{fig:mat_comp_39b_indivdiual} shows the volume fractions, $V_{\rm s}/V_{\rm tot}$,  of the individual material condensates at the substellar, antistellar, and equatorial morning and evening terminator. The composition of the cloud layers is shown grouped into silicates (s=\ce{MgSiO3}[s], \ce{Mg2SiO4}[s], \ce{Fe2SiO4}[s], \ce{CaSiO3}[s]), metal oxides (s=\ce{SiO}[s], \ce{SiO2}[s], \ce{MgO}[s], \ce{FeO}[s], \ce{Fe2O3}[s]), high temperature condensates (s=\ce{TiO2}[s], \ce{Fe}[s], \ce{FeS}[s], \ce{Al2O3}[s], \ce{CaTiO3}[s]), and salts (s=\ce{KCl}[s], \ce{NaCl}[s]) for the morning and evening terminators as 2D slice plots in Fig.~\ref{VsVtot}. 

In the upper atmosphere ($p_{\rm gas}\lesssim10^{-2}~{\rm bar}$) there is no clear dominant group of material condensates globally, with both silicates and metal oxides representing the bulk composition and the relative fraction of each varying between profiles. At the morning terminator, silicates and metal oxides represent almost equal fractions of the cloud particle composition with $\sim50\%$ and $\sim40\%$ respectively. Over the same pressure region for the evening terminator, the silicates dominate slightly over the metal oxides, each comprising $\sim60\%$ and $\sim30\%$ respectively. The remaining $\sim10\%$ of the cloud particle volume is comprised of high temperature condensates. The dominant silicate condensates are by \ce{MgSiO3}[s], \ce{Mg2SiO4}[s], and \ce{Fe2SiO4}[s]. The dominant metal oxide condensates are \ce{SiO}[s], \ce{SiO2}[s], and \ce{MgO}[s]. The mixed silicate and metal oxide cloud layer extends until the metal oxides become thermally unstable at $p_{\rm gas}\sim10^{-3}~{\rm bar}$ for the evening terminator and $p_{\rm gas}\sim10^{-2}~{\rm bar}$ for the morning terminator. The thermally unstable material evaporates releasing Mg, Si, and O back into the gas phase which permits further silicate condensation. The increase in the silicate fraction is initially due to an increase in the fraction of \ce{MgSiO3}[s] from $\sim16-17\%$ to $\sim45\%$ of the cloud volume. When \ce{MgSiO3}[s] evaporates,  \ce{Mg2SiO4}[s] becomes the dominant silicate at $\sim61\%$ of the total cloud particle volume. The maximum contribution of silicates is $\sim81\%$ occurring by $p_{\rm gas}\sim10^{-2}~{\rm bar}$ at the morning terminator and $p_{\rm gas}\sim10^{-1}~{\rm bar}$ at the evening terminator. The deepest layers of the atmosphere are dominated by high temperature condensates as all other materials considered here are thermally unstable. The general trends in material composition outlined for the terminators also apply for the substellar and antistellar points. Hence, the clouds on WASP-39b are expected to be patchy in terms of mixed composition in the upper atmosphere, with an extended silicate dominated layer until $p_{\rm gas}\sim10^{2}~{\rm bar}$.


\subsubsection{Effect of global metallicty on cloud formation}\label{ss:metal}

\begin{figure*}
    \includegraphics[width=1.\textwidth]{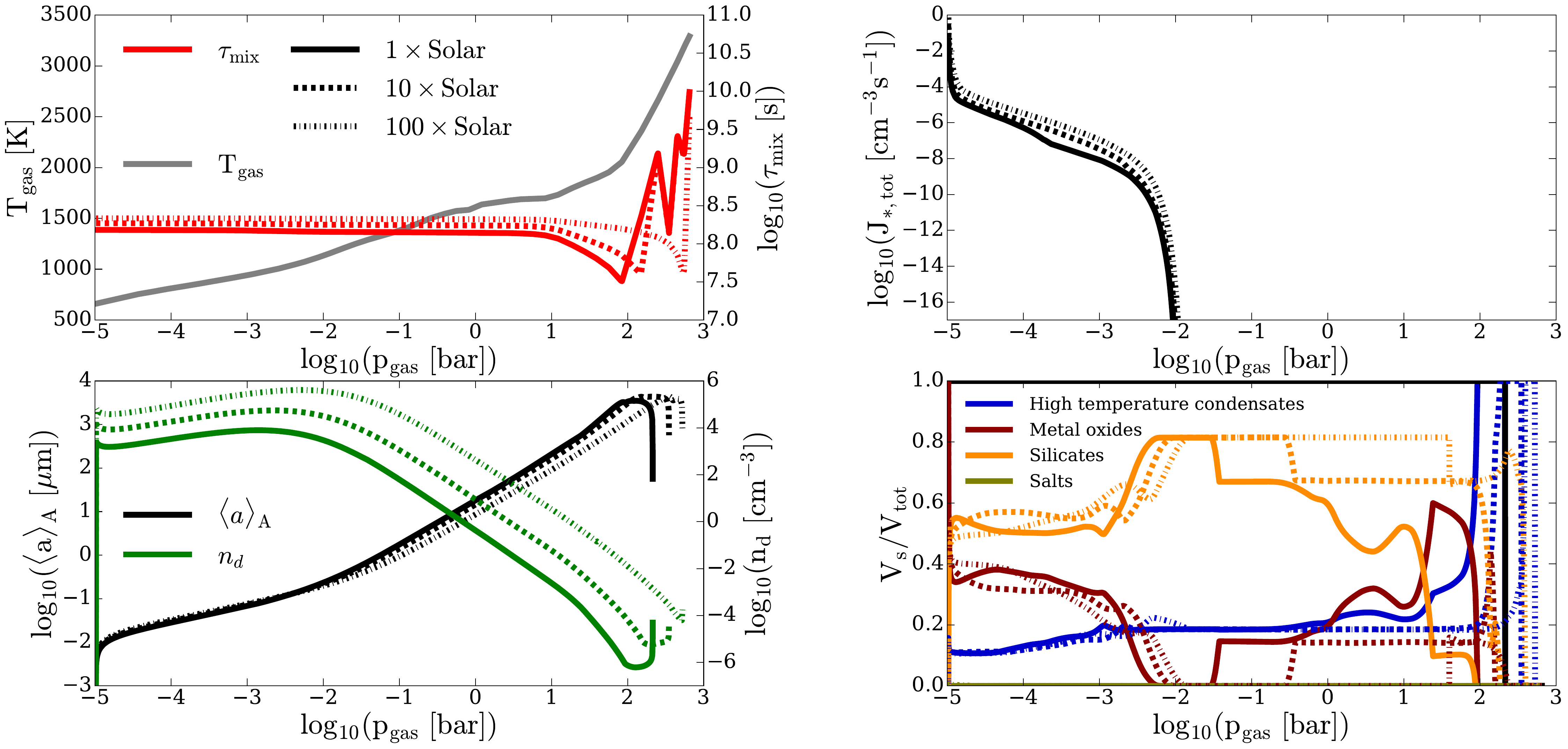}
    \caption{Metallicity affect on cloud properties shown for the same evening terminator equator (T$_{\rm gas}$, p$_{\rm gas}$)-profile (top left, black line) for WASP-39b. An increased global amount of heavy elements increases the thermal stability of the silicate materials (orange) at higher temperatures where p$_{\rm gas}>10^{-2}$bar.  {\bf Top right:} Total nucleation rate, $J_{\rm *, tot}$ [cm$^{-3}$ s$^{-1}$], {\bf Bottom left:} Surface averaged mean particle size (black) and cloud particle number density, $n_{\rm d}$ [cm$^{-3}$] (green), {\bf Bottom right:} Material volume fractions, $V_{\rm s}/V_{\rm tot}$, for four material groups (high temperature condensates, metal oxides, silicates, salts).}
    \label{fig:comp_39_met}
\end{figure*}
Previous works before JWST data were obtained have reported a wide range of different values for the metallicity of WASP-39b: From solar \citep{Nikolov_2016} to moderately super-solar \citep{Pinhas2019} to very high values of $151^{+48}_{-46}\times\varepsilon_{\rm solar}$ metallicity \citep{2018AJ....155...29W}. 

Notably, the derived atmospheric metallicity changed, depending on cloud model being used for retrieval for the same data. Retrieval with a simple cloud prescription as used by \citet{2018AJ....155...29W} favoured a cloud-free, very high metallicity composition. ARCiS yielded $\sim15\times\varepsilon_{\rm solar}$ metallicity \citep{20MiOrCh.arcis} with a more complex cloud model, which is in accordance with recently obtained JWST data \citep{JWST_39b_CO2,https://doi.org/10.48550/arxiv.2211.10487,https://doi.org/10.48550/arxiv.2211.10488,https://doi.org/10.48550/arxiv.2211.10489,Feinstein2022_JWST_NIRISS}. For this work $10\times\varepsilon_{\rm solar}$ metallicity was adopted in the nominal model in accordance of newly obtained JWST data, in contrast to previous work on the similarly warm exo-Saturn WASP-96b, for which solar metallictiy was assumed \citep{Samra2022_96b}.

While \citet{20MiOrCh.arcis} constrained metallicity for WASP-39b with a cloud model from the retrieval side, it is worthwhile to also explore the impact of different metallicities with forward modelling using a fully microphysical cloud model. The implications of different assumptions of the atmospheric metallicity on the formation of clouds on WASP-39b is explored here. Three different metallicity values are tested for the same equatorial evening terminator (T$_{\rm gas}$, p$_{\rm gas}$)-profile: $1\times$, $10\times$, and $100\times\varepsilon_{\rm solar}$ abundances . The mean molecular weight, $\mu$, varies for a given (T$_{\rm gas}$, p$_{\rm gas}$)-profile. The values of $\mu=2.3,2.4,~{\rm and}~4.75$ are adopted for the 1$\times$, 10$\times$, and 100$\times\varepsilon_{\rm solar}$ cases, respectively. A value of $\mu = 2.3$ is expected for a solar \ce{H2}/He dominated atmosphere. The choice of $\mu = 4.75$ is motivated by equilibrium chemistry calculations for the equatorial evening terminator (T$_{\rm gas}$, p$_{\rm gas}$)-profile using GGChem \citep{Woitke2018_GGChem} which show $\mu=4.73\,\ldots\,4.81$ throughout the atmosphere. The 10$\times\varepsilon_{\rm solar}$ case with $\mu=2.4$ is the same as presented in the previous sections. The change in metallicity is applied only in the cloud formation simulation, therefore, any impact on the (T$_{\rm gas}$, p$_{\rm gas}$)-profile which may arise from a different metallicity is not included here. 

The differences in the total nucleation rate ($J_{*,{\rm tot}}$), total particle number density ($n_d$), surface averaged particle size ($\langle a \rangle_{\rm A}$), and material composition of cloud particles ($V_{\rm s}/V_{\rm tot}$) for each case are presented in Fig.~\ref{fig:comp_39_met}. The nucleation efficiency increases slightly with increased metallicity, however, the pressure range over which nucleation occurs does not change significantly. The increased availability of condensable elements results in approximately an order of magnitude increase in the total cloud particle number density for each factor of 10 increase in metallicty. In the upper atmosphere, $p_{\rm gas}\lesssim10^{-2}~{\rm bar}$, there is little difference in $\langle a \rangle_{\rm A}$. At higher pressures, the slightly increased nucleation rate in the upper atmosphere manifests in a slightly smaller $\langle a \rangle_{\rm A}$ for $p_{\rm gas}>10^{-2}~{\rm bar}$.

The general trend in which condensates dominate the cloud particle composition remains consistent between in the three cases. In general, the upper atmosphere has a 55\%, 40\%, 15\% mix of silicates, metal oxides, and high temperature condensates, respectively. The most significant difference between the three cases is the extent of deep atmosphere ($p_{\rm gas}\gtrsim10^{-2.5}~{\rm bar}$) silicate cloud layer. The increased global abundance of heavy elements increases the thermal stability of the silicate materials at higher pressures. Consequently, the pressure level at which \ce{MgSiO3}[s] evaporates increases from $p_{\rm gas}\sim10^{-1.5}~{\rm bar}$ in the $1\times\varepsilon_{\rm solar}$ case to $p_{\rm gas}\sim10^{-0.5}~{\rm bar}$ and $p_{\rm gas}\sim10^{1.5}~{\rm bar}$ in the $10\times\varepsilon_{\rm solar}$ and $100\times\varepsilon_{\rm solar}$ cases, respectively.
\subsection{Atmospheric gas composition}\label{s:gas}

The formation of cloud particles affects the composition of the observed atmospheric gas by depleting those elements that form the respective cloud materials ((Mg, Si, Ti, O, Fe, Al, Ca, S, K, Cl, Na). The most abundant elements (O) are the least affected. Therefore, it is important to explore the composition of the gas phase which results from the formation of clouds.
\subsubsection{Dominant gas-phase species}

\begin{figure*}
    \centering
    \includegraphics[width=0.48\linewidth]{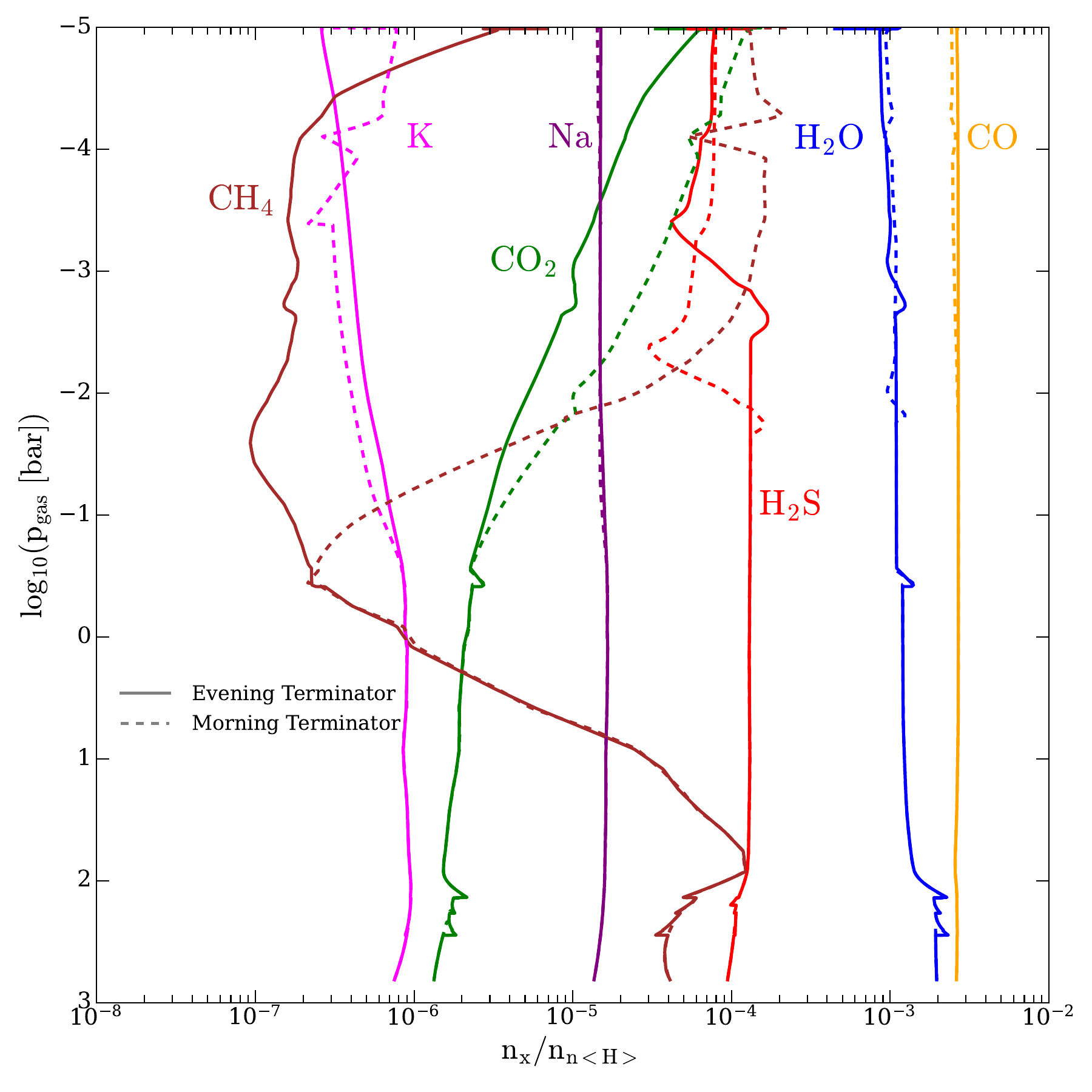}
    \includegraphics[width=0.48\linewidth]{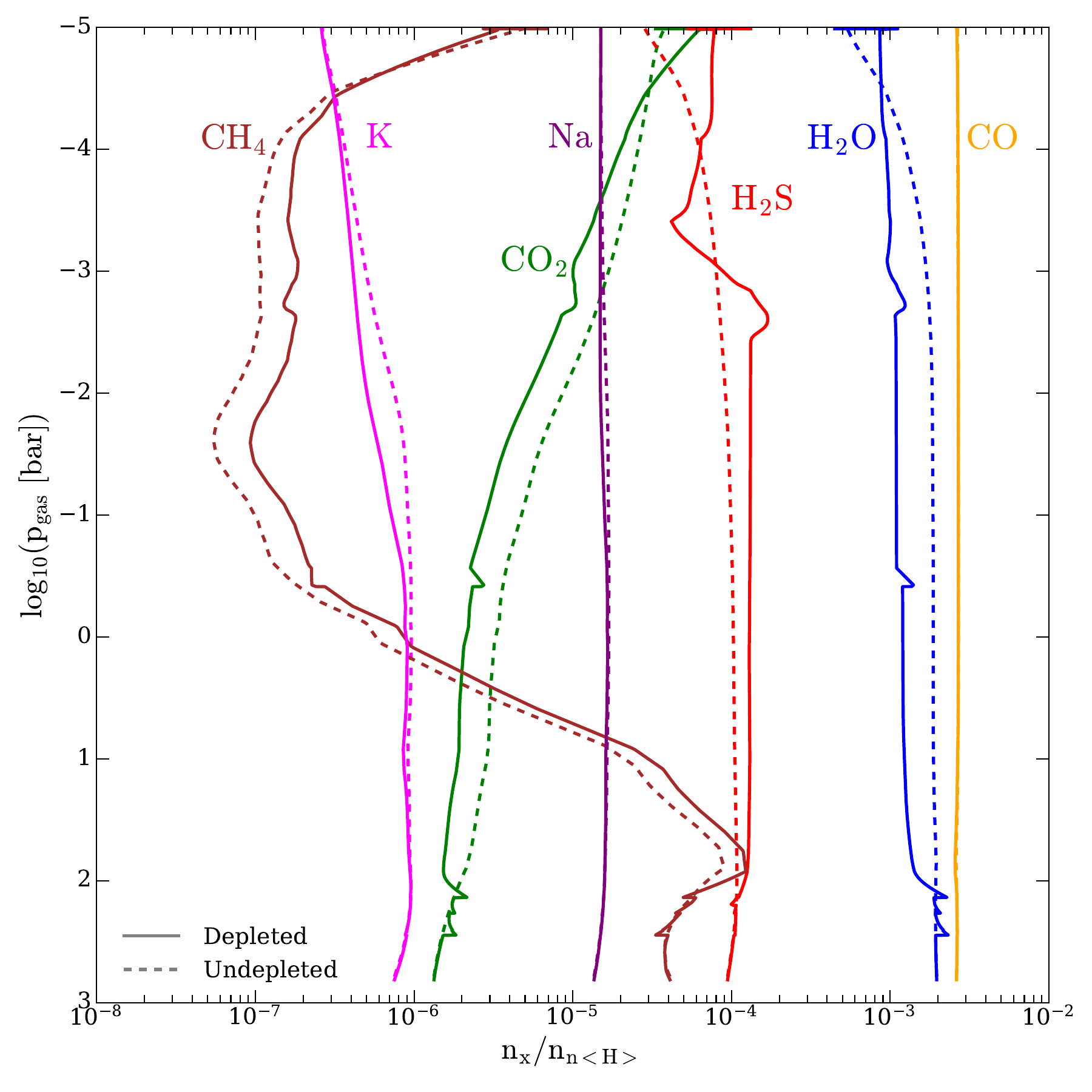}
    \caption{Gas phase concentrations ($\rm n_{x}/n_{<H>}$) for selected molecules at the WASP-39b equatorial morning and evening terminators in chemical equilibrium. \textbf{Left:} Comparing morning and evening terminator. \textbf{Right:} Comparing results for depleted (after cloud formation) and undepleted element abundances
    at the equatorial evening terminator.}
    \label{fig:gas_comp}
\end{figure*}

In Figure~\ref{fig:gas_comp} (left) the concentrations of the dominant gas species, excluding \ce{H2}/\ce{He}, for the equatorial morning and evening terminators are shown. The most dominant gas phase species include CO and H$_2$O, both with concentrations greater than $10^{-3}$, as well as H$_2$S, CO$_2$, \ce{CH4}, Na, and K. For most of the species shown the concentrations are generally broadly similar between the two profiles and the concentrations of these species generally do not change significantly through the atmospheres, with the exception of \ce{CO2} and \ce{CH4}. The \ce{CO2} concentration slightly increases by approximately an order of magnitude between $p_{\rm gas}\sim10^{-1}~{\rm bar}$ and $p_{\rm gas}\sim10^{-5}~{\rm bar}$. The major exception, however, is \ce{CH4}, with the morning terminator concentration exceeding that of the evening terminator by approximately 3 orders of magnitude in the upper atmosphere ($p_{\rm gas}\lesssim10^{-2~{\rm bar}}$) in equilibrium chemistry. At the cooler gas temperatures, the \ce{CO} reacts with \ce{H2} to form \ce{CH4} resulting in an increase in the concentration of \ce{CH4}, with the O liberated in this reaction serving to increase the \ce{H2O} concentration \citep{2007ApJS..168..140S}. 

The concentration of \ce{CH4} at the morning terminator in the upper atmosphere of approximately $n_{\rm CH_4}/n_{<H>}\sim10^{-4}$ is in principle detectable by HST/WFC3 and JWST \citep[e.g.][]{Kreidberg2018,Carone2021}. So far, however, methane has not been detected with low resolution spectroscopy for 
warm exoplanets  ($T_{\rm eq}\lesssim 1200~{\rm K}$) like WASP-39b. The absence of spectral \ce{CH4} features in these planets suggests that methane abundances may be affected by disequilibrium chemistry \citep{2020AJ....160..288F} via vertical \citep{moses2011disequilibrium,venot2012chemical} and maybe also by horizontal mixing \citep{agundez2014,2021MNRAS.505.5603B}. Both would quench the amount of \ce{CH4} below the observation limit (VMR $<10^{-5}$). Hence, it is reasonable to assume the concentration of \ce{CH4} will be globally homogeneous and below the detectability threshold. Thus, \ce{CH4} is not included as an opacity species in Sect.~\ref{s:spectra}. 


Figure~\ref{fig:gas_comp} (right) compares the concentrations of the dominant gas phase species at the evening terminator between the cloud model and the atmosphere with equilibrium chemistry but without condensation. The concentrations of each species do not differ significantly due to the element depletion associated with the cloud formation.

\subsubsection{The role of sulphur in clouds}\label{sec:sulphur_gas}

Elements such as K, Na, and Cl  are not significantly affected by cloud formation because their possible condensate materials are thermally unstable in the atmosphere of WASP-39b. In addition, the formation of sulphur-containing  materials is considerably less  favoured  in comparison to the Fe/Mg-containing silicates. Figure 5 in \cite{2019AREPS..47..583H} demonstrates that materials like S[s], MgS[s], and FeS[s] would reach a sweet spot of maximum volume contribution of $\sim 15\%$ when C/O=0.99\,\ldots\,1.10 and if the sulphur abundance is enriched above the solar value for a $T_{\rm{eff}}=1200$~K exoplanet. \citet{2017MNRAS.472..447M} (Table 3) demonstrate that FeS[s] would contribute with $<1\%$ to cloud particle in gas giants and with $<10\%$ in hot rocky planets like 55 Cnc~e. If sulphur compounds do not condense, the sulphur needs to remain in the gas phase of exoplanet atmosphere. Therefore, the S/O ratio would remain near-solar as demonstrated in Fig.~4 in \cite{2019AREPS..47..583H}. This negative result for the role of sulphur contributing to the cloud mass in exoplanets (and brown dwarfs) is supported by similar findings for AGB stars. The study of post-AGB stars shows a lack of sulphur depletion (\citealt{1991A&A...251..495W,2007A&A...463L...1R}) which is interpreted as a lack of sulphure condensation into dust grain in AGB stars (\citealt{2018A&A...617A.132D}). 

AGB stars, which undergo strong condensation events (dust formation), enrich the ISM for the next generation of stars and planets to form. Sulphur, is however, not nuclearsynthesised in AGB stars but rather in type II supernovae \citep{Colavitti2009,Perdigon2021}, and is the 10th most abundant element in the Universe, as well as an important element for life on Earth. Oxygen-rich AGB stars are observed to have SO and SO$_2$  (\citealt{2020MNRAS.494.1323D}) but H$_2$S is most likely not a parent molecule since it decays rapidly \citep{2016A&A...588A.119D}. H$_2$S has been found in high-mass loss oxygen-rich stars and is argued to account for a significant fraction of the sulphur abundance in these objects (\citealt{2017A&A...606A.124D}).

\cite{2016A&A...585A...6G} point out also the chemical link between CS, CN, SH and H$_2$S in carbon-rich AGB stars. Here, SH can combine with O to SO which further may form SO$_2$;  this would also be relevant for oxygen-rich environments or the upper exoplanet atmospheres where photochemistry enables to formation of CS and CN.  However, the SH reservoir may be depleted through the formation of H$_2$S such that it indirectly affects the presence of SO and SO$_2$. The SH/H$_2$S chemistry is further dictating the formation of CS and CN which then may continue to form HCN (their Eqs.~5 to 16). 
The research question is therefore which gas-phase constituent holds the sulphur reservoir if sulphur is not depleted by dust formation in AGB stars or in cloud particles in extrasolar planets / brown dwarfs.  \lc{Recently,} \cite{https://doi.org/10.48550/arxiv.2211.10490} suggested that \ce{H2S} is a precursor molecule which gives rise to SO$_2$ on WASP-39b via gas-phase non-equilibrium in combination with a solar element over-abundance of $10\times \varepsilon_{\rm solar}$.

To gather a first impression of the exoplanet sulphur reservoir,  the most abundant sulphur-binding gas species in the WASP-39b atmosphere, H$_2$S, is included  into the comparison of major gas species in Fig.~\ref{fig:gas_comp}. These are the gas species for which a radiative transfer solution is presented to fit the observational date for WASP-39b in Sect.~\ref{s:spectra}.





%


\subsubsection{Mineral ratios Si/O, Mg/O, Fe/O, S/O, C/O}

\begin{figure}[h!]
    \includegraphics[width=1.01\linewidth]{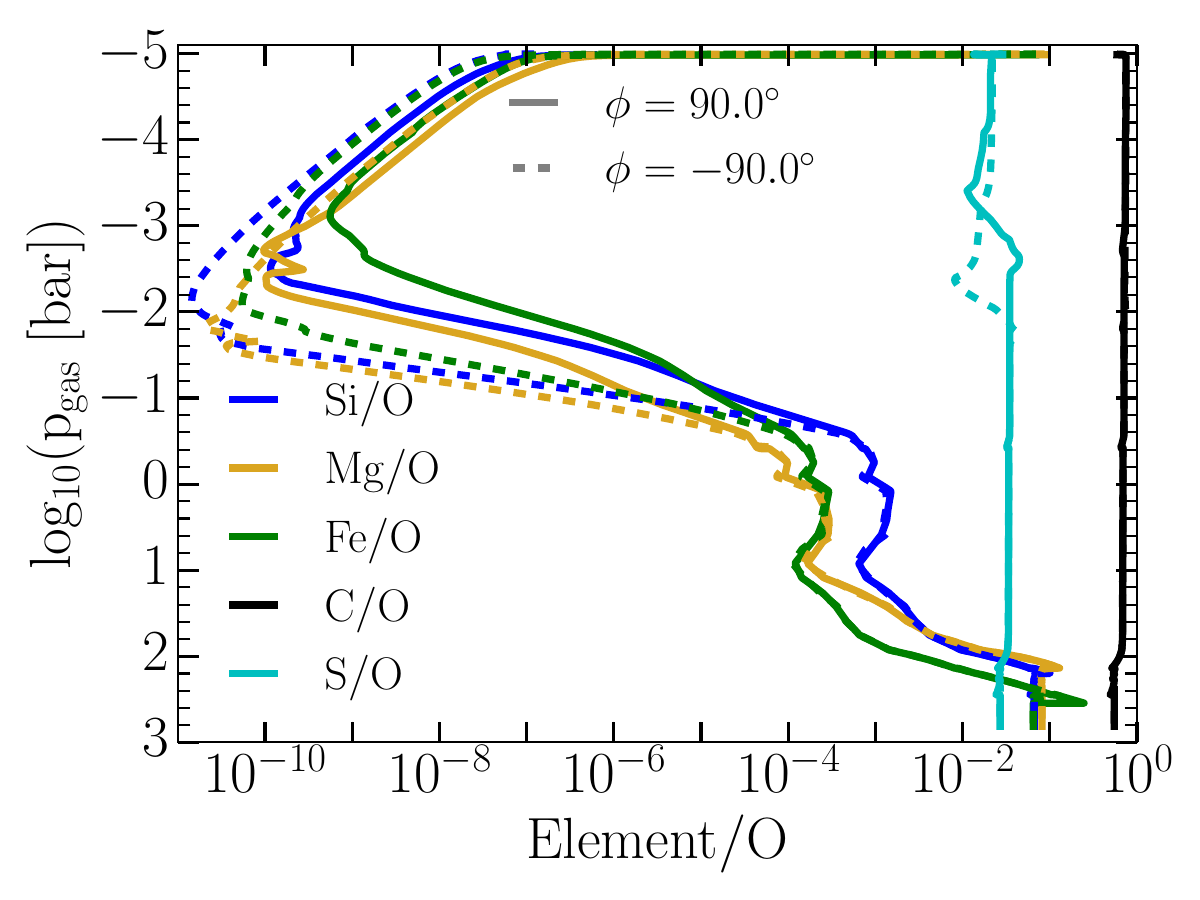}\\*[-0.3cm]
    \includegraphics[width=1.01\linewidth]{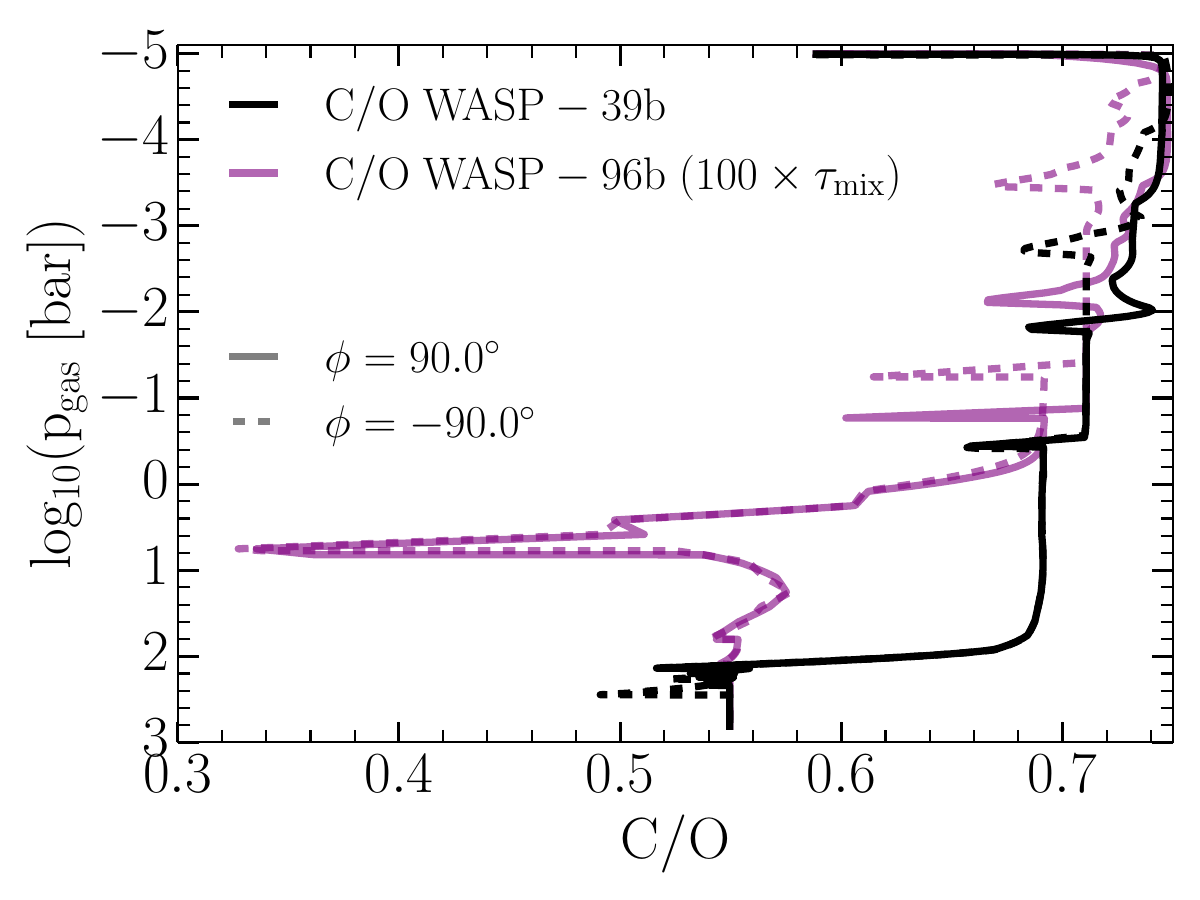}
    \caption{Gas phase element abundance ratios, for equatorial morning ($\phi = -90.0$, dashed lines) and evening ($\phi = 90.0$, solid lines) terminators. 
    {\bf Top:} The mineral ratios Si/O, Mg/O, Fe/O, S/O, and C/O for WASP-39b. {\bf Bottom:} C/O for both WASP-39b and WASP-96b.}
    \label{fig:Gas_Abunds}
\end{figure}

The Si/O, Mg/O, Fe/O, S/O and C/O element abundance ratios for the equatorial morning and evening terminator points are shown in Fig.~\ref{fig:Gas_Abunds}. The cloud formation process reduces the abundances of Si, Mg, and Fe by several orders of magnitude in comparison to oxygen. The reduction is seen for both the morning and evening terminators as cloud formation occurs at both points, however, the morning terminator abundances are reduced more than the evening terminator. The maximum difference in the element ratios between the two terminator points is approximately 1 order magnitude occurring over a pressure range of $p_{\rm gas}\leq10^{-2}-10^{-1}$ bar. 

No substantial reduction is seen for sulphur which confirms previous results \citep{2017MNRAS.472..447M,2019AREPS..47..583H}. This indicates that potential reaction partners Fe, Mg, Si are stronger bound by other materials and leads to the conclusion that sulphur gas species may provide means to determine the primordial abundances and hence, to link to planet formation processes. This finding is relevant for all exoplanets.

In contrast to the other element ratios, in regions of cloud formation the C/O is increased from the solar C/O $=0.54$ as the oxygen is depleted from the gas phase. The maximum value of C/O $\sim$ 0.75 occurs where $p_{\rm gas}=10^{-3}$ bar in the upper atmosphere. In Figure~\ref{fig:Gas_Abunds} (bottom), the equatorial terminator C/O of WASP-39b is compared to that of the reduced mixing efficiency case of WASP-96b from \citep{Samra2022_96b}. The C/O ratio for WASP-96b drops to C/O $\sim 0.31$ for both the equatorial morning and evening terminators at $p_{\rm gas}\sim10$ bar due to the evaporation of \ce{Mg2SiO4}[s] resulting from the increased gas temperature (see Figs. 2 and 3 in \citet{Samra2022_96b}). The reduced mixing efficiency serves to maintain the sub-solar C/O at this pressure level as the oxygen is not efficiently removed from the evaporation edge of the cloud base.



\section{The value of simplistic cloud models for observations}
\label{sec:simplistic value}



Cloud models of varying complexity are used to represent observational data for extrasolar planets. The self-consistent, complex cloud model used in this work  yields detailed cloud properties. In atmosphere retrievals, more simplified cloud models in form of a grey cloud prescription is generally used. In this section, both cloud model approaches are applied to understand how the complex cloud model can aid atmosphere retrieval to realise the full potential of JWST observational data. Section~\ref{subsec:full_opt_depth} identifies two distinct wavelength regimes longer and shorter than $\sim 4~\mu$m based on the disperse, mixed material cloud results from our kinetic model. Section~\ref{s:spectra} presents a synthetic spectrum for WASP-39b for $\lambda\lesssim4~\rm \mu m$ where the cloud may be reasonably well represented by a grey opacity, and Sect.~\ref{ss:os} discusses the effect of a homo-disperse, homo-material cloud on the cloud opacity which is particularly strong for $\lambda>4\mu$m.

\subsection{Disperse, mixed material cloud opacity}
\label{subsec:full_opt_depth}


Following \cite{Helling2022} and \citet{Samra2022_96b}, the atmospheric gas pressure, $p_{\rm gas}$, is explored where the cloud reaches optical depth $\tau(\lambda)=1$ as function of wavelength, $\lambda$.

The optically thick pressure level aligns with the cloud top pressure which many of the simple grey cloud deck approaches use for fitting spectra. Further, the pressure level at which the clouds become optically thick is a result of the complex cloud model.

\begin{figure}
    \centering
    \includegraphics[width=1.01\linewidth]{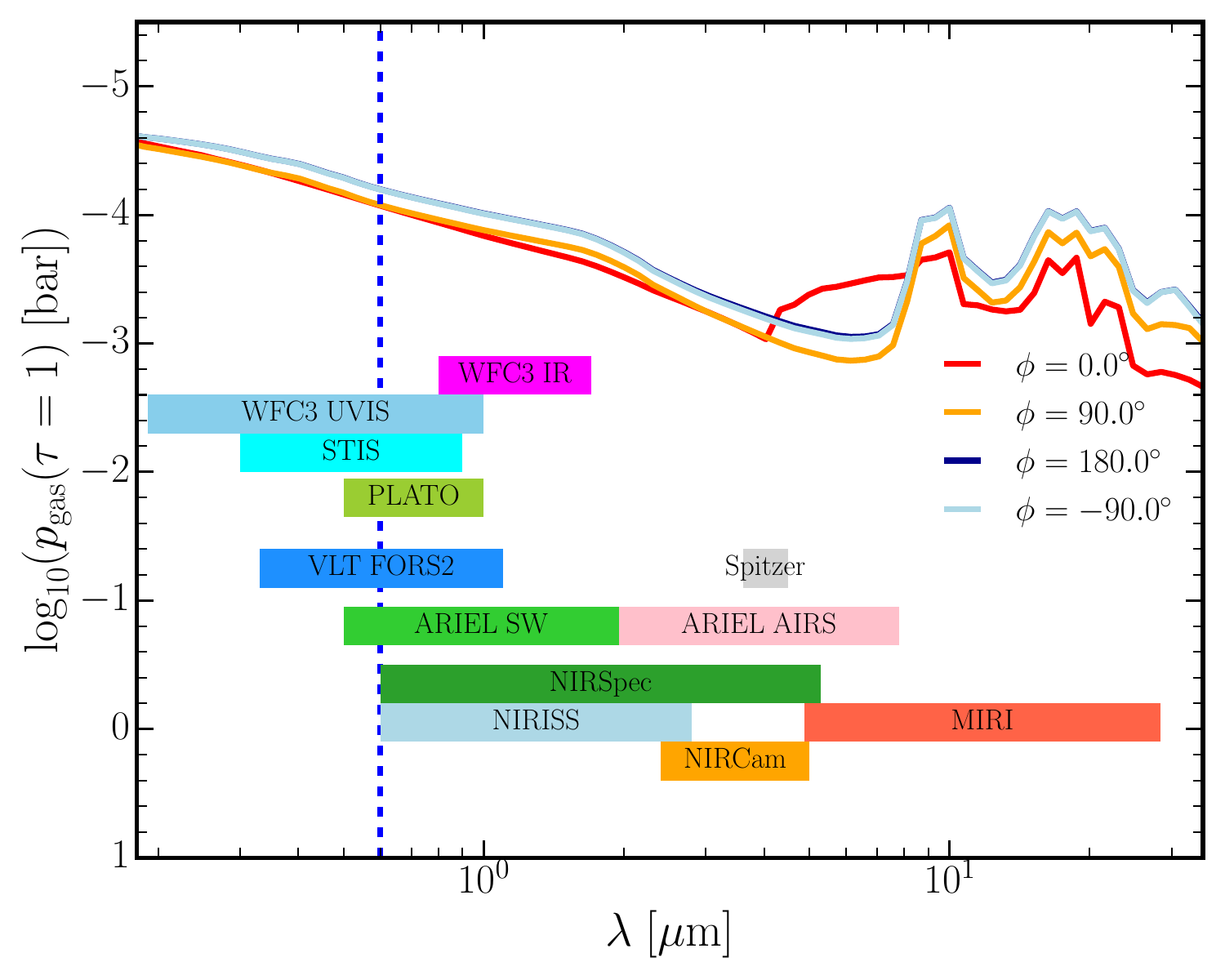}\\
    \includegraphics[width=1.01\linewidth]{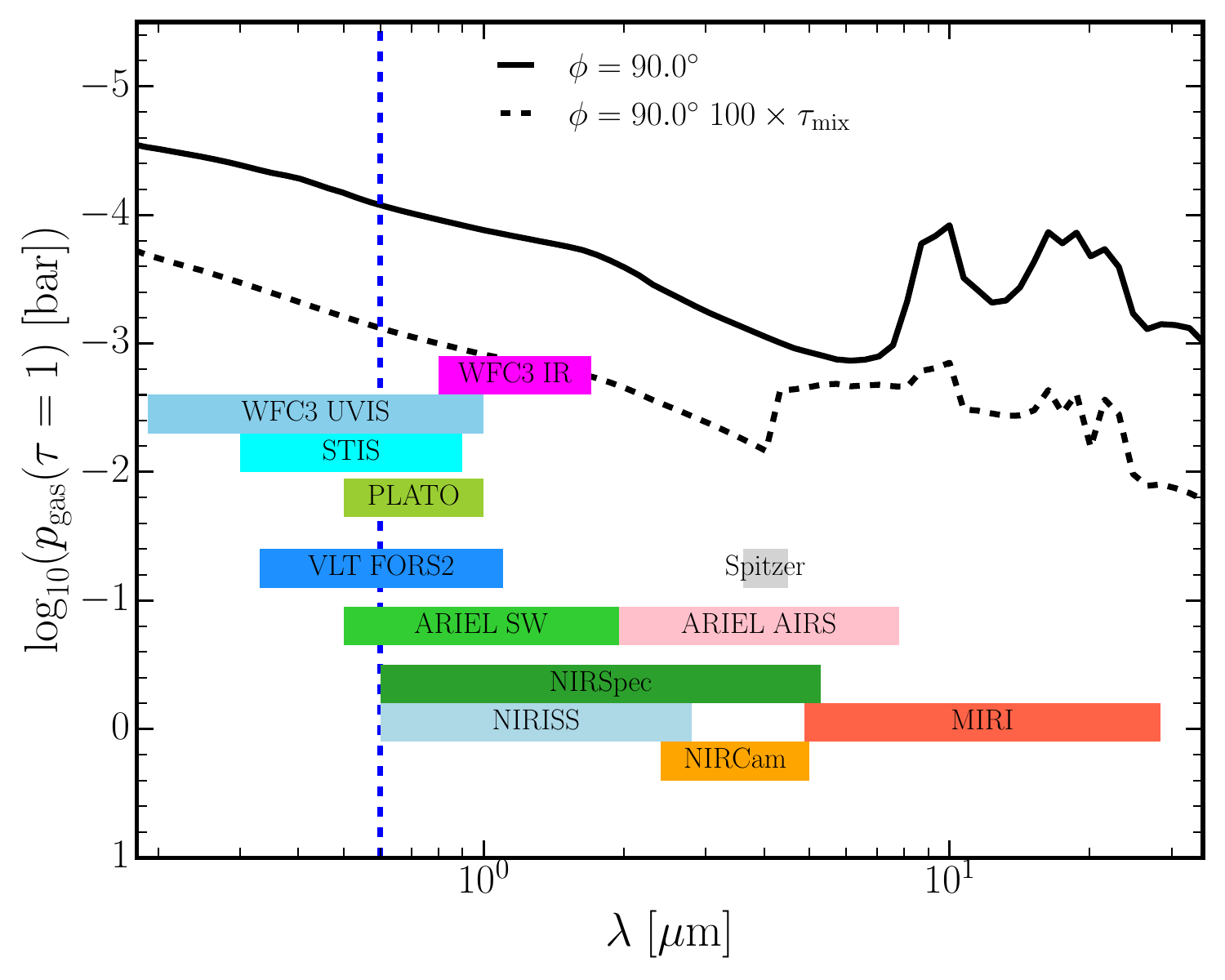}\\*[-0.3cm]
    \caption{WASP-39b pressure levels, $p_{\rm gas}(\tau(\lambda)=1)$ [bar],  at which the atmosphere becomes optically thick due to cloud opacity. \textbf{Top:} for  the substellar point ($\phi=0\degree$, red), antistellar point ($\phi=180\degree$, dark blue) and the equatorial morning ($\phi=-90\degree$, light blue) and equatorial evening terminator ($\phi=90\degree$, yellow). \textbf{Bottom:} for the equatorial evening terminator compared to the same profile but with the mixing timescale $100\times \tau_{\rm mix}$. The dashed vertical blue line is the location of Na, coloured bars show wavelength ranges of various instruments. }
    \label{fig:keypoints_optdepth}
\end{figure}

Figure~\ref{fig:keypoints_optdepth} demonstrates how the $\tau(\lambda)=1$ pressure level changes with wavelength if the cloud opacity is calculated using the full microphysical cloud model for WASP-39b as presented in Sect.~\ref{sec:cps}. The optically thick cloud pressure level, $p_{\rm gas}(\tau(\lambda)=1)$, is the result of the varying composition, particle size, and number density of cloud particles with pressure. Two distinct wavelength regimes may be identified in Fig.~\ref{fig:keypoints_optdepth}: $\lambda < 4\mu$m with no solid-material spectral features but a slope, and $\lambda > 4\mu$m where solid-material spectral features occur.

\smallskip
For $\lambda < 4\mu$m, $p_{\rm gas}(\tau(\lambda)=1)$ varies by $\sim$ 1.5 orders of magnitude in the wavelength bands that are accessible by HST/WFC3, HST/STIS, VLT/FORS2, JWST/NIRSpec, JWST/NIRCam and future missions like PLATO and Ariel. The pressure range $p_{\rm gas}\sim10^{-4.5}\ldots 10^{-3}$ bar that is probed at these wavelengths is characterised by cloud particles made of a mix of materials as shown in Table~\ref{tab:cloud_mix}. The dominating materials are Mg/Si/O and Fe/Si/O silicates with inclusions from various  materials, including further iron compounds like Fe[s] and FeS[s].


For $\lambda > 4\mu$m, the pressure range $p_{\rm gas} \sim 10^{-3}$bar is probed. This coincides, for the substellar point and evening terminator profiles, with chemically very active regions of the atmosphere, namely where the iron-silicates (for example, Fe$_2$SiO$_4$[s]), MgO[s] and SiO$_2$[s]) become thermally unstable and evaporate. Instead, Fe[s], SiO$_2$[s], Mg$_2$SiO$_4$[s] and MgSiO$_3$[s] increase their respective volume fractions considerably.This is the reason for the increase in the optical depth of the clouds for only the substellar point (between $\lambda = 4 \ldots 8\, {\rm \mu m}$). Therefore, assuming cloud particles observed at these wavelengths are made of only silicates is questionable.

While the assumption of uniform cloud composition is called into question here, the concept of optical depth $\tau=1$ is still useful to compare between the complex models, observational data and cloud parameterisation.

\subsection{Synthetic transmission spectra of WASP-39b}\label{s:spectra}

\begin{figure*}
    \centering
    \includegraphics[width=\linewidth]{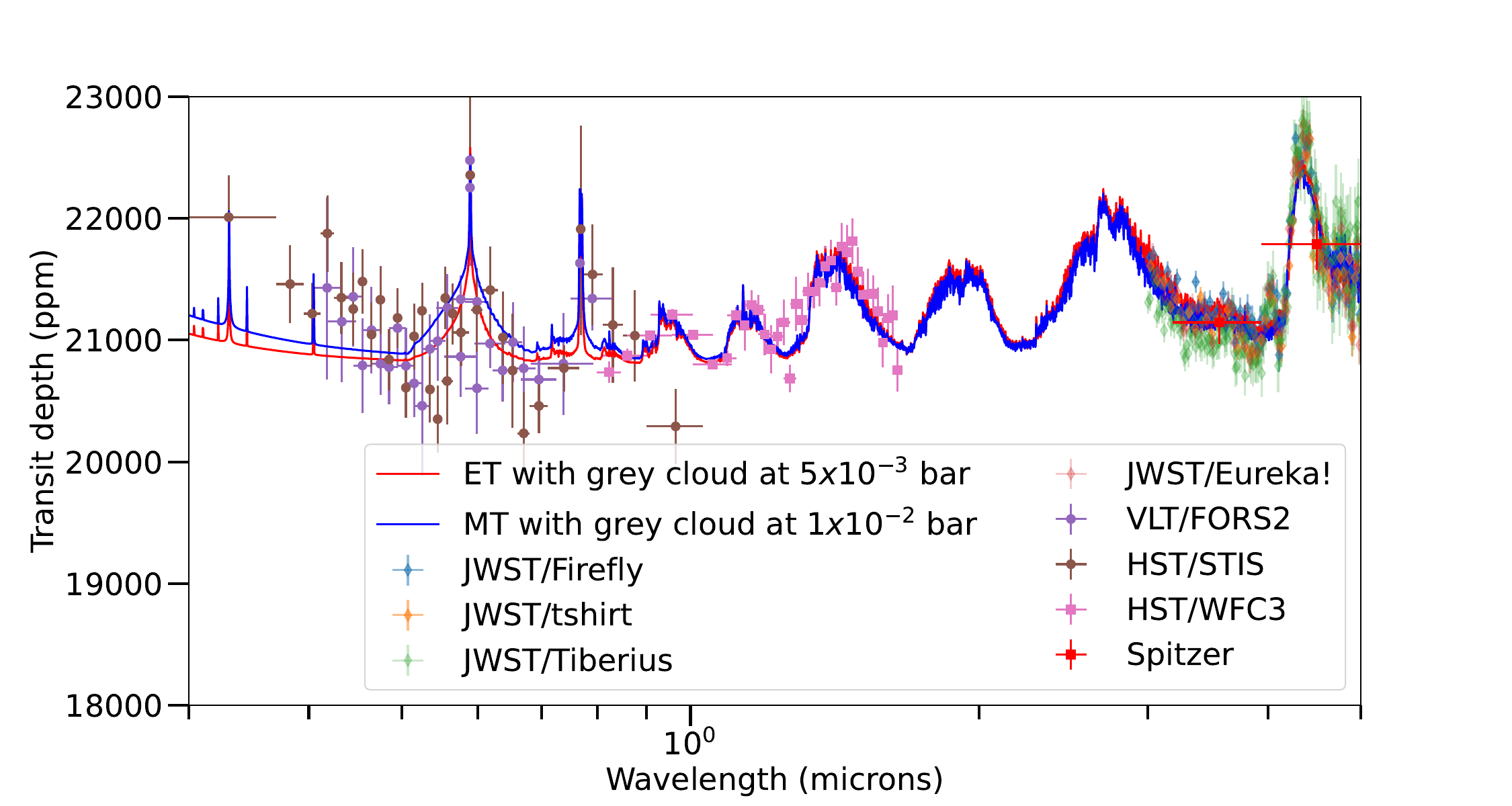}
    \caption{Synthetic spectra for the equatorial morning and evening terminators of WASP-39b compared to both pre-JWST and JWST observation for $\lambda=0.3\,\ldots\sim5.0~{\rm \mu m}$,
    computed with \textit{petitRADTRANS}. Opacities are considered using the concentrations of the dominant gas phase species output from the kinetic cloud model (\ce{H2O}, \ce{CO2}, \ce{CO}, \ce{H2S}, \ce{Na}, \ce{K})}
    \label{fig:spectra_gas}
\end{figure*}

A first exploration of the available observational data for WASP-39b for $\lambda<5~{\rm\mu m}$ is undertaken with the current, publicly available data, using a grey cloud model for the evening and morning terminator, respectively to derive insights between our complex cloud model and observations (Fig.~\ref{fig:spectra_gas}). The synthetic spectra shown in Fig.~\ref{fig:spectra_gas} are computed using \textit{petitRADTRANS} \citep{2019A&A...627A..67M} for the GCM (T$_{\rm gas}$, p$_{\rm gas}$)-profile for the equatorial morning and evening terminators, and their respective cloud-depleted gas-phase concentrations (see Fig.~\ref{fig:gas_comp}). The line opacities used for the spectra are \ce{CO} \citep{HITEMP2010}, \ce{CO2} \citep{20YuMeFr.co2}, \ce{H2O} \citep{HITEMP2010}, \ce{H2S} \citep{16Az.h2s}, Na \citep{19AlSpLe.broad}, and K \citep{16AlSpKi.broad}. A simple grey cloud deck is applied to both terminators separately. The synthetic spectra are compared to both pre-JWST observations and all the four different pipeline reduction of the JWST observations from \citet{JWST_39b_CO2}.  Systematic effects that lead to offsets between data make it difficult to compare observations from different telescopes.  
Here, this effect was taken into account by adding 100 - 150 ppm for individual data sets to achieve best fit with the model. \textit{Spitzer} data was treated as suggested in  \citet{JWST_39b_CO2}.

The grey cloud model produces generally a good fit with the data available for $\lambda=0.3\,\ldots\, 5\mu$m, but is unable to reproduce the slope in the optical which is associated with small cloud particles. The qualitative impact of the cloud opacity can be estimated by examining the parameter $x = 2\pi r/\lambda$, where $r$ is the radius of a spherical cloud particle and $\lambda$ is the observation wavelength. For $x\ll1$ Rayleigh scattering is the dominant. Using Fig.~\ref{fig:CI_nuc_rate} as a guide, taking $r\sim\langle \langle a \rangle_{\rm A} \rangle \sim 10^{-1.5}~{\rm \mu m}$ (representative of the morning terminator) and an observing wavelength of $\lambda=0.3~{\rm \mu m}$ yields a value of $x\approx0.66$, hence, a Rayleigh scattering slope would be expected in the optical wavelength regime based on the complex cloud model (see  Fig.~\ref{fig:keypoints_optdepth}). Figure~\ref{fig:spaghetti_clouds} shows that the population of cloud particles is expected to be smaller than the column integrated value in the upper atmosphere, and thus supports the expectation of an optical slope.

The location of the grey cloud to  enables another comparison to  results from the complex cloud model. Grey cloud decks at $p_{\rm gas}\sim10^{-2}~{\rm bar}$ and $p_{\rm gas}\sim5\times10^{-3}~{\rm bar}$ for the morning and evening terminators, respectively, appear to reproduce the pressure broadening required for the Na line in Fig.~\ref{fig:spectra_gas}. There is good agreement between both the synthetic terminator spectra and the observed \ce{CO2} feature at $4.3~\rm \mu m$, and a reasonable agreement with the observed \ce{H2O} features. New work of \citet{https://doi.org/10.48550/arxiv.2211.10488,https://doi.org/10.48550/arxiv.2211.10487,https://doi.org/10.48550/arxiv.2211.10489,https://doi.org/10.48550/arxiv.2211.10490}  derive a cloud deck between $3\times 10^{-4} \ldots 10^{-2}$~bar qualitatively agreeing with the range of cloud decks derived here. These authors further noted that varying vertical opacity contribution is required. Horizontal differences in cloud coverage between the limbs is also mentioned as a possibility. As has been noted in Section~\ref{sec:column_int_props}, average cloud particle sizes can indeed differ between morning and evening terminators.

In Section~\ref{s:gas}, it was shown that H2S may be present in chemical equilibrium in significant quantities for WASP-39b. Thus. \ce{H2S} was included as an additional opacity source, but no clear impact on the simulated spectrum could be found for $\lambda < 5~\rm\mu m$. However, other molecules for which \ce{H2S} is a precursor may impact observations \citep{Zahnle_2009} as was discussed for AGB stars (see Sect.~\ref{sec:sulphur_gas}). The important role of \ce{H2S} and sulphur chemistry in WASP-39b has been recently confirmed by \citet{https://doi.org/10.48550/arxiv.2211.10490} who present a photochemical pathway from \ce{H2S} to explain the recent detection of \ce{SO2}.


 The inferred cloud deck pressure level from the complex cloud model can be derived for the Na feature at $\lambda\sim0.6~{\rm\mu m}$ that lies in the optical slope region (Fig.~\ref{fig:keypoints_optdepth}). The cloud deck is slightly higher in the atmosphere at $p_{\rm gas}\sim10^{-4.4}~{\rm bar}$ for the morning terminator than the evening terminator at $p_{\rm gas}\sim10^{-4.2}~{\rm bar}$. The width of the sodium feature seen in observations suggests, similar to WASP-96b, that the actual cloud deck must be deeper in the atmosphere than shown in Fig.~\ref{fig:keypoints_optdepth}. This also agrees with the synthetic  spectra fits, which require a deeper cloud deck, namely $p_{\rm gas}\sim10^{-2}\ldots 5\times10^{-3}~{\rm bar}$, to match observations.
 
To apply the lessons learned from WASP-96b to WASP-39b, it has to be taken into account that for WASP-96b solar metallicity is assumed whereas for WASP-39b a higher metallicity of 10 times solar is assumed. \citet{Samra2022_96b} illustrate in their Fig.~5 that an enhanced atmosphere metallicity of $10\times\varepsilon_{\rm solar}$ raises the cloud deck to higher altitudes by almost half an order of magnitude for their models of WASP-96b. They show that the altitude of the cloud deck is reduced by an order of magnitude when the mixing efficiency is reduced by a factor of 100. For WASP-39b, hence, the location of the grey cloud deck required to fit the evening terminator spectrum is broadly consistent with a required factor of $\sim100\times$ reduction in mixing efficiency (compare to Fig.\ref{fig:keypoints_optdepth}, bottom). Further the cloud deck for WASP-39b is slightly higher compared to WASP-96b which can be entirely explained by its 10 times higher metallicity. Thus, the cloud model is capable of meeting observations of both planets, by adjusting only one factor: vertical mixing. 

\subsection{The effect of over-simplification}
\label{ss:os}


It has been shown here that clouds are expected to form in WASP-39b given its temperature of $T_{\rm{eq}}\sim 1100$~K. A higher cloud mass is expected in WASP-39b compared to WASP-96b because a higher metallicity is assumed. Thus, retrieval approaches that treat atmosphere metallicity and cloud formation as competing processes, when in reality they are intrinsically related may lead to unrealistically high metallicity values (Sect.~\ref{ss:metal}). Further, cloud formation in the temperature range of $T_{\rm{eq}}\sim 1200$~K always tends to raise the C/O ratio in the remaining gas phase due to the silicate and metal oxides removing oxygen \citep[see e.g.][]{Helling2022}. Here, assuming a solar C/O ratio, the observed C/O ratio in the gas phase can be raised to C/O$ \sim 0.7$ (Fig.~\ref{fig:Gas_Abunds}). Thus, again, treating C/O ratios and cloud formation as unrelated to each other, may result in unrealistically low C/O ratios. Retrievals thus tend to  produce diverging results with respect to the metallicity and the C/O in exoplanet data interpretation. However, it is challenging to find the optimal degree of simplifications for complex processes like cloud formation. 

Again, $p_{\rm gas}(\tau(\lambda)=1)$ is used as a tool to demonstrate how assumptions like constant cloud particle sizes and homogeneous cloud particle composition may bias retrieval results. Figure~\ref{fig:keypoints_optdepth} illustrates why the use of over-simplified cloud models is problematic: the cloud particles in the optically thin region (above $p_{\rm gas}(\tau(\lambda) =1)$) are highly mixed, with no single condensate species contributing more than $\sim 20\%$.

At different wavelength ranges, different pressure level and thus different cloud materials are probed. This is readily apparent for the substellar point, which shows an excess in cloud opacity between 5 and 8 micron (Fig.~\ref{fig:keypoints_optdepth}).
Furthermore, as described in Section~\ref{s:spectra} the cloud top of grey clouds needs to be between $p_{\rm gas} = 10^{-2} \ldots 5\times 10^{-3}$ bar in order to fit the Na line pressure broadening. In the case of such a deeper cloud top, a different cloud composition becomes visible by observations: a mixed composition of MgSiO$_3$[s]/Mg$_2$SiO$_4$[s] ($\sim 45$\% and $\sim 30$\% respectively) at morning terminator. Similarly, lowering the cloud top will allow different local gas-phase chemistry to be probed.

\begin{figure}
    \centering
    \includegraphics[width=1.01\linewidth]{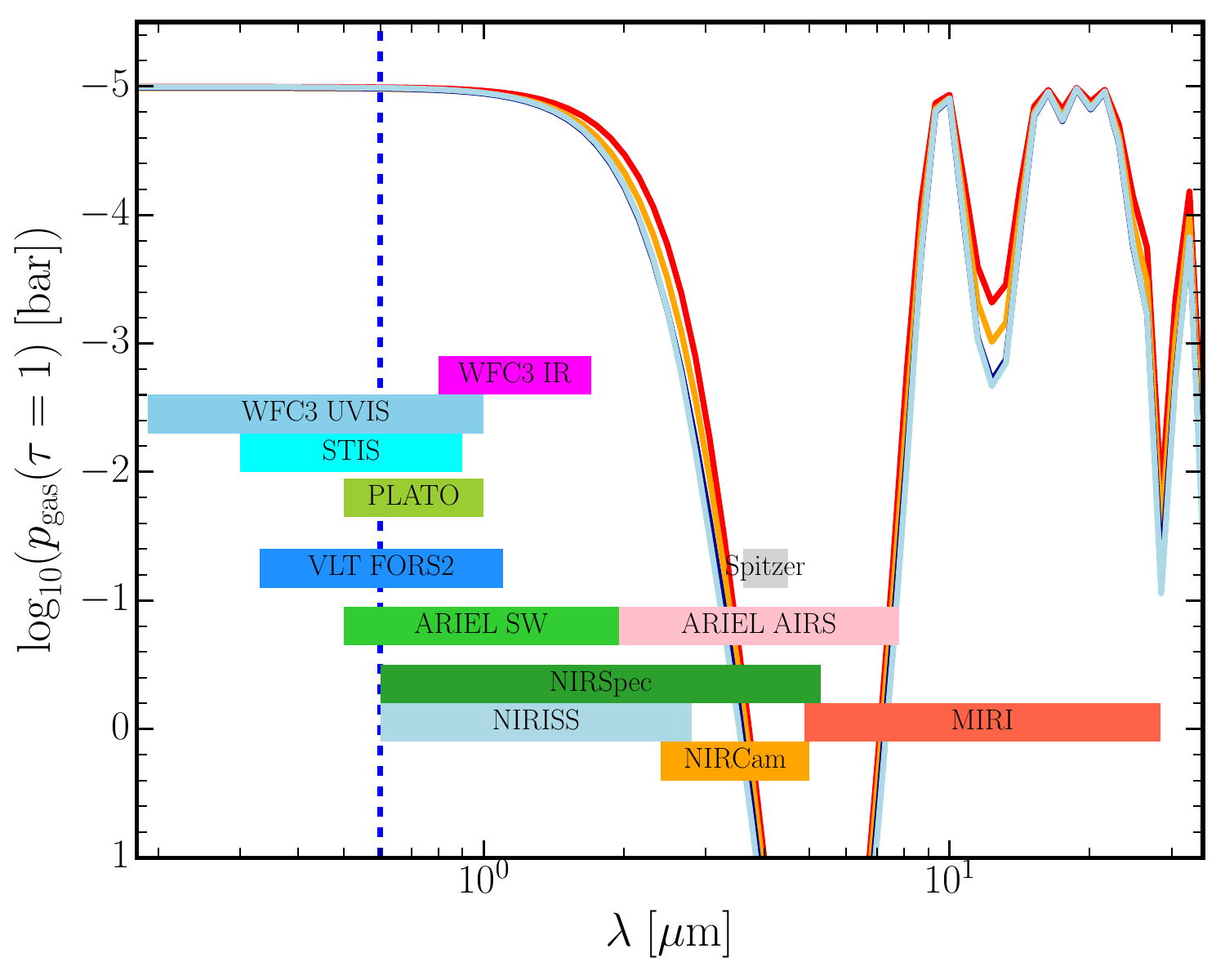}\\
    \includegraphics[width=1.01\linewidth]{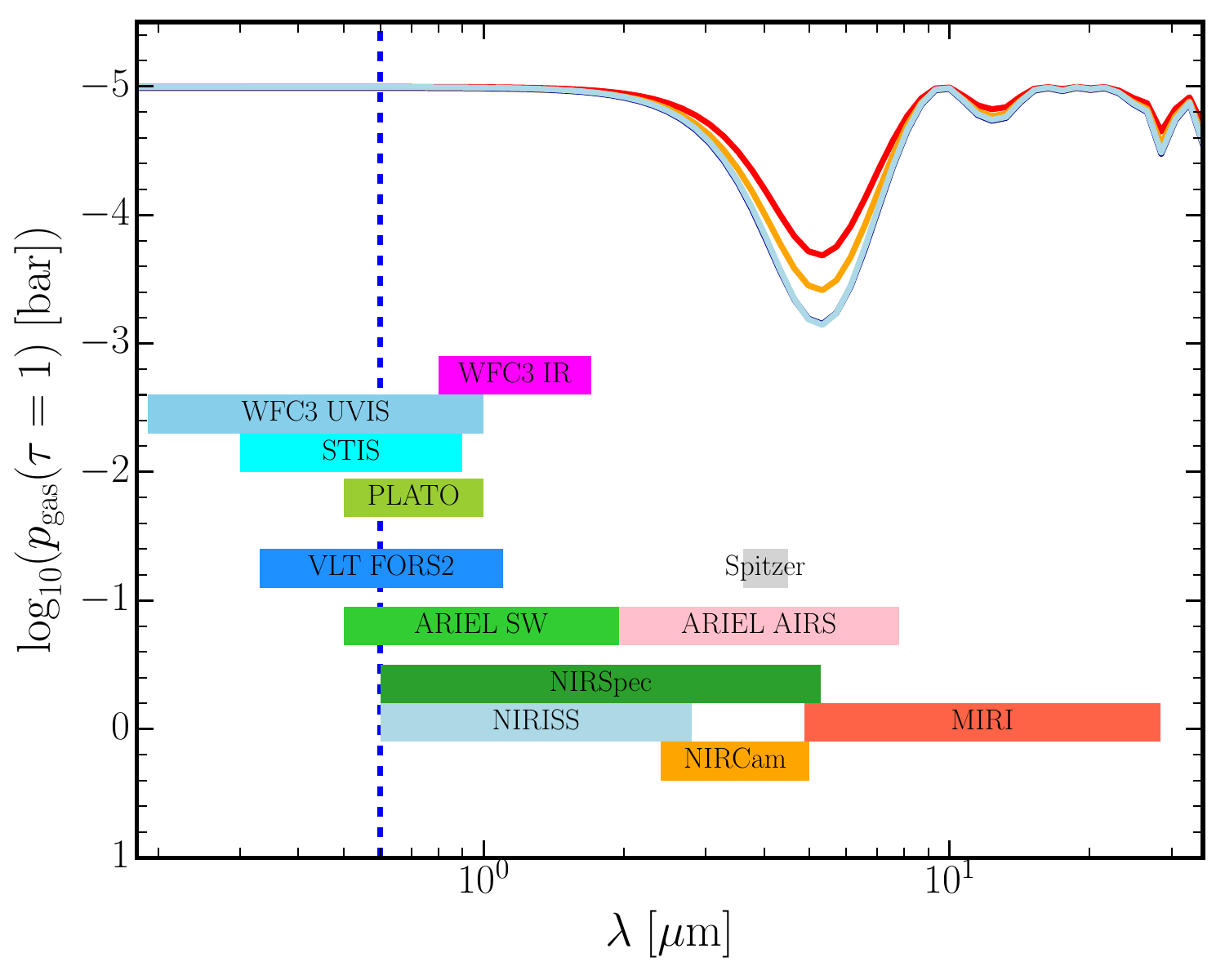}\\
    \includegraphics[width=1.01\linewidth]{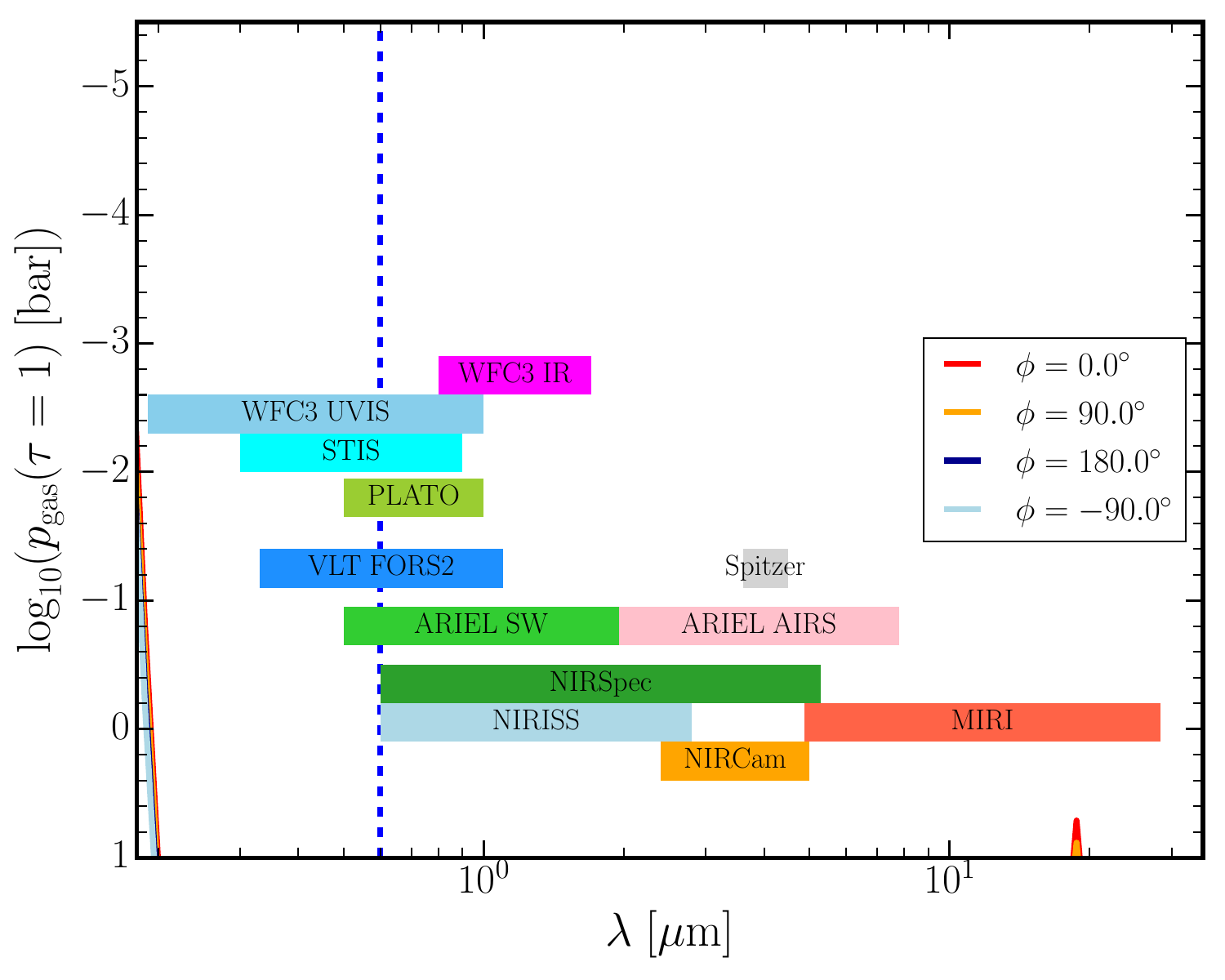}
    \caption{
    Pressure levels, $p_{\rm gas}(\tau(\lambda)=1)$ [bar],  at which the atmosphere becomes optically thick due to cloud opacity for WASP-39b, assuming constant, mono-disperse, cloud particles made of forsterite (\ce{Mg2SiO4}[s]). \textbf{Top:}  $\langle a \rangle_{A}=10^{-1}\,\mu{\rm m}$, $n_{\rm d}=10^{4}\,{\rm cm^{-3}}$, 
    \textbf{Middle:} $\langle a \rangle_{A}=10^{-1}\,\mu{\rm m}$, $n_{\rm d}=10^{5}\,{\rm cm^{-3}}$, \textbf{Bottom:} $\langle a \rangle_{A}=10^{-2}\,\mu{\rm m}$, $n_{\rm d}=10^{4}\,{\rm cm^{-3}}$ leaving the cloud optically thin.}
    \label{fig:forsterite_opt_depth}
\end{figure}

\newcommand{\Mixtable}[2]{ #1 & #2\\}
\begin{table}[!ht]
    \centering
    \caption{Material volume fractions, $V_{s}/V_{\rm tot}$,  for the evening terminator WASP-39b clouds in decreasing order where $p_{\rm gas} \approx 2\times 10^{-4}\, {\rm bar}$ representing where the cloud regions becomes optically thin.}
    \begin{tabular}{rl}
        \hline
        \hline
         \Mixtable{material}{$V_{s}/V_{\rm tot}$}
         \hline
         \Mixtable{Mg$_2$SiO$_4$[s]}{$\sim19$\%}
         \Mixtable{MgSiO$_3$[s]}{$\sim16$\%}
         \Mixtable{Fe$_2$SiO$_4$[s]}{$\sim16$\%}
         \Mixtable{MgO[s]}{$\sim13$\%}
         \Mixtable{FeS[s]}{$\sim7$\%}
         \Mixtable{SiO$_2$[s]}{$\sim7$\%}
         \Mixtable{SiO[s]}{$\sim6$\%}
         \Mixtable{FeO[s]}{$\sim5$\%}
         \Mixtable{CaSiO$_3$[s]}{$\sim5$\%}
         \Mixtable{Fe[s]}{$\sim3$\%}
         \Mixtable{Al$_2$O$_3$[s]}{$\sim2$\%}
         \Mixtable{Fe$_2$O$_3$[s]}{$\sim1$\%}
         \Mixtable{TiO$_2$[s]}{$<0.1$\% (Trace)}
         \Mixtable{CaTiO$_3$[s]}{$<0.1$\% (Trace)}
         \Mixtable{NaCl[s]}{None}
         \Mixtable{KCl[s]}{None}
        \hline
        \hline
    \end{tabular}
    \label{tab:cloud_mix}
\end{table}





\smallskip
How oversimplified cloud models effect the cloud opacity\ds{, particularly }for $\lambda>2\mu$m is shown in Fig.~\ref{fig:forsterite_opt_depth}. The effect of  simplifications in terms of constant particles sizes and homogeneous material properties is demonstrated. For this simplified model, two particle number densities are used ($n_{\rm d}=10^{4},\,10^{5}\,{\rm cm^{-3}}$), and two cloud particle sizes ($\langle a \rangle_{A}=10^{-2},\,10^{-1}\,{\rm \mu m}$). These values are based on the minimum and maximum values of cloud particle properties in the full microphysical model (Fig.~\ref{fig:comp_39_met}) in the optically thin pressure range ($10^{-4.5} \ldots 10^{-3}\,{\rm bar}$, Fig~\ref{fig:keypoints_optdepth}). Forsterite (Mg$_2$SiO$_4$[s]) was chosen as it is the largest volume constituent of the cloud particles in the optically thin pressure range (Fig.~\ref{fig:mat_comp_39b_indivdiual}).

The spectral fingerprint of the single material Mg$_2$SiO$_4$[s] becomes very apparent for small mean particle sizes of $0.1\mu$m a small cloud particle number density of $10^4$cm$^{-3}$ and diminishes with increasing number density of cloud particles (middle, $10^5$cm$^{-3}$). 


With $\langle a \rangle_{A}=10^{-1}\,{\rm \mu m}$, clouds are more optically thick than in the full microphysical case explored so far. However, even in such an optically thick case, 'windows' that allow to observe deeper parts of the atmosphere can appear for different cloud particle compositions. For example, by assuming a full forsterite (Mg$_2$SiO$_4$[s]) composition, there are optically thinner `windows' both between the 8 and 18~${\rm \mu m}$ silicate features. But also a completely optically thin (down to at least $10^{2}\,{\rm bar}$, not shown) window in the near- and mid-infrared ($\sim 4 \dots 7\,{\rm \mu m}$) for the lower number density case ($n_{\rm d}=10^{4}\,{\rm cm^{-3}}$).

\begin{figure}
    \centering
    \includegraphics[width=1.01\linewidth]{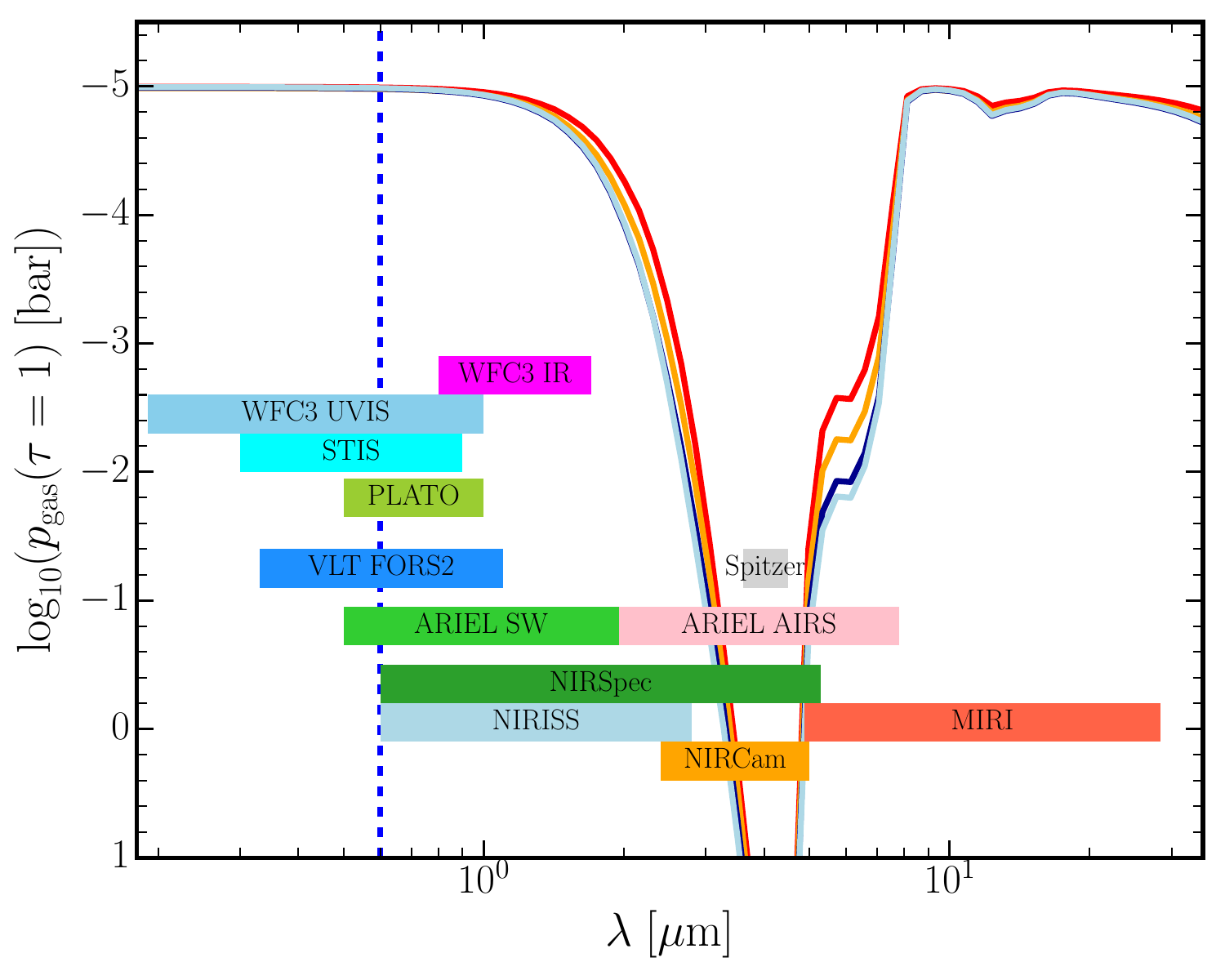}\\
    \includegraphics[width=1.01\linewidth]{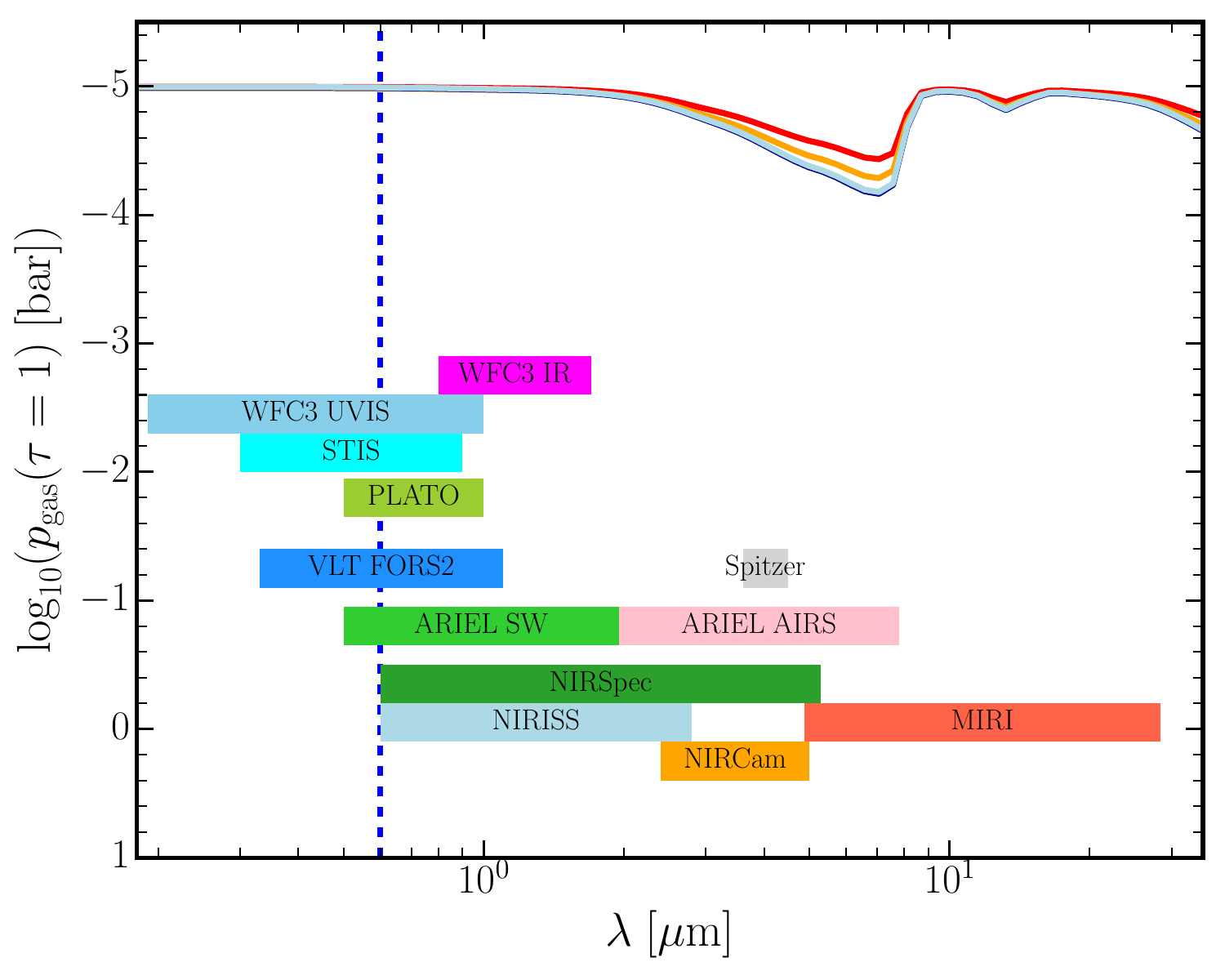}\\
    \includegraphics[width=1.01\linewidth]{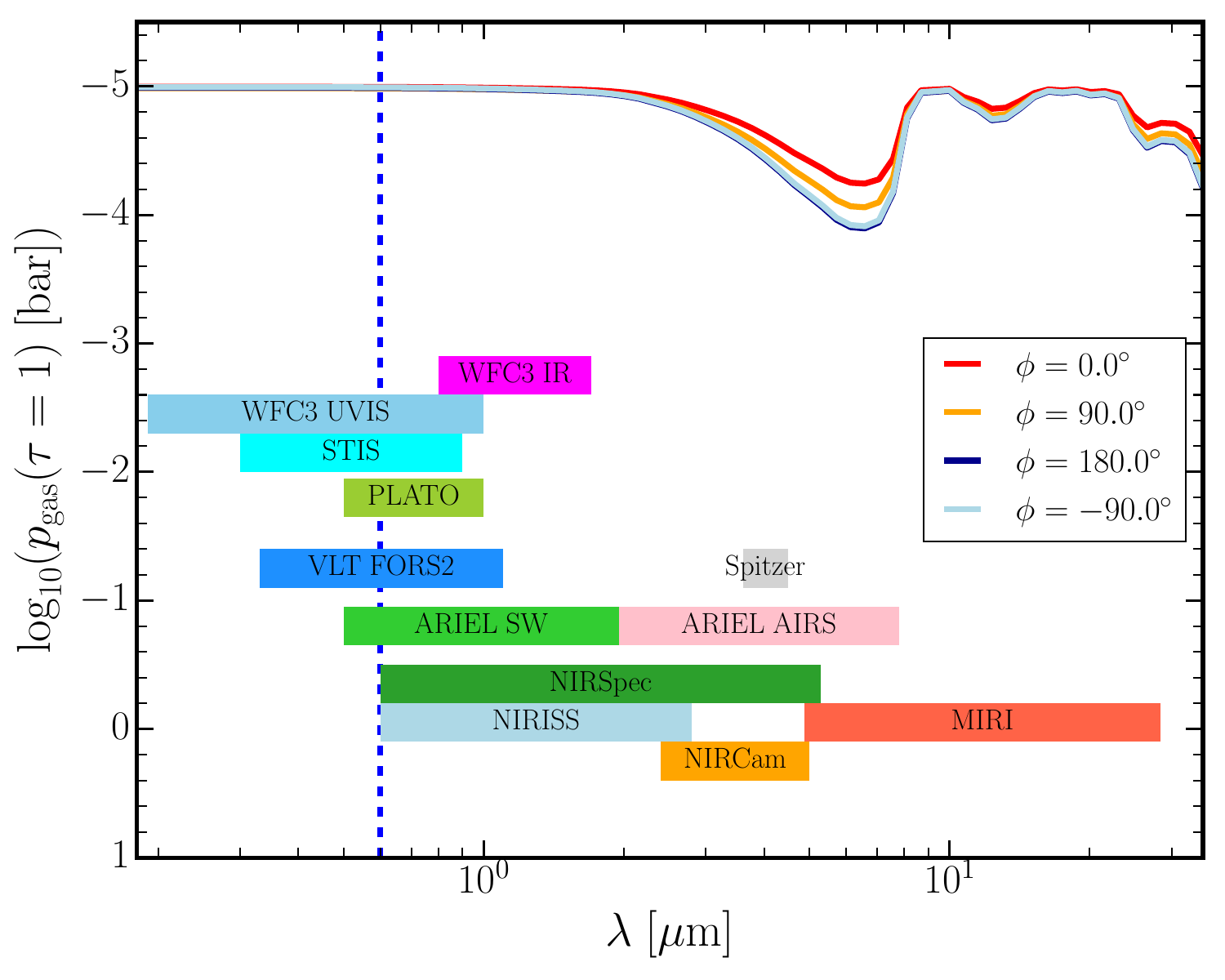}
    \caption{Pressure levels, $p_{\rm gas}(\tau(\lambda)=1)$ [bar],  at which the atmosphere becomes optically thick due to cloud opacity for WASP-39b, assuming constant, mono-disperse, cloud particles for various materials, where $\langle a \rangle_{A}=10^{-1}\,\mu{\rm m}$, $n_{\rm d}=10^{4}\,{\rm cm^{-3}}$ (comparable to the top of Fig.~\ref{fig:forsterite_opt_depth}). \textbf{Top:} For cloud particles made of MgSiO$_3$[s].
    \textbf{Middle:} For cloud particles made of Fe$_2$SiO$_4$[s]. \textbf{Bottom:} Mixed cloud composition based on $p_{\rm gas} = 2\times10^{-4}\,{\rm bar}$, see Table~\ref{tab:cloud_mix}.}
    \label{fig:materials_opt_depth}
\end{figure}

If instead a pure Fe$_2$SiO$_4$[s] or pure MgSiO$_3$[s] composition is assumed, the infrared spectral features would be very different (Fig.~\ref{fig:materials_opt_depth}) because of the clearly different refractory indices (Fig~\ref{fig:silicate_refind}). In particular, pure Fe$_2$SiO$_4$[s] cloud particles do not show the optically thin window at $\sim 4 \dots 7\,{\rm \mu m}$, as is the case for Mg$_2$SiO$_4$[s] and MgSiO$_3$[s]. This wavelength regime is especially sensitive to cloud material composition, size, and number density. There is a particular sensitivity between iron and magnesium silicates.

\ds{In addition, for wavelengths $>7\,{\rm \mu m}$ there is substantial differences between the features of all of the materials.} \ds{As the cloud particles in the observable atmosphere are highly mixed (see Table~\ref{tab:cloud_mix}), a simple model with this constant composition is also shown (Fig.~\ref{fig:materials_opt_depth}, bottom). From this, the mixing of infrared features between all the materials involved (including Fe[s]) can be readily seen, and hence the lack of an optically thin window in the full material composition optical depth (Fig.~\ref{fig:keypoints_optdepth}).}

The bottom of Fig.~\ref{fig:forsterite_opt_depth} shows that the $\tau = 1$ level of the clouds is also very sensitive to particle size ($\langle a \rangle_{A}\sim10^{-2}~{\rm \mu m}$). In this instance because of both the changing total cloud mass in this simplified model, but also Rayleigh scattering as discussed in Section~\ref{s:spectra}. The changing particle size with pressure is what causes the slope in optically thick pressure level; at different wavelengths, different depths in the atmosphere are where the cloud particles change scattering regime (from Rayleigh to Mie).


Taken together, what this simplified approach shows is the biases that are introduced when reducing the complexity of cloud particle size, number density, and material composition. As observational data across a broader wavelength range becomes available, problems may occur with the patchiness of the of the clouds. The sensitivity of near- and mid-infrared observations to the Fe/Mg composition of silicate cloud particles, which is being discussed for substellar atmospheres \citep[e.g.][]{Wakeford2015,Luna2021,Burningham2021}, is also difficult to assess without taking into account the possibility of highly-mixed composition. Fits using parameterised cloud opacities for material composition \citep{Kitzmann2018,Taylor2021} will be challenged by the mixed material composition of clouds, which changes with height in the atmosphere. Furthermore, retrievals of the same JWST/NIRSpec G395H ($3\ldots5~{\rm \mu m}$) observations of WASP-39b yield differing values of metallicity when specific condensate clouds are modelled in comparison to a grey cloud deck \citep{https://doi.org/10.48550/arxiv.2211.10488}.

\section{Conclusion}
WASP-39b, similar to WASP-96b, is cool enough that clouds are expected to form globally. A cloud-free atmosphere with $>100 \times\varepsilon_{\rm solar}$, as implied by previous studies for this planet \citep{2018AJ....155...29W}, is thus not consistent with the high efficiency of cloud formation in this temperature range. A cloudy and $\sim10\times\varepsilon_{\rm solar}$ atmosphere provides a better fit, consistent with recent JWST observations. We thus suggest that retrievals should add a cloud model by default. 

Application of a non-equilibrium cloud formation model further elucidates that simple grey cloud models - as used in atmosphere retrieval - do not capture the mixed composition cloud particles which are expected to form in the WASP-39b atmosphere. The cloud composition will vary throughout the atmosphere in response to the changing local thermodynamic conditions. Inclusion of the varying composition of clouds in exoplanet atmospheres may be required to fully interpret the wealth of observational data that is offered by current and future JWST observations.

The following points summarise the findings on the atmosphere structure and cloud composition of WASP-39b, as well as highlighting the care that is needed in interpreting simple cloud models used in retrievals:

\begin{itemize}
\item Hydrodynamic redistribution of irradiated heating is efficient enough in such cool objects that day-night temperature differences are not very large, therefore, WASP-39b is expected to be homogeneously covered in clouds.
\item The terminators inherit a temperature similar to either one hemisphere, the morning terminator is similar to the nightside and the evening terminator is similar to the dayside. Hence, trends in cloud properties at the terminators of WASP-39b are divergent between the terminators and similar to the influencing hemisphere. 
\item Cooler temperatures in Rossby vorticies at the nightside increase cloud formation compared to other locations at the same pressure level.  
\item The cloud composition in this temperature regime can be very heterogeneous, leading to vertical patchiness in terms of material composition. The cloud deck is characterised by an almost equal mixture of silicates and metal oxides, with a small fraction of high temperature condensates. The deeper atmosphere is dominated by an extended silicate cloud layer and a high temperature condensate cloud base.
\item Increased atmospheric metallicity enhances cloud mass and stabilises the deep extended silicate cloud layer to higher pressures. 
\item Increased metallicity does not qualitatively change the expectation of mixed composition upper cloud layers.
\item  Sulphur may be used to trace planet formation processes more easily than Fe, Mg, Si or even O since it is considerably less affected by condensation processes.
\item Similar to WASP-96b \citep{Samra2022_96b}, a reduced vertical mixing by approximately two orders of magnitude may be required to explain a cloud deck between $ 5 \times 10^{-3}$ bar and $ 1 \times 10^{-2}$ bar in a $10 \times\varepsilon_{\rm solar}$ metallicity WASP-39b atmosphere. 
\item Simplification of cloud microphysical properties can lead to biases in retrieval. These simplifications are: neglecting the pressure dependence of cloud particle size and number density. Further, a highly mixed material composition has a profound impact on cloud optical depth and observations. Understanding these biases will be especially crucial for near- and mid-infrared observations with JWST.

\end{itemize}

It is promising that adjustment of vertical mixing alone by the same factor yields for the complex model already good agreement with both, the WASP-96b and here the WASP-39b data. Thus, this work indicates that the microphysical cloud model can be adjusted using the new JWST data to yield better predictions of future observations and to act as a physically motivated background model to guide atmospheric retrievals. 

\begin{acknowledgements}
D.A.L. and D.S. acknowledge financial support from the Austrian Academy of Sciences. Ch.H., L.C. and A.D.S. acknowledge funding from the European Union H2020-MSCA-ITN-2019 under Grant Agreement no. 860470 (CHAMELEON). 
\end{acknowledgements}

\bibliographystyle{aa}
\bibliography{reference.bib}

\begin{appendix}


\section{GCM parameters and additional details}

\newcommand{\GCMtable}[2]{ #1 & #2\\}
\begin{table}[!ht]
    \centering
    \caption{Model parameters for the GCM used to produce 1D profiles for WASP-39b}
    \begin{tabular}{cc}
        \hline
        \hline
         \GCMtable{Parameter}{Value}
         \hline
         \GCMtable{Dynamical time-step $\Delta t$}{$25\,{\rm s}$}
         \GCMtable{Radiative time-step $\Delta t_{\mathrm{rad}}$}{$100\,{\rm s}$}
         \GCMtable{Stellar Temperature$^{1}$ ($T_{*}$)}{$5400\,{\rm K}$}
         \GCMtable{Stellar Radius$^{1}$ ($R_{*}$)}{$0.895\,R_{\rm sun}$}
         \GCMtable{Semi-major axis$^{1}$ ($a_{\rm p}$)}{$0.0486\,{\rm au}$}
         \GCMtable{Substellar irradiation temperature$^{2}$ ($T_{\rm irr}$)}{$1580\,{\rm K}$}
         \GCMtable{Equilibrium temperature$^{2}$ ($T_{\rm eq}$)}{$1117\,{\rm K}$}
         \GCMtable{Planetary Radius$^{1}$ ($R_{\rm p}$)}{$1.27\,R_{\rm Jup}$}
         \GCMtable{Specific heat capacity at constant pressure$^{3}$ ($c_{\rm p}$)}{$13165\,{\rm J\,kg^{-1}\,K^{-1}}$}
         \GCMtable{Specific gas constant$^{3}$ ($R$)}{$3456\,{\rm J\,kg^{-1}\,K^{-1}}$}
         \GCMtable{Rotation period ($P_{\rm rot}$)}{$4.06\,{\rm days}$}
         \GCMtable{Surface gravity ($g$)}{$430\,{\rm cm\,s^{-2}}$}
         \GCMtable{Lowest pressure ($p_{\rm top}$)}{$10^{-5}\,{\rm bar}$}
         \GCMtable{Highest pressure ($p_{\rm bottom}$)}{$700\,{\rm bar}$}
         \GCMtable{Vertical Resolution ($N_{\rm layers}$)}{$47$}
         \GCMtable{Wavelength Resolution$^{4}$ (S1)}{$11$}
        \hline
        \hline
    \end{tabular}
    \\ Notes: 1. Values taken from \cite{39b_discovery}, 2. Calculated using Eq.~1 from \citet{10Guillot.exo}, 3. Inferred using petitRADTRANS equilibrium package for $10\times$ solar metallicity, 4. Same as \cite{Kataria2013} benchmarked in Appendix B. \cite{2022arXiv220209183S}
    \label{tab:GCM_params}
\end{table}


\begin{figure}
    \includegraphics[width=0.5\textwidth]{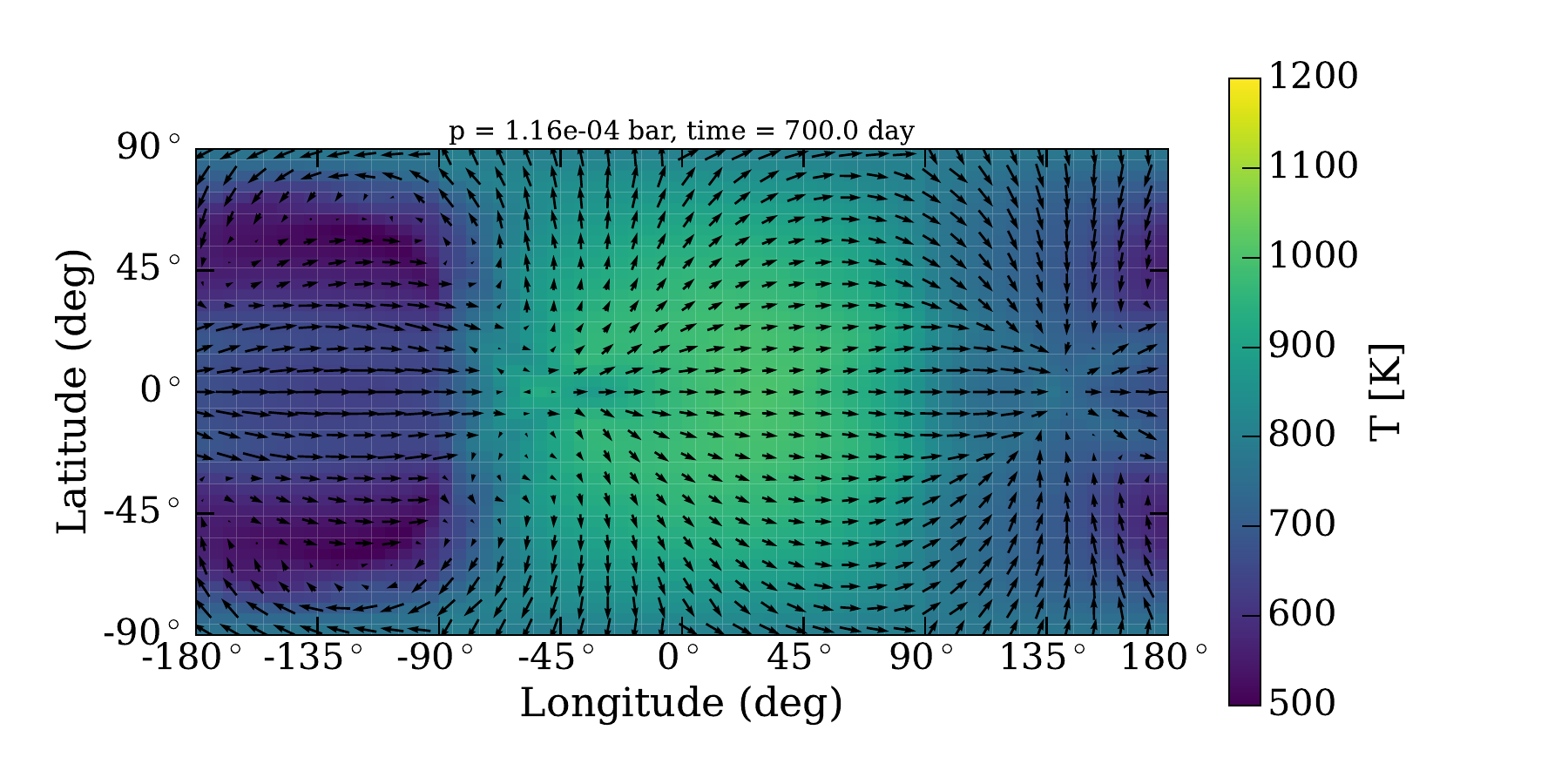}\\
    \includegraphics[width=0.5\textwidth]{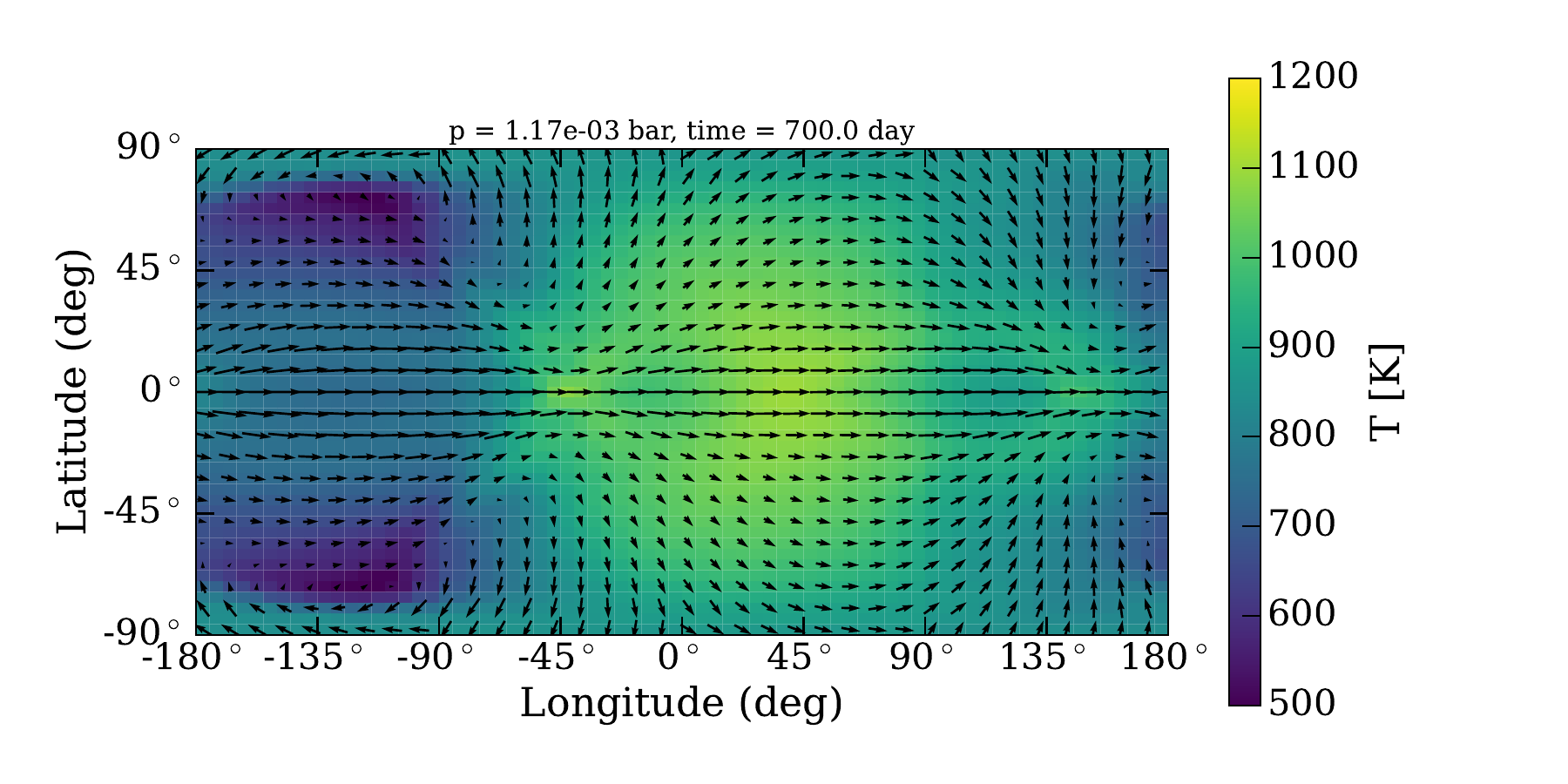}\\
    \includegraphics[width=0.5\textwidth]{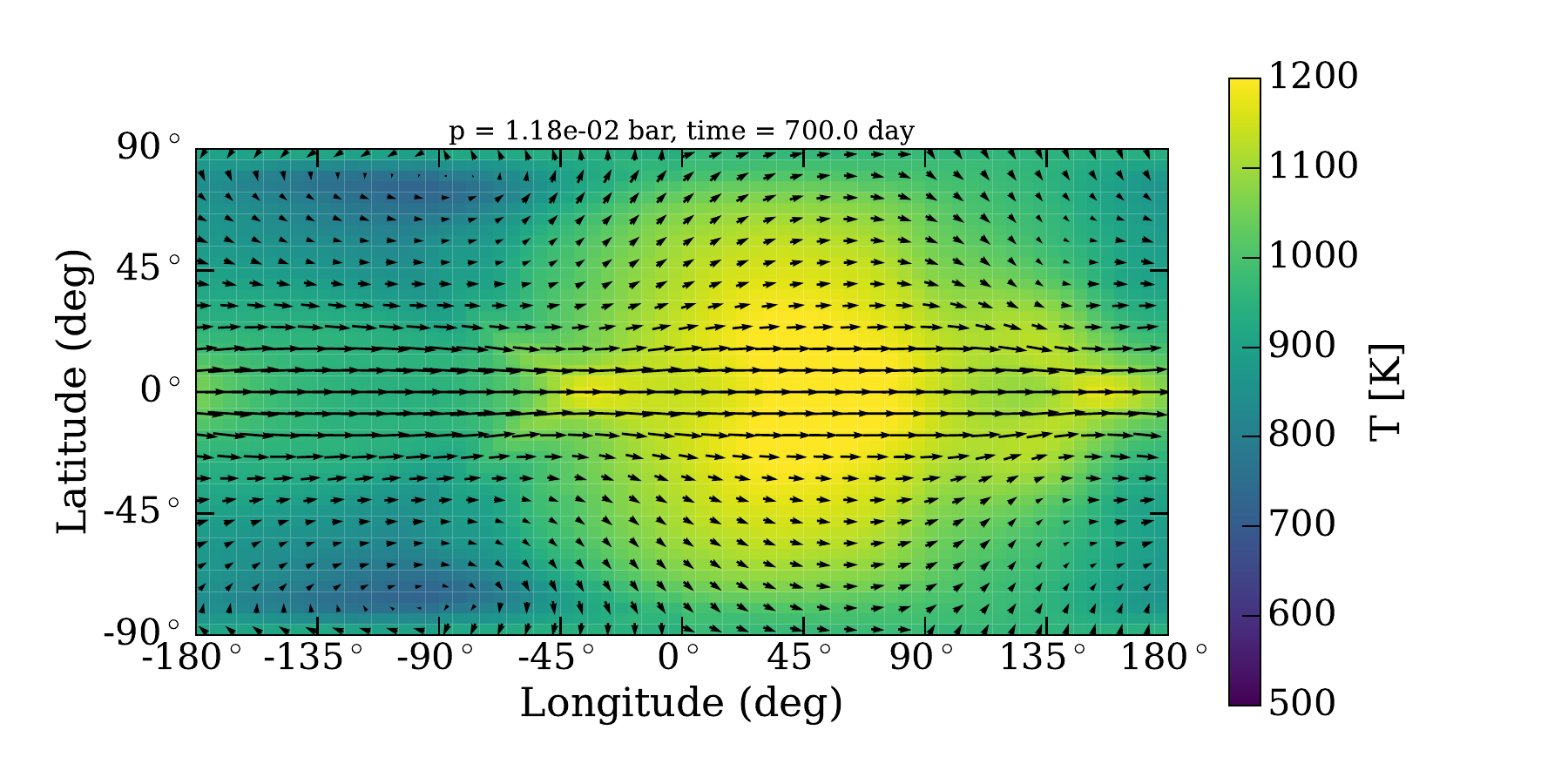}\\
    \caption{Isobaric maps of average gas temperature for $p_{\rm gas}\sim10^{-4},10^{-3},~{\rm and}~10^{-2}~{\rm bar}$. The Rossby vortices are seen clearly on the nightside around latitude = 68$^{\degree}$ at the $p_{\rm gas}\sim10^{-4}~{\rm and}~10^{-3}$ bar levels. \textbf{Top:} $p_{\rm gas}\sim10^{-4}$~bar. \textbf{Middle:} $p_{\rm gas}\sim10^{-3}$~bar. \textbf{Bottom:} $p_{\rm gas}\sim10^{-2}$~bar}
    \label{fig:gcm_maps}
\end{figure}

\begin{figure*}
    \centering
    \includegraphics[width=\textwidth]{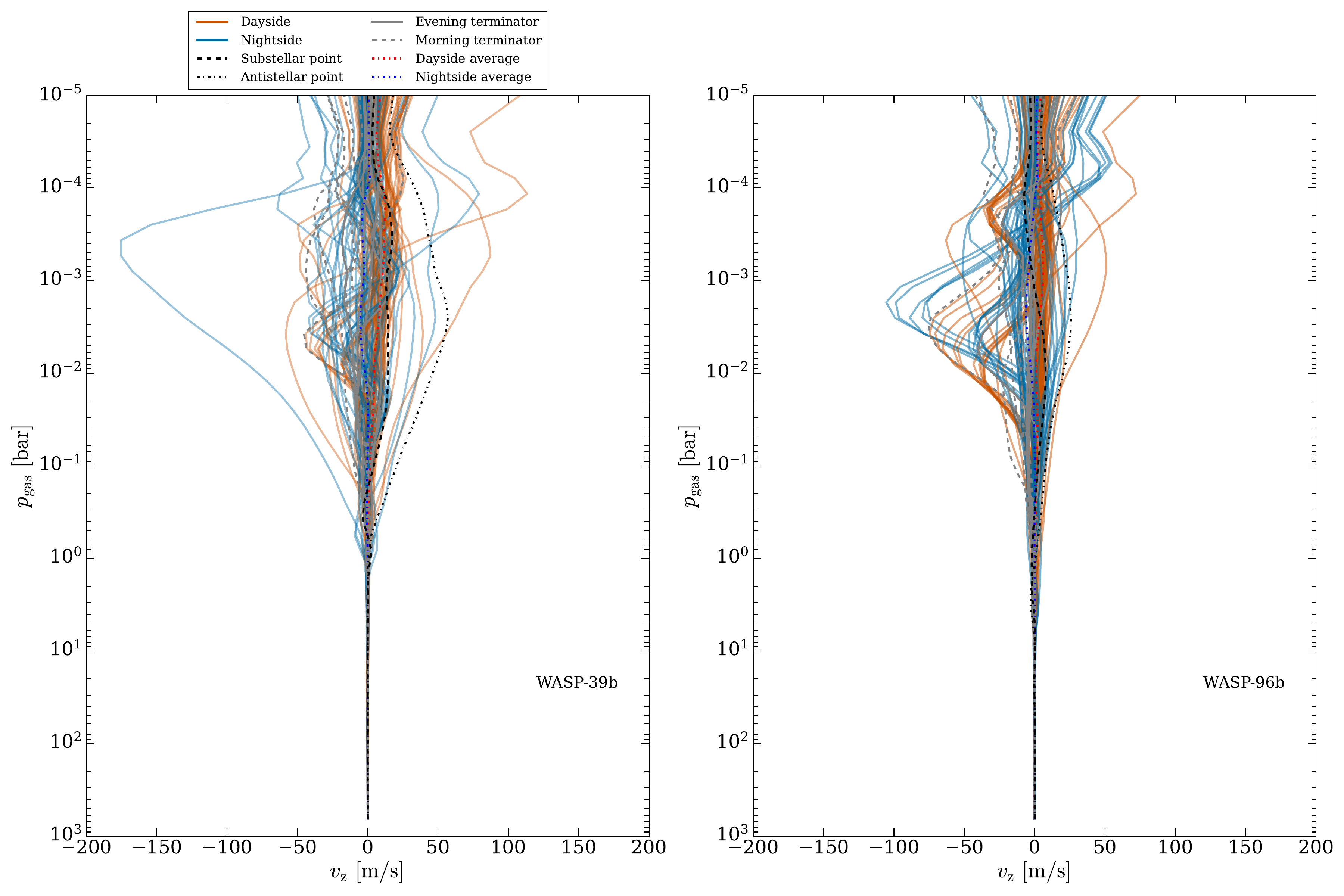}
    \caption{($v_{\rm z}$, p$_{\rm gas}$)-profiles extracted from the 3D GCM. \textbf{Left:} WASP-39b. \textbf{Right:} WASP-96b}
    \label{fig:Vvel_comp1}
\end{figure*}

\section{Cloud Condensate Refractive Indices}

\begin{table*}[t]
  \centering
   \caption{Cloud condensate refractive indices wavelength coverage and sources}
  \begin{tabular}{ccc}
     \hline\hline 
     Material Species & Reference & Wavelength Range $\rm (\mu m)$\\ \hline \hline
    \ce{TiO2}[s] (rutile) & \citet{TiO2_opt} & $0.47$--$36$\\
    \ce{SiO2}[s] (alpha-Quartz) & \citet{Palik1985}, \citet{SiO2_opt} & $0.00012$--$10000$\\
    \ce{SiO}[s] (polycrystalline)& Philipp in \citet{Palik1985} & $0.0015$--$14$\\
    \ce{MgSiO3}[s] (glass)& \citet{MgSiO3_opt} & $0.20$--$500$\\
    \ce{Mg2SiO4}[s] (crystalline)& \citet{Suto2006} & $0.10$--$1000$\\
    \ce{MgO}[s] (cubic)& \citet{Palik1985} & $0.017$--$625$\\
    \ce{Fe}[s] (metallic)& \citet{Palik1985} & $0.00012$--$285$\\
    \ce{FeO}[s] (amorphous)& \citet{FeO_opt} & $0.20$--$500$\\
    \ce{Fe2O3}[s] (amorphous)& Amaury H.M.J. Triaud (unpublished) & $0.10$--$1000$\\
    \ce{Fe2SiO4}[s] (amorphous)& \citet{MgSiO3_opt} & $0.20$--$500$ \\
    \ce{FeS}[s] (amorphous)& Henning (unpublished) & $0.10$--$100000$\\
    \ce{CaTiO3}[s] (amorphous)& \citet{Posch2003} & $2$--$5843$\\
    \ce{CaSiO3}[s] & No data - treated as vacuum & N/A\\
    \ce{Al2O3}[s] (glass)& \citet{Begemann1997} & $0.10$--$200$ \\
    \ce{C}[s] (graphite)& \citet{Palik1985} & $0.20$--$794$\\\hline\hline
  \end{tabular}
  \\Notes: 
  \label{tab:opt}
\end{table*}

\begin{figure*}
    \centering   
    \includegraphics[page=2,width=0.49\linewidth]{Figures/simple_optical_depths/SW_Material_Refractive_Indexes_APPENDIX_logscale_32mum_labelled.pdf}
    \includegraphics[page=14,width=0.49\linewidth]{Figures/simple_optical_depths/SW_Material_Refractive_Indexes_APPENDIX_logscale_32mum_labelled.pdf}\\
    \includegraphics[page=15,width=0.49\linewidth]{Figures/simple_optical_depths/SW_Material_Refractive_Indexes_APPENDIX_logscale_32mum_labelled.pdf}
    \includegraphics[width=0.49\linewidth]{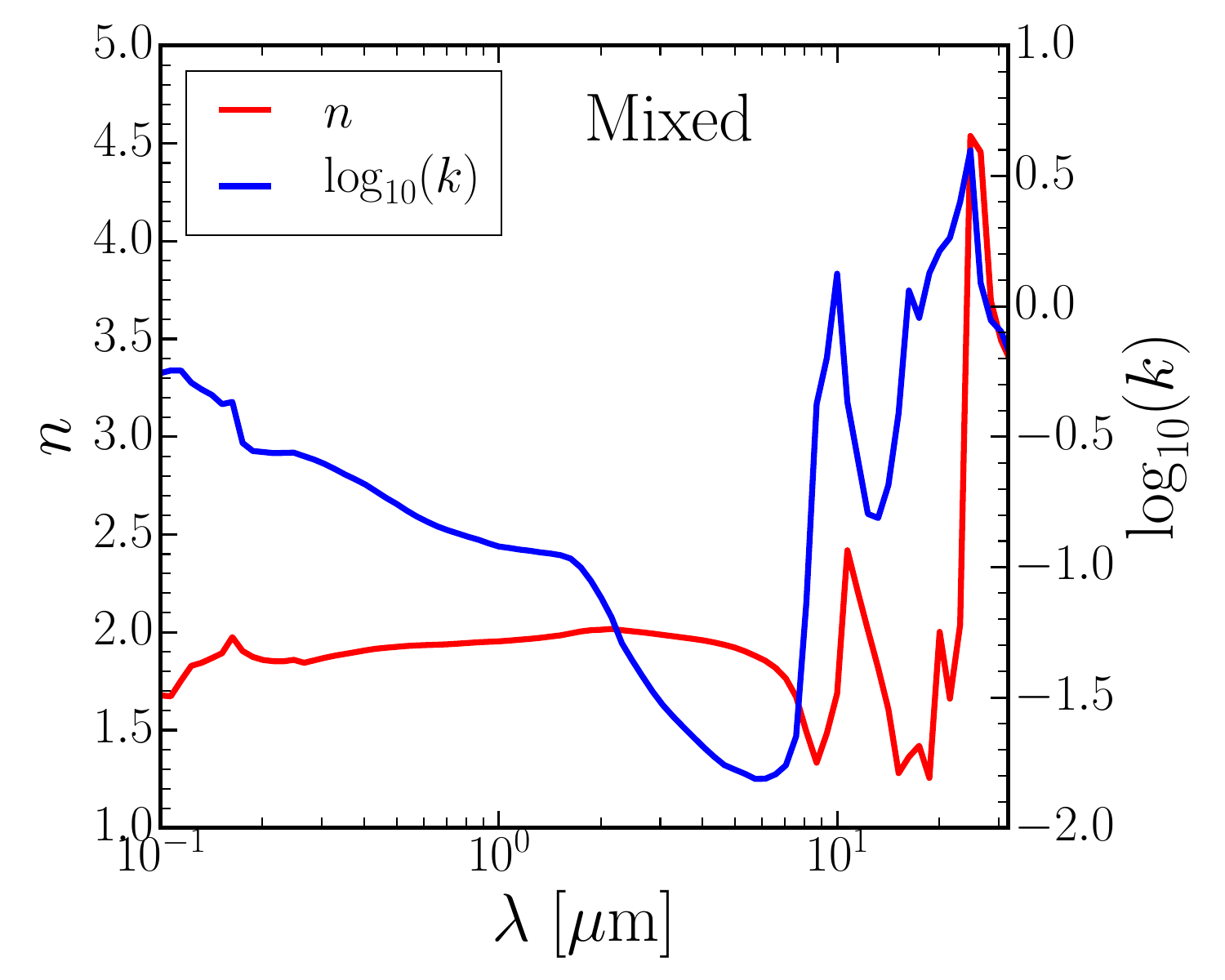}
    \caption{Refractive indices of silicate materials used \textbf{Top Left:} forsterite (\ce{Mg2SiO4}[s]), \textbf{Top right:} enstatite (MgSiO$_3$), \textbf{Bottom left:} fayalite (\ce{Fe2SiO4}[s]), \textbf{Bottom right:} mixed composition cloud based on $p_{\rm gas} = 2\times10^{-4}\,{\rm bar}$, see Table~\ref{tab:cloud_mix}. Dashed lines indicate regions for which refractive indices were extrapolated. Extrapolation was done as according to \cite{Lee2016}.}
    \label{fig:silicate_refind}
\end{figure*}
\end{appendix}

\clearpage

\end{document}